\documentclass[11pt]{article}
\pdfoutput=1

\usepackage{jheppubmod}
\usepackage{hyperref}
\usepackage{mathtools}
\usepackage{psfrag}
\usepackage{array}
\usepackage{amssymb}
\usepackage{amsmath}
\usepackage{cancel}
\usepackage{braket}
\allowdisplaybreaks
\usepackage{mathrsfs}
\usepackage{wrapfig}
\usepackage{multicol}
\usepackage{feynmp-auto}
\usepackage[bb=pazo]{mathalpha}

\usepackage{amsthm}
\usepackage{graphicx}
\usepackage{caption}
\usepackage{paralist}
\usepackage{comment}
\usepackage[sorting=none]{biblatex}
\usepackage{dsfont}
\usepackage{tikz}

\newcommand{\secref}[1]{Section \ref{#1}}
\newcommand{\appref}[1]{Appendix \ref{#1}}
\newcommand{\figref}[1]{Fig. \ref{#1}}

\newcommand{\marrow}[5]{%
\fmfcmd{style_def marrow#1
expr p = drawarrow subpath (1/4, 3/4) of p shifted 6 #2 withpen pencircle scaled 0.4;
label.#3(btex #4 etex, point 0.5 of p shifted 6 #2);
enddef;}
\fmf{marrow#1,tension=0}{#5}}

\title{The black hole IR triangle for scalar QED}
\author{Fabiano Feleppa$^{\star}$, }
\author{Nava Gaddam$^{\dagger}$, }
\author{Nico Groenenboom$^{\ddag}$} 
\affiliation{$^{\star}$Dipartimento di Fisica ``E.R. Caianiello'', Universit\`a di Salerno, Via Giovanni Paolo II 132, I-84084 Fisciano, Italy \\ Istituto Nazionale di Fisica Nucleare, Sezione di Napoli, Via Cintia, 80126 Napoli, Italy \\ Center for Fundamental Physics, School of Physical Science and Technology, ShanghaiTech University, 393 Middle Huaxia Road, Shanghai 201210, China}

\affiliation{$^{\dagger, \ddag}$Institute for Theoretical Physics and Center for Extreme Matter and Emergent Phenomena,
Utrecht University, 3508 TD Utrecht, The Netherlands.}
\emailAdd{ffeleppa@unisa.it}
\emailAdd{ n.k.gaddam@uu.nl }
\emailAdd{ n.groenenboom@uu.nl }

\date{\today}

\abstract{We construct the black hole infrared triangle of scalar QED on a stretched Schwarzschild horizon at tree level. In the covariant phase space of the near-horizon Maxwell field, the soft charge is a corner term. The memory is the zero mode of the gauge-invariant radiative field along the generators. It is conjugate to the boundary gauge mode that labels the degenerate horizon vacua, and Gauss’ law fixes it by the flux of charge through the horizon. The Noether flux of the residual gauge transformation through the stretched horizon is not conserved by the scattering, because charged matter crosses the stretched horizon in both directions. The flux with its matter part removed is the soft charge, which generates the large gauge transformation. Its Ward identity is equivalent to a soft photon theorem in every radiative partial wave, which we derive from the near-horizon Feynman rules and, where these do not control the emission, from the field equations at leading order. The main difference from null infinity is that on a stretched horizon both radial branches of the field are independent boundary data. The hard charge is a radial flux. With both branches present it weights each leg by the direction of its radial motion, and the soft factor is a sign per leg. The endpoint data on the two branches give two boundary observables, the time-integrated transverse magnetic and electric fields. The field radiated by a scattering process contributes only to the first, where it cancels the change in the Coulomb memory of the charges, so that the soft charge is conserved. A charge that crosses the horizon enters both. We also discuss the infrared divergence, the dressing of the external states, and the fate of the memory in the Hartle--Hawking state, where it performs a random walk driven by the Hawking radiation.}

\begin{document}
\maketitle

\section{Introduction and Summary}\label{sec:intro}

Over the last decade the infrared structure of gauge theory and gravity in asymptotically flat spacetime has been reorganised around a single triangular equivalence \cite{Strominger:2017zoo}, whose three corners had each been known, in isolation, for a very long time. Soft theorems go back to the observation of Bloch and Nordsieck that the emission of arbitrarily long-wavelength quanta factorises off the hard process \cite{Bloch:1937pw}, sharpened by Low \cite{Low:1958sn} and given their universal form by Weinberg \cite{Weinberg:1965nx}. Asymptotic symmetries go back to the discovery of Bondi, van der Burg, Metzner and Sachs that the symmetry group of asymptotically flat gravity at null infinity is an infinite-dimensional enhancement of the Poincar\'e group \cite{Bondi:1962px, Sachs:1962wk}, enlarged further by superrotations \cite{Barnich:2009se}. The memory effect goes back to the realisation that a burst of gravitational radiation leaves a permanent relative displacement of freely falling detectors \cite{Zeldovich:1974gvh, Braginsky:1987kwo, Christodoulou:1991cr}, with an electromagnetic counterpart in the net velocity kick imparted to test charges \cite{Bieri:2013hqa}. That these are three faces of one structure was established by relating the supertranslation Ward identity to Weinberg's theorem and thence to memory \cite{Strominger:2013jfa, He:2014laa, Strominger:2014pwa, Pasterski:2015tva}, and in gauge theory by relating large gauge transformations at null infinity to the soft photon theorem \cite{He:2014cra, Kapec:2015ena}; the pattern extends to subleading orders \cite{Cachazo:2014fwa, Kapec:2014opa, Campiglia:2019wxe} and now underpins the celestial holography programme \cite{Pasterski:2016qvg, Raclariu:2021zjz, Pasterski:2021rjz}. Lurking behind all three is the much older fact that the $S$-matrix of these theories is infrared divergent, and even the resolution of that, by dressing the external states \cite{Chung:1965zza, Kibble:1968sfb, Kulish:1970ut}, has come to be seen as the same triangle from another side: the $S$-matrix between Fock states vanishes because the soft charge has to change, and the dressing is what changes it \cite{Kapec:2017tkm}.

It is natural to ask whether the same structure exists at the horizon of a black hole, the other null boundary of the exterior. In this paper we construct one at tree level: a black hole infrared triangle that has a new soft theorem near the horizon, Ward identities associated with asymptotic symmetries near the horizon, and a memory on the horizon at its three corners. Several aspects of the horizon have been studied in the literature. Symmetries and charges have been written down near horizons in several forms \cite{Hawking:2016msc, Hawking:2016sgy, Donnay:2015abr, Donnay:2016ejv, Averin:2016ybl, Chandrasekaran:2018aop, Grumiller:2019fmp, Adami:2020amw, Ashtekar:2021wld, Sheikh-Jabbari:2022mqi, Aggarwal:2023qwl, Ashtekar:2024bpi} and some proposals for horizon memory have been made \cite{Donnay:2018ckb, Rahman:2019bmk, Adami:2021nnf}. Ward identities corresponding to symmetries of the horizon have been studied and soft theorems that these Ward identities would imply have been postulated \cite{Cheng:2022xyr, Cheng:2022xgm, Mao:2023rca, Mao:2023dsy, Mao:2024phf}.\footnote{For soft amplitudes computed in other curved backgrounds see \cite{Klisch:2025mxn, Chattopadhyay:2025ano}.} However, soft theorems are a result of dynamics. Without a direct amplitude computation, there is no way to tell whether a horizon symmetry is a statement about scattering or about a foliation that no dynamics is obliged to respect. This is also why the question of whether the charges constitute soft hair recording what fell in has remained an argument about symmetry \cite{Hawking:2016msc, Mirbabayi:2016axw, Strominger:2017aeh, Bousso:2017dny}. We derive a new soft theorem in a black hole background directly from the dynamics and demonstrate its equivalence to the Ward identity derived from asymptotic symmetries near the horizon. Working in scalar QED on Schwarzschild, using the near-horizon effective theory of \cite{Gaddam:2020mwe, Gaddam:2020rxb, Gaddam:2021zka, Gaddam:2022pnb, Feleppa:2023eoi}, we construct the covariant phase space of the Maxwell sector on the horizon and find that the soft charge is its corner term and the memory the zero mode conjugate to the vacuum label; we compute the single soft photon amplitude from the near-horizon dynamics and find that it is the Ward identity of the soft charge, the generator of the large gauge transformation on the horizon.

\begin{figure}[h!]
    \centering
    \includegraphics[width=1\linewidth]{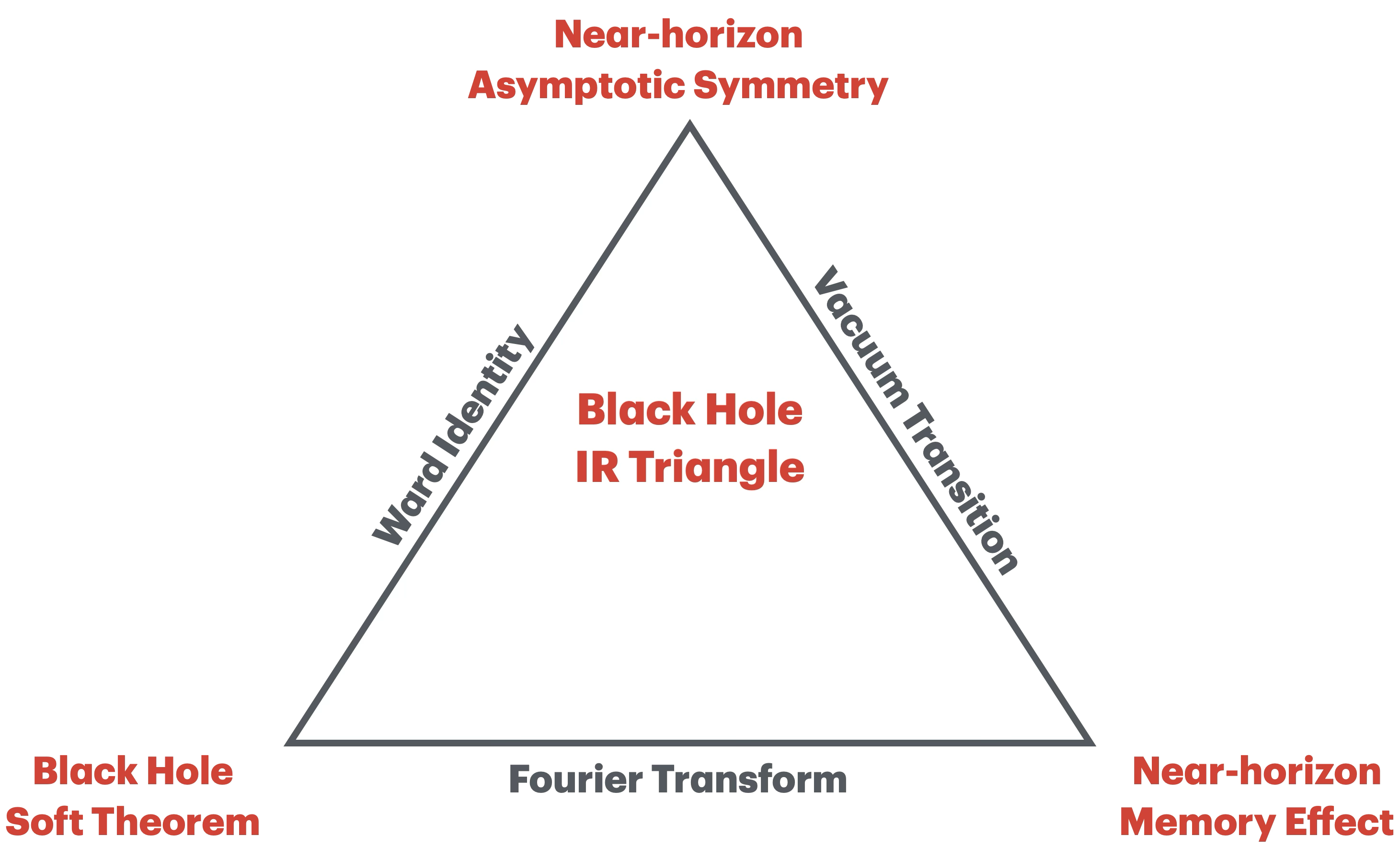}
    \caption{The black hole infrared triangle. Its corners are the near-horizon large gauge transformations of \secref{sec:symmetries}, with charges \eqref{eqn:softCharge} and \eqref{eqn:hardCharge}; the soft photon theorem \eqref{eqn:finalAmp}; and the horizon memory \eqref{eqn:memoryOperator}.}
    \label{fig:bhIRtriangle}
\end{figure}

We call this the \emph{black hole infrared triangle} in direct analogy to the infrared triangle in flat space (see \figref{fig:bhIRtriangle}). However, two features on which the flat-space construction is usually built are not available here. One is the soft pole, which belongs to the plane-wave basis and not to the boundary. The other is that at null infinity every external leg moves in one radial direction, which is a property of the boundary. The rest of this introduction is about what takes their place.

Running through the flat-space triangle is the soft pole. Weinberg's soft factor is a simple pole in the frequency of the emitted quantum, the memory is its residue, and in four dimensions the soft phase-space integral of the squared amplitude turns the pole into the logarithmic divergence of the $S$-matrix. The pole and the divergence are different things, though: in $D$ dimensions the soft integral goes as $\int \mathrm{d}\omega \, \omega^{D-5}$ and diverges only at $D \leq 4$, whereas the triangle does not depend on the divergence\footnote{For even dimensions see \cite{Kapec:2015vwa, Hollands:2016oma, Aggarwal:2018ilg}; the soft theorem and its Ward identity hold in all dimensions \cite{He:2019jjk}. In odd dimensions the memory was found to vanish \cite{Satishchandran:2017pek, Satishchandran:2019pyc}, and \cite{Raclariu:2026mfs} has recently argued that it is present at Coulombic order once the memory integral is taken before the large-$r$ limit.} and the soft factor $\varepsilon \cdot p_n / k \cdot p_n$ has its pole in all of them. The Ward identity is derived by extracting its residue, and the equivalence has accordingly come to look like a statement about the analytic structure of amplitudes. That appearance is tied to the plane-wave basis and the covariant normalisation of its oscillators. In the partial-wave basis in which the near-horizon theory is formulated, the one-photon state carries a factor of $\omega$ from the angular measure, the soft factor is a pure number per leg, and the amplitude carries only the canonical $\omega^{-1/2}$ of the external state. In this basis, the infrared divergence is the norm of the zero mode alone. What does distinguish null infinity from the black hole is an obvious property of the $S$-matrix: external states at past null infinity $\mathcal{I}^{-}$ are purely ingoing whereas those at future null infinity $\mathcal{I}^{+}$ are purely outgoing. Near the horizon, external legs come in both radial directions. Modes may travel along the horizon. But they may also fall into or emerge from the horizon.  

To see what the radial direction has to do with the boundary, reduce a massless field along the null direction and look at one frequency. There are two radial branches; relative to any null surface, one of them varies along the generators and meets the surface in its cuts, the other is constant along them. At $\mathcal{I}^+$ the branch along the surface is the ingoing one, a function of advanced time, which the limit that defines $\mathcal{I}^+$ pushes to infinity. So it carries no data there, and nothing is lost by this because regularity at the origin of flat space ties the two branches together.\footnote{Said differently: an ingoing mode reflects off the origin and becomes an outgoing one, so whatever it carries arrives at $\mathcal{I}^+$ on the outgoing branch.} On the true horizon the branch along it meets the same fate, but the near-horizon region is at finite proper distance from the horizon and the phase of that branch grows only logarithmically in it. So, on any surface just outside the horizon both branches are boundary data. With no regular origin behind the horizon, they are independent. Half of the radiative phase space on such a stretched horizon has no counterpart at null infinity, and the whole construction lives there.

In this article, we reduce the Maxwell field to partial waves, and each parity sector then carries one gauge-invariant master variable, the Moncrief variable $\Psi_{\ell m}$ \cite{Moncrief:1974am} in the even sector and $A^-_{\ell m}$ in the odd one, both obeying the same two-dimensional wave equation with a curvature-induced mass $\mu^2_\ell = f\left(r\right)\ell\left(\ell+1\right)/r^2$ that vanishes on the horizon. We build the covariant phase space (using the standard formalism \cite{Lee:1990nz, Iyer:1994ys, Wald:1999wa, Barnich:2001jy}) on a timelike surface of constant $r^*$ and take the horizon limit afterwards. This is how the corner terms survive. We derive the residual symmetry and its charge, split the charge into soft and hard parts, quantise the radiative mode on both branches, and compute the amplitude. One point needs care. In the half of the phase space where photon and hard leg move on the same radial branch, both numerator and denominator of the momentum-space soft factor vanish at zero mass, and the curvature masses fix the ratio to a value that does not depend on them. The masses depend on the radius, and holding them fixed in the computation is justified only when the collinear emission is completed over a distance short compared with the size of the hole, which fails close to the horizon and for frequencies below the Hawking scale $1/R$. Where it fails, the momentum-space computation does not control this half of the phase space. There we compute the radiated field from the field equation for the radial electric flux, at leading order near the process, and it gives the same soft factor on each branch with no reference to the masses. A complete derivation from the classical field equations in the full potential is left to future work.

The bulk symplectic form of the reduced Maxwell theory turns out to be invariant under the $SO(2)$ rotation of the two master variables into one another, and the whole asymmetry between the electric and magnetic sectors sits in a corner term at the two ends of the horizon, pairing the Moncrief variable against the boundary value of the pure-gauge mode; the odd sector has no gauge freedom and no such term. That corner term is the soft charge, and it identifies the memory as the zero mode $N_{\ell m} = \Delta\Psi_{\ell m}$ of the radiative data, conjugate to the boundary gauge coordinate $\varphi_{\ell m}$, one pair per radiative partial wave, labelling a family of degenerate horizon vacua. Gauss' law then fixes the memory by the change in the radial electric flux between the two ends of the horizon and the flux-weighted charge that crossed in between, on any state and in any sector. The Noether flux of the residual gauge transformation through the stretched horizon, with the Maxwell and the matter sector included, is the difference of the radial electric flux between the two ends of the horizon. At null infinity the corresponding flux yields a conserved charge, because the cut charge is taken to vanish at the far end. The stretched horizon is a timelike surface which charged matter crosses in both directions, and the flux through it is not conserved by the scattering. What is conserved is the flux with its matter part, the hard charge, removed. That is the soft charge, which generates the shift of the vacuum label. Its Ward identity is the statement that the memory of the in-state is the memory of the out-state. On states of hard scalars it equates the change in the flux-weighted charge, which the dynamics produces by turning charges around, to minus the change in the radial electric flux, which is carried by the soft photons the process radiates, and this is the soft theorem. As at null infinity, it is the soft theorem that establishes the identity, and with it the invariance of the near-horizon $S$-matrix under the shift. At $\ell = 0$ there is no soft charge, the flux is the hard charge alone, and the identity is trivial. No matching across the bifurcation sphere is assumed. The magnetic charge, consistently, has no hard component, and the magnetic soft theorem vanishes near the horizon.

The hard charge is the radial current integrated along the generators. It is this charge, not the charge on a slice, that enters the memory through Gauss' law. It acts on each hard leg with a sign set by the direction of that leg's radial motion, where a bulk gauge rotation restricted to the horizon treats every leg alike. The two differ by one sign per leg that is invisible at null infinity because every leg there is outgoing. Near the horizon the scattering does not conserve the flux-weighted charge, and the soft photons it radiates carry exactly the change. The memory accordingly has two parts, the zero mode of the radiated field and the Coulomb memory of the charges, and a scattering process changes them by equal and opposite amounts: free propagation changes neither, whereas a process that turns a charge around changes both and leaves their sum unchanged. The Coulomb memory is what a charge leaves by merely crossing the horizon, the time-integrated transverse field of its own electromagnetic shock wave. This is the electromagnetic counterpart of the Dray--'t Hooft shift \cite{Dray:1984ha, tHooft:1996rdg}, left by the crossing. Since the shock sits on the same branch the charge moves on, it separates the two boundary observables the horizon carries. The branch-symmetric combination of the endpoint data is the time-integrated transverse magnetic field, and it is the soft charge; the antisymmetric combination is the time-integrated transverse electric field, the velocity kick on a near-horizon charge, and it is not a charge at all. At null infinity, where the branch along the surface is empty, the two coincide and the memory is the kick \cite{Bieri:2013hqa, Pasterski:2015zua}. On the horizon the field a scattering process radiates, summed over its legs, has equal zero modes on the two branches, by charge conservation, and enters only the magnetic observable, while a crossing charge enters both. A detector near the horizon is kicked by whatever fell past it. The scattering registers no impulse on it, and is seen instead as a change in the radial electric flux through a cut of the horizon.

The dynamical soft factor, the vertex numerator over the propagator denominator, is a sign per leg and carries no power of the frequency; the $\omega^{-1}$ of the plane-wave factor is the Jacobian to partial waves. The infrared divergence does not depend on the basis: the squared soft factor has lost two powers of frequency, and the partial-wave measure has lost the same two. The spectrum on the horizon is gapless, and the photon number diverges logarithmically exactly as in four-dimensional flat space. In the partial wave basis, the divergence is nothing but the norm of the zero mode, the memory counted in photons. The horizon supplies a scale, the transverse mass $\sqrt{\ell\left(\ell+1\right)}/R$ of a partial wave on a sphere of finite area, which on a stretched horizon sets the resolution of the count of radiated soft photons and makes it finite and independent of the stretching, but it supplies no gap. The dressing that removes the divergence of the external states at leading soft order is a Wilson line in the boundary value of the pure-gauge mode, which gives each state a radiative zero mode equal and opposite to its Coulomb memory. And its photon number, summed over partial waves, turns the soft logarithm into a Sudakov double logarithm. The near-horizon charges are effectively massless and the transverse mass serves as the collinear regulator. In the plane-wave basis at null infinity the memory is the residue of the pole and the divergence its integral. In the partial-wave basis the amplitude has no simple pole at either boundary. Both are properties of the corner of the phase space, where the memory is the zero mode and the divergence is the orthogonality of the vacua it distinguishes, and that corner is the same at both boundaries.

There is a natural timescale associated with the memory depending on the state. For large parts of the paper, we work in the Boulware vacuum. The Ward identity and the soft theorem are operator statements and hold in any state. However, in the Hartle--Hawking state, the soft sector is thermal at the Hawking temperature.\footnote{\label{fn:unruh}For a black hole formed by collapse the appropriate state is the Unruh state, in which the modes emerging from the past horizon are thermal while those arriving from $\mathcal{I}^-$ are not. Near the horizon these are not the two branches of \secref{sec:externalstates}: the outgoing modes are partly reflected by the Schwarzschild potential back towards the horizon, so that the population of each branch depends on the reflection and transmission coefficients of the barrier. We do not treat this here. Statements in this paper about a thermal state are made for the Hartle--Hawking state.} The divergence of the undressed $S$-matrix becomes linear instead of logarithmic, and the memory performs a random walk with diffusion constant $T_H/\ell\left(\ell+1\right)$. What this means for the memory depends on its size. The memory of a single elementary charge is a displacement of the zero mode by $q/\ell\left(\ell+1\right)$ against quantum fluctuations of order $\left[\ell\left(\ell+1\right)\right]^{-1/2}$: in the Boulware state a single measurement resolves it only at order $q^2/\ell\left(\ell+1\right)$, which is the statement that one charge radiates fewer than one soft photon per partial wave over any observation time that is not exponentially long, and the bath lowers this further. The memory of a macroscopic coherent flux, $\left|\Delta\mathcal{Q}_{\ell m}\right|^2 \gg \ell\left(\ell+1\right)$, is resolvable, and in the Hartle--Hawking state it stays resolvable against the random walk for a Killing time of order $R\left|\Delta\mathcal{Q}_{\ell m}\right|^2/\ell\left(\ell+1\right)$, which is long compared with the thermal time by the same condition. 

The soft-hair proposal \cite{Hawking:2016msc, Hawking:2016sgy} and the dressed-state arguments against it \cite{Mirbabayi:2016axw, Bousso:2017dny} are both arguments about symmetry at a horizon modelled on null infinity. The horizon lacks two ingredients the flat-space argument needs, the identification of the two branches and a cold vacuum. Our results show that the dynamics, the branch structure and the state differ near the horizon in comparison to null infinity. Memory-carrying states at null infinity have been placed beyond the Fock representation \cite{Prabhu:2022zcr}, and on a Killing horizon the memory of a Maxwell field has recently been represented on an extension of the horizon Weyl algebra, in the Killing-invariant state, together with the decoherence it induces \cite{Danielson:2025aji}. That extension contains the memory and the unitaries that shift it. It does not contain the generator of the shift, which is not an operator in the vacuum representation, so the commutator of the memory with its conjugate enters only in exponentiated form. The corner pair found here is the classical version of that commutator: the memory and the vacuum label, with their Poisson bracket. Its quantum version is the memory together with the dressing, the Wilson line in the vacuum label. In the thermal state the random walk described above acts on this pair, and the time for which the memory remains resolvable from outside follows from it. The crossed-product algebras of the exterior and of the static patch \cite{Witten:2021unn, Chandrasekaran:2022eqq, Chandrasekaran:2022cip, Kudler-Flam:2023qfl}, and their extension to the boost supertranslations of the horizon and to the angle-dependent gauge transformations of long-range fields \cite{Klinger:2026tws}, restrict the radiative data to vanishing memory. What changes once the memory is admitted is a question we can now state precisely in the electromagnetic sector. We do not answer it here.

The paper is organised as follows. \secref{sec:maxwell} reduces Maxwell theory on Schwarzschild to partial waves and isolates the radiative mode. \secref{sec:symmetries} derives the horizon symmetries and their charges, splits the charge into soft and hard parts, writes down the Ward identity, and identifies the two radial branches of the external states. \secref{sec:softtheorem} computes the soft photon theorem from the near-horizon Feynman rules and checks it against the Ward identity, and \secref{sec:gaugeinvariance} contrasts the horizon symmetry with the restriction of the bulk one. \secref{sec:memory} identifies the memory and the two boundary observables, separates the scattering contribution from the Coulomb memory\footnote{We refer to the charge's own field as Coulomb memory and not to the field at a specific order in the $1/r$ expansion. These notions are compatible in the sense that Coulombic memory (of \cite{Raclariu:2026mfs} for instance) is flux-sourced at finite radius with no change in the radial electric field.} of the crossing charges, and compares the whole structure with null infinity. \secref{sec:irbehaviour} treats the infrared behaviour: the radiative window on a stretched horizon, the zero mode, the dressing, and the thermal state. We end in \secref{sec:discussion} with conclusions and outlook. The appendices contain the covariant phase space of the Maxwell sector, the quantisation of the hard and radiative modes, and the near-horizon Feynman rules.

\section{Maxwell fields in Schwarzschild spacetime}\label{sec:maxwell}
In this article, we study soft modes of Maxwell fields in a Schwarzschild background on the past and future horizons of the black hole. We will work with the Eddington-Finkelstein coordinates for a black hole of mass $M$
\begin{align}\label{eqn:metric}
    \mathrm{d}s^2 ~ &= ~ f\left(r\right) \left( - \mathrm{d}w^2\right) + 2 \eta_{wr^*} \mathrm{d}w \mathrm{d}r + r^2 \mathrm{d}\Omega^2 \quad \text{with} \quad f\left(r\right) ~ = ~ 1 - \dfrac{R}{r} \nonumber \\
    &= ~ f\left(r\right) \left( - \mathrm{d}w^2 + 2 \eta_{w r^*} \mathrm{d}w \mathrm{d}r^*\right) + r^2 \mathrm{d}\Omega^2  \quad \text{with} \quad \dfrac{\mathrm{d}r^*}{\mathrm{d}r} ~ = ~ \dfrac{1}{f\left(r\right)} \, .
\end{align}
Choosing $w = v$ or $w = u$ along with $\eta_{v r^*} = +1$ and $\eta_{u r^*} = - 1$ allows us to switch between ingoing and outgoing Eddington-Finkelstein coordinates respectively. Thereafter, an appropriate limit in $r$ allows us to further switch between the horizon ($r\rightarrow 2 G M$) and null-infinity ($r\rightarrow \infty$). Of course, setting $R = 2 G M = 0$ gives us flat space.

\subsection{Vector spherical harmonics and gauge freedom}

The gauge field can be expanded in vector spherical harmonics\footnote{In this article, we will use real spherical harmonics as opposed to the complex ones.} as
\begin{align}\label{eqn:gaugefieldharmonics}
    A_a ~ = ~ \sum\limits_{\ell m}A_a^{\ell m} Y_{\ell m} \qquad \text{and} \qquad A_A ~ = ~ \sum\limits_{\ell m}\left(A^+ _{\ell m} \eta_A^{\ell m} + A^-_{\ell m} \bar{\eta}_A^{\ell m} \right) \, ,
\end{align}
where the lowercase index $a$ runs over the longitudinal coordinates $w$ and $r$ or $r^{*}$, the uppercase index $A$ runs over the transverse sphere directions. Furthermore, $\eta_A$ and $\bar{\eta}_A$ are even and odd parity eigenvectors, respectively, of the sphere Laplacian operator. They are given by
\begin{align}
    \eta_A^{\ell m} ~ = ~ \partial_A Y_{\ell m} \qquad \text{and} \qquad \bar{\eta}_A^{\ell m} ~ = ~ - \epsilon_A^{~ B}\partial_B Y_{\ell m}  \, .
\end{align}
Both vanish for $\ell = 0$, since $Y_{00}$ is constant. The sum in $A_A$ therefore starts at $\ell = 1$. There is no $A^{\pm}_{00}$, and at $\ell = 0$ the gauge field has only the longitudinal components $A^{00}_a$. We have that the unit two-sphere Laplacian acts on these eigenvectors as
\begin{align}
    \Delta_{\Omega} \eta_A^{\ell m} ~ = ~ - \ell \left(\ell + 1\right) \eta_A^{\ell m} \qquad \text{and} \qquad \Delta_{\Omega} \bar{\eta}_A^{\ell m} ~ = ~ - \ell \left(\ell + 1\right) \bar{\eta}_A^{\ell m} \, .
\end{align}
A gauge transformation on the vector field reads as $\delta A_{\mu} ~ = ~ \partial_{\mu} \lambda$. Expanding $\lambda$ in spherical harmonics
\begin{align}\label{eqn:gaugeparamharmonics}
    \lambda ~ = ~ \sum \limits_{\ell m} \lambda_{\ell m} Y_{\ell m} \, ,
\end{align}
we find that the variations of the partial wave components of the vector field transform as
\begin{align}\label{eqn:gaugetransfo}
    \delta A_a^{\ell m} ~ = ~ \partial_a \lambda_{\ell m} \, , \quad \delta A^+_{\ell m} ~ = ~ \lambda_{\ell m} \quad \text{and} \quad \delta A^-_{\ell m} ~ = ~ 0 \, .
\end{align}
The first relation holds for every $\ell \geq 0$, the other two for $\ell \geq 1$. We see that the odd mode does not contain any gauge degrees of freedom. In what follows, we will write down the equations of motion for the fields in partial waves. And to identify radiative data, we will employ the analogue of the radiation gauge for gravitons \cite{Martel:2005ir, Gaddam:2022pnb}. In this gauge, $A^{\ell m}_w = 0$ and therefore the gauge parameter is constant in null time.

\subsection{Equations of motion}
The equations of motion are given by the inhomogeneous Maxwell equations in terms of the field strength tensor $F_{\mu\nu} = \partial_\mu A_\nu - \partial_\nu A_\mu$. We begin with the transverse components which can be written as
\begin{align}
    J_A ~ &= ~ \dfrac{g_{A \nu}}{\sqrt{-g}} \partial_\mu \left(\sqrt{-g} g^{\mu \rho} g^{\nu \sigma} F_{\rho \sigma}\right) \nonumber \\
    &= ~ \dfrac{\gamma_{A B}}{f\left(r\right) \sqrt{\gamma}} \partial_\mu \left(f\left(r\right) \sqrt{\gamma} \, g^{\mu \rho} \gamma^{B C} F_{\rho C}\right) \, ,
\end{align}
where we made use of the fact that $\sqrt{-g} = f\left(r\right) r^2 \sqrt{\gamma}$. This may further be simplified as
\begin{align}
    f\left(r\right) J_A ~ &= ~ \eta^{a b} \partial_a \left(F_{b A}\right) + \dfrac{f\left(r\right)}{r^2 } \hat{\nabla}^B \left(F_{B A}\right) \nonumber \\
    &= ~ \eta^{w r^*} \partial_{r^*} \left(\partial_w A_A - \partial_A A_w\right) + \eta^{r^* r^*} \partial_{r^*} \left(\partial_{r^*} A_A - \partial_A A_{r^*}\right) \nonumber \\
    &\qquad + \eta^{w r^*} \partial_w \left(\partial_{r^*} A_A - \partial_A A_{r^*}\right) + \dfrac{f\left(r\right)}{r^2} \hat{\nabla}^{B} \left(\hat{\nabla}_B A_A - \hat{\nabla}_A A_B\right) \, ,
\end{align}
where $\hat{\nabla}$ is the covariant derivative on the transverse two-sphere. We now expand the current and the gauge field in partial waves to find
\begin{align}\label{eqn:transverseEoM}
    f\left(r\right) J^+_{\ell m} ~ &= ~ \eta^{w r^*} \left(\partial_{r^*} \partial_w A^+_{\ell m} - \partial_{r^*} A_w\right) + \eta^{w r^*} \left(\partial_w \partial_{r^*} A^+_{\ell m} - \partial_w A_{r^*} \right) \nonumber \\
    &\qquad + \eta^{r^* r^*} \left(\partial_{r^*} \partial_{r^*} A^+_{\ell m} - \partial_{r^*} A_{r^*}\right) \nonumber \\
    &= ~ \eta^{a b} \partial_a \left(\partial_b A^+_{\ell m} - A^{\ell m}_b\right) \, ,
\end{align}
in the even-parity sector. Similarly, in the odd parity sector, we have a corresponding equation which reads:
\begin{align}\label{eqn:oddEoM}
    f\left(r\right) J^-_{\ell m} ~ &= ~ \left(\eta^{w r^*} \partial_{r^*} \partial_w  + \eta^{r^* r^*} \partial_{r^*} \partial_{r^*} + \eta^{w r^*} \partial_w \partial_{r^*}\right) A^-_{\ell m} {\epsilon_A}^B \partial_B \nonumber \\
    &\qquad + \dfrac{f\left(r\right)}{r^2} A^-_{\ell m} \hat{\nabla}^{B} \left(\hat{\nabla}_B {\epsilon_A}^C \partial_C - \hat{\nabla}_A {\epsilon_B}^C \partial_C\right) Y_{\ell m} \nonumber \\
    &= ~ \eta^{a b} \partial_{a} \partial_b A^-_{\ell m} - \dfrac{\ell \left(\ell + 1\right)}{r^2} f\left(r\right) A^-_{\ell m} \, .
\end{align}
In arriving at the second line, we used the covariant constancy and the anti-symmetry of the Levi-Civita symbol ${\epsilon_A}^B$ on the sphere. 

From the transverse even parity equation of motion \eqref{eqn:transverseEoM}, we see that its solution can be written as
\begin{equation}\label{eqn:AaSoln}
    A^{\ell m}_a ~ = ~ \partial_a A^+_{\ell m} + \chi^{\ell m}_a \quad \text{with} \quad \eta^{ab} \partial_a \chi_b ~ = ~ - f\left(r\right) J^+_{\ell m} \, .
\end{equation} 
The second constraint equation above ensures that the mode $\chi^{\ell m}_a$ only contains a single degree of freedom. To identify this degree of freedom, we introduce $\epsilon_{ab}$ as the Levi-Civita tensor of the flat two-dimensional metric $\eta_{ab}$, with $\epsilon_{w r^*} = + 1$, and all $a,b$ indices are raised and lowered with $\eta_{ab}$. Since $\det \eta = -1$, we have
\begin{equation}\label{eqn:epsilonIdentities}
    \epsilon^{ab} \epsilon_{cd} ~ = ~ - \left(\delta^a_c \delta^b_d - \delta^a_d \delta^b_c\right)
    \, , \qquad
    \epsilon_{ac}\epsilon^{bc} ~ = ~ - \delta^b_a \, ,
    \qquad \text{and} \qquad
    {\epsilon_a}^{b} ~ \coloneqq ~ \eta^{bc}\epsilon_{ac} \, .
\end{equation}
It is also convenient to abbreviate $\Box_\eta \coloneqq \eta^{ab}\partial_a \partial_b
= 2 \eta^{w r^*} \partial_w \partial_{r^*} + \partial_{r^*}^2$.

Then, the solution to the constraint in \eqref{eqn:AaSoln} can be written as
\begin{equation}\label{eqn:chiaSoln}
    \chi^{\ell m}_a ~ = ~ \epsilon_{ab} \partial^b \Psi_{\ell m} + \partial_a \rho_{\ell m} \quad \text{with} \quad \eta^{ab} \partial_a \partial_b \rho_{\ell m} ~ = ~ - f\left(r\right) J^+_{\ell m} \, .
\end{equation} 
It is now easy to check that
\begin{equation}\label{eqn:PsiIdentities}
    \eta^{ab}\partial_a \chi^{\ell m}_b ~ = ~ \Box_\eta \rho_{\ell m} \, , \qquad
    \epsilon^{ab}\partial_a \chi^{\ell m}_b ~ = ~ \Box_\eta \Psi_{\ell m} \, , \qquad
    \partial_a \chi^{\ell m}_b - \partial_b \chi^{\ell m}_a ~ = ~ - \epsilon_{ab} \, \Box_\eta \Psi_{\ell m} \, .
\end{equation}
The scalar $\Psi_{\ell m}$ is the Moncrief master variable of the even sector. The decomposition \eqref{eqn:chiaSoln} admits a homogeneous ambiguity: locally, a solution of $\Box_\eta h_{\ell m}=0$ may be added to $\Psi_{\ell m}$ and compensated by a corresponding homogeneous shift of $\rho_{\ell m}$, leaving $\chi_a^{\ell m}$ unchanged. We may fix this freedom by choosing
\begin{equation}
\rho_{\ell m} ~ = ~ - G_{\mathrm{ret}} \, \left[ f \left(r\right) J_{\ell m}^{+} \right] \, ,
\end{equation}
where $G_{\mathrm{ret}}$ is a fixed retarded Green operator with specified boundary conditions, and no independent homogeneous solution is added. With this prescription, $\partial_a\rho_{\ell m}$ is fixed by the matter current, and the even-sector Moncrief master variable $\Psi_{\ell m}$ is determined by \eqref{eqn:chiaSoln} up to an additive constant. This constant drops out of $\chi_a^{\ell m}$. All independently specifiable homogeneous data in this decomposition are thereby assigned to $\Psi_{\ell m}$.

Next, we consider the longitudinal component $J_w$:
\begin{align}
    J_a ~ &= ~ g_{a\nu} \nabla_{\mu} F^{\mu\nu} ~ = ~ \dfrac{g_{a\nu}}{\sqrt{-g}} \partial_{\mu} \left(\sqrt{-g} g^{\mu\rho}g^{\nu\sigma} F_{\rho\sigma}\right) \nonumber \\
    &= ~ \dfrac{1}{r^2} \partial_{d} \left(\dfrac{r^2}{f\left(r\right)} \eta^{d e} F_{e a}\right) + \dfrac{1}{r^2} \nabla^A \partial_A A_a - \dfrac{1}{r^2} \partial_a \nabla^A A_A \, .
\end{align}
In the partial wave basis, using the solution \eqref{eqn:AaSoln}, substituting \eqref{eqn:chiaSoln} and using \eqref{eqn:PsiIdentities},  we find that the above equation reduces to
\begin{align}\label{eqn:eomJa}
    J^{\ell m}_a ~ &= ~ \dfrac{1}{r^2} \partial_d \left[\dfrac{r^2}{f\left(r\right)} \eta^{d e} \left(\partial_e \chi^{\ell m}_a - \partial_a \chi^{\ell m}_{e}\right)\right] - \dfrac{\ell \left(\ell + 1\right)}{r^2} \chi^{\ell m}_a \nonumber \\
    &= ~ \dfrac{{\epsilon_a}^{d}}{r^2} \, \partial_d \left[\dfrac{r^2}{f\left(r\right)} \Box_\eta \Psi_{\ell m} -\ell\left(\ell+1\right) \Psi_{\ell m}\right] - \dfrac{\ell\left(\ell+1\right)}{r^2} \partial_a \rho_{\ell m} \, ,
\end{align}
where we used \eqref{eqn:PsiIdentities} and the constancy of ${\epsilon_a}^{d}$. Contracting \eqref{eqn:eomJa} with $\epsilon^{ba}$ and using
${\epsilon^{ba}}{\epsilon_a}^{d} = \eta^{bd}$, the longitudinal equation can be rewritten as
\begin{align}\label{eqn:MoncriefMaster}
    \partial_a \left[\dfrac{r^2}{f\left(r\right)} \Box_\eta \Psi_{\ell m} - \ell\left(\ell+1\right)\Psi_{\ell m}\right] ~ = ~ {\epsilon_a}^{b}\left(r^2 J^{\ell m}_b + \ell\left(\ell+1\right)\partial_b \rho_{\ell m}\right) \, .
\end{align}
To see what this equation implies, we define
\begin{equation}\label{eqn:radialFlux}
    \mathcal{E}_{\ell m} ~ \coloneqq ~ \dfrac{r^2}{f\left(r\right)} \, \Box_\eta \Psi_{\ell m} \, ,
\end{equation}
so that the square bracket in \eqref{eqn:MoncriefMaster} is $\mathcal{E}_{\ell m} - \ell\left(\ell+1\right)\Psi_{\ell m}$. Where the current vanishes, the right-hand side of \eqref{eqn:MoncriefMaster} vanishes and the bracket is a constant $c_{\ell m}$,
\begin{equation}
    \mathcal{E}_{\ell m} - \ell\left(\ell+1\right)\Psi_{\ell m} ~ = ~ c_{\ell m} \, .
\end{equation}
The constant is not the same everywhere: it takes different values in two regions separated by the support of the current. We now apply $\Box_\eta$ to both sides. The constant on the right gives zero, so
\begin{equation}
    \Box_\eta\mathcal{E}_{\ell m} - \ell\left(\ell+1\right)\Box_\eta\Psi_{\ell m} ~ = ~ 0 \, .
\end{equation}
The definition \eqref{eqn:radialFlux} says that $\Box_\eta\Psi_{\ell m} = f\left(r\right)\mathcal{E}_{\ell m}/r^2$. Substituting this into the second term gives
\begin{align}\label{eqn:MoncriefEoM}
    \Box_\eta \mathcal{E}_{\ell m} - \dfrac{\ell\left(\ell+1\right) f\left(r\right)}{r^2}\,\mathcal{E}_{\ell m} ~ = ~ 0 \, ,
\end{align}
which is identical in form to the odd-parity equation \eqref{eqn:oddEoM} under $\mathcal{E}_{\ell m} \leftrightarrow A^-_{\ell m}$. The field that obeys the free equation is therefore $\mathcal{E}_{\ell m}/\ell\left(\ell+1\right)$ and not $\Psi_{\ell m}$ itself. The two differ by the piecewise constant $c_{\ell m}/\ell\left(\ell+1\right)$. For the charge computation below we will only need the $w$-component of \eqref{eqn:MoncriefMaster}, which reads
\begin{align}\label{eqn:eomLongitudinalCombination}
    \partial_w \left[\dfrac{r^2}{f\left(r\right)}\Box_\eta \Psi_{\ell m} - \ell\left(\ell+1\right)\Psi_{\ell m}\right] ~ = ~ r^2 \, {\epsilon_w}^{b} J^{\ell m}_b + \ell\left(\ell+1\right) \partial^{r^*}\!\rho_{\ell m} \, ,
\end{align}
where ${\epsilon_w}^{b} J_b = \eta^{w r^*} J_w + J_{r^*}$.

\subsubsection{The radiative mode}\label{sec:radMode}
Under a gauge transformation \eqref{eqn:gaugetransfo}, the solution \eqref{eqn:AaSoln} transforms as
\begin{equation}
    A'^{\ell m}_a ~ = ~ \partial_a A'^{+}_{\ell m} + \chi'^{\ell m}_a \, , \text{ which is the same as } A^{\ell m}_a + \cancel{\partial_a \lambda_{\ell m}} ~ = ~ \partial_a A^+_{\ell m} + \cancel{\partial_a \lambda_{\ell m}} + \chi^{\ell m}_a \, .
\end{equation}
Therefore, we see that the mode $\chi^{\ell m}_a$ is indeed gauge invariant. Since $\chi^{\ell m}_a$ determines $\Psi_{\ell m}$ up to an additive constant, the Moncrief variable is gauge invariant as well, and it is the minimal such datum: where $\chi^{\ell m}_a$ is a two-component object subject to the constraint in \eqref{eqn:AaSoln}, $\Psi_{\ell m}$ is a single unconstrained scalar, on the same footing as the odd-parity variable $A^-_{\ell m}$. The field strength can be written entirely in terms of this mode in the even sector using \eqref{eqn:AaSoln}:
\begin{subequations}\label{eqn:FinTermsofChi_a}
    \begin{align}
    F_{a b} ~ &= ~ - \sum_{\ell m} \epsilon_{ab} \, \Box_\eta \Psi_{\ell m} \, Y_{\ell m} \\
    F_{a A} ~ &= ~ - \sum_{\ell m} \left({\epsilon_a}^{b}\partial_b \Psi_{\ell m} + \partial_a \rho_{\ell m}\right) \partial_A Y_{\ell m} - \sum_{\ell m} \partial_a A^-_{\ell m} {\epsilon_A}^B \partial_B Y_{\ell m} \\
    F_{A B} ~ &= ~ - \sum_{\ell m} A^-_{\ell m} \left({\epsilon_B}^C \hat{\nabla}_A - {\epsilon_A}^C \hat{\nabla}_B \right) \partial_C Y_{\ell m} \, .
\end{align}
\end{subequations}

\section{New symmetries of scalar QED near the horizon}\label{sec:symmetries}

From the gauge transformation of the vector field \eqref{eqn:gaugetransfo} we see, in the $A_w = 0$ gauge, that $\lambda_{\ell m}$ is independent of the null coordinate
\begin{equation}\label{lambda}
    \lambda_{\ell m} \left(w , r^*\right) ~ = ~ \lambda_{\ell m}\left(r^*\right) \, .
\end{equation}
In order to find the residual asymptotic symmetries near the horizon, we assume that the gauge parameter admits a limit to the horizon 
\begin{equation}
	\lambda^{(0)}_{\ell m} ~ \coloneqq ~ \lim_{r^{*} \to - \infty} \lambda_{\ell m} \left(r^{*}\right) \, .
\end{equation}
Note that this is a weaker requirement than a smooth near-horizon expansion. For $\ell \geq 1$ the transverse even component of the vector field then also admits a similar limit, and \eqref{eqn:gaugetransfo} implies that
\begin{equation}\label{eqn:AplusNearHor}
    \delta A^{+,(0)}_{\ell m} ~ = ~ \lambda^{(0)}_{\ell m} \, .
\end{equation}
Therefore, at leading order, we have a different constant $\lambda^{(0)}_{\ell m}$ for each partial wave, amounting to an arbitrary function on the sphere. All such functions are residual symmetries of the theory. The $\ell = 0$ member is the constant gauge parameter. There is no $A^{+}_{00}$ for it to shift, and it acts on the charged matter alone. These symmetries result in covariant charges which we turn to next.

\subsection{Covariant charge}\label{sec:covariantCharge}

We will use the standard covariant phase space formalism to derive the charge corresponding to the asymptotic symmetries of interest. We start with the Maxwell action
\begin{align}\label{eqn:MaxwellAction}
    S_{\text{Maxwell}} ~ = ~ \int \mathrm{d}^4 x \mathcal{L}_{\text{Maxwell}} ~ = ~ - \dfrac{1}{4} \int \mathrm{d}^4 x \sqrt{-g} \, F_{\mu \nu} F^{\mu \nu} \, .
\end{align}
The variation of this action under a gauge parameter, say $\lambda$, can be written as
\begin{align}
    \delta_\lambda S_{\text{Maxwell}} ~ = ~ \int \mathrm{d}^4 x \delta_\lambda \mathcal{L}_{\text{Maxwell}} ~ = ~ \int \mathrm{d}^4 x \Bigr[ \sqrt{-g} \, \nabla_{\nu} F^{\nu\mu} \delta_{\lambda} A_{\mu} - \partial_{\nu} \left(\sqrt{-g} \, F^{\nu\mu} \nabla_{\mu} \lambda \right) \Bigr] \, .
\end{align}
Therefore, we may immediately identify the quantity $E^\mu$ and the pre-symplectic potential $\Theta^\mu\left(\delta_\lambda\right)$ as
\begin{equation}
    E^{\mu} ~ = ~ \nabla_{\nu} F^{\nu\mu} \quad \text{and} \quad \Theta^{\nu}\left(\delta_\lambda\right) ~ = ~ - F^{\nu\mu} \nabla_{\mu} \lambda \, .
\end{equation}
Since the Lagrangian is invariant under gauge transformations and $\delta_{\lambda} A_{\mu} = \nabla_\mu \lambda$, we have
\begin{equation}
    0 ~ = ~ E^{\mu} \nabla_{\mu} \lambda + \nabla_{\nu} \Theta^{\nu}\left(\delta_\lambda\right) ~ = ~ \nabla_{\mu} \left( \lambda E^{\mu} + \Theta^{\mu}\left(\delta_\lambda\right) \right) ~ \eqqcolon ~ \nabla_\mu J_{\lambda}^{\mu} \, ,
\end{equation}
where we used that $\nabla_{\mu}E^{\mu} = 0$ and defined the Noether current $J_{\lambda}^{\mu}$. This current can be written as a total derivative in the following way
\begin{align}
    J_{\lambda}^{\nu} ~ &= ~ \lambda \, \nabla_{\mu} F^{\mu\nu} + \, F^{\mu\nu} \nabla_{\mu} \lambda ~ = ~ \nabla_{\mu} \left( \lambda \, F^{\mu\nu} \right) ~ \eqqcolon ~ \nabla_{\mu} Q^{\mu\nu} \, .
\end{align}
This allows us to identify the surface charge density as
\begin{equation}
    k^{\mu\nu} ~ = ~ \delta Q^{\mu\nu} ~ = ~ \lambda \delta F^{\mu\nu} \, ,
\end{equation}
which can be integrated on the surface of interest to find the infinitesimal surface charge
\begin{equation}
    \delta Q\left[\lambda\right] ~ = ~ \int \left(\mathrm{d}^2 x\right)_{\mu\nu} \, \lambda \delta F^{\mu\nu} \quad \text{with} \quad \left(d^2 x\right)_{\mu\nu} ~ = ~ \dfrac{1}{4} \sqrt{-g} \, \epsilon_{\mu\nu\rho\sigma} \mathrm{d} x^\rho \wedge \mathrm{d} x^\sigma \, .
\end{equation}
Integrating on the space of solutions, we have that
\begin{align}\label{eqn:electricCharge}
    Q\left[\lambda\right] ~ &= ~ \int \left(\mathrm{d}^2 x\right)_{\mu\nu} \, \lambda \, g^{\mu\rho}g^{\nu\sigma} F_{\rho \sigma} \, .
\end{align}
To find the flux on the sphere, we now restrict the integral above to be on the transverse coordinates as
\begin{align}
    Q\left[\lambda\right] ~ &= ~ \dfrac{1}{4} \int \mathrm{d} x^A \wedge \mathrm{d} x^B \epsilon_{\mu\nu AB} \sqrt{-g} \, \lambda \, g^{\mu\rho} g^{\nu\sigma} F_{\rho \sigma} \nonumber \\
    &= ~ \dfrac{1}{2} \int \mathrm{d} x^A \wedge \mathrm{d} x^B \epsilon_{w r^*} \epsilon_{AB} \sqrt{-g} \, \lambda \, g^{w r^*} g^{r^* w} F_{w r^*} \nonumber \\
    &= ~ \phantom{-} \sum_{\ell m} \, \lambda_{\ell m} \, \dfrac{r^2}{f\left(r\right)} \, \Box_\eta \Psi_{\ell m} \, ,
\end{align}
where we used the third identity in \eqref{eqn:PsiIdentities}. In the last line, we have expanded the gauge parameter in partial waves to integrate over the sphere. This is the charge at each cut on the horizon. 

On the surface of constant $r^*$, say $\Sigma$, on which the charges are evaluated before the horizon limit (\appref{app:MaxwellPhaseSpace}), the null coordinate runs over the whole real line. We orient every null integral toward the future, on both horizons,
\begin{equation}\label{eqn:orientation}
    \int \mathrm{d}w ~ \coloneqq ~ \int_{-\infty}^{+\infty} \mathrm{d}w \, , \quad \Delta X ~ \coloneqq ~ X\big|_{w \rightarrow +\infty} - X\big|_{w \rightarrow -\infty} \, , \quad \text{so that} \quad \int \mathrm{d}w \, \partial_w X ~ = ~ \, \Delta X \, ,
\end{equation}
for $w = u$ on $\mathcal{H}^-$ as for $w = v$ on $\mathcal{H}^+$. The early end $w \rightarrow -\infty$ is the bifurcation sphere on $\mathcal{H}^+$ and the past endpoint on $\mathcal{H}^-$; the late end $w \rightarrow +\infty$ is the future endpoint on $\mathcal{H}^+$ and the bifurcation sphere on $\mathcal{H}^-$. The sign of $\Delta$ is a convention, and we take the late value minus the early one. What is not a convention is that the two horizons are oriented alike. The $u$- and $v$-charts parametrise one and the same surface, whose horizon limit at fixed $u$ is $\mathcal{H}^-$ and at fixed $v$ is $\mathcal{H}^+$, and Killing time increases in the same direction along both. The charge integrals on the two horizons are the two horizon limits of one integral over that surface, evaluated with the free-field expansions valid at its early and late ends. The cut charge then changes along the surface by
\begin{align}\label{eqn:endpointContributions}
    Q\left[\lambda\right]\left|_{w=+\infty} - Q\left[\lambda\right]\right|_{w=-\infty} ~ &= ~ \sum_{\ell m} \lambda_{\ell m} \int \mathrm{d}w \, \partial_w \left(\dfrac{r^2}{f\left(r\right)} \, \Box_\eta \Psi_{\ell m}\right) \nonumber \\
    &= ~ + \sum_{\ell m} \ell\left(\ell+1\right) \lambda_{\ell m} \int \mathrm{d}w \, \partial_w \Psi_{\ell m} \nonumber \\
    &\phantom{=} ~~ + \, \sum_{\ell m} \lambda_{\ell m} \, \int \mathrm{d}w \Bigg(r^2 \, {\epsilon_w}^{b} J^{\ell m}_b  + \, \ell\left(\ell+1\right) \, \partial^{r^*}\!\rho_{\ell m}\Bigg) \, ,
\end{align}
where the second line uses \eqref{eqn:eomLongitudinalCombination}. By Stokes' theorem the left-hand side is the flux of the Noether current through $\Sigma$, $\Delta Q\left[\lambda\right] = \int_\Sigma \mathrm{d}\Sigma_\nu \, J^\nu_\lambda$, with the normal of \appref{app:MaxwellPresymplectic}. The negative of this flux\footnote{In the rest of this paper, we will refer to this as simply ``the flux'', with the minus sign implicitly understood. Similarly, we will also refer to the minus of the endpoint difference also as ``the endpoint difference''. We keep track of the signs in all equations correctly.} is the generator of the residual transformation. We denote it by $Q_{\text{cut}}$. For a single partial wave, the first equality of \eqref{eqn:endpointContributions} gives
\begin{align}\label{eqn:totalCharge}
        Q_{\text{cut}}\left[\lambda_{\ell m}\right] ~ \coloneqq ~ - \, \lambda_{\ell m} \int \mathrm{d}w \, \partial_w \mathcal{E}_{\ell m} ~ = ~ - \, \lambda_{\ell m} \, \Delta \mathcal{E}_{\ell m} \, .
\end{align}
The last line of \eqref{eqn:endpointContributions} is the part of this flux that is carried by the matter, and minus that line is the hard charge
\begin{align}\label{eqn:hardCharge}
    Q_{\text{hard}}\left[\lambda_{\ell m}\right] ~ &= ~ - r^2 \lambda_{\ell m} \int \mathrm{d}w \, {\epsilon_w}^{b} J^{\ell m}_b - \ell\left(\ell+1\right) \lambda_{\ell m} \int \mathrm{d}w \, \partial^{r^*}\!\rho_{\ell m} \nonumber \\
    &= ~ - r^2 \lambda_{\ell m} \int \mathrm{d}w \, \Bigg( \eta^{wr^*} J^{\ell m}_w + J^{\ell m}_{r^*}\Bigg) - \ell\left(\ell+1\right) \lambda_{\ell m} \, \int \mathrm{d}w \, \partial^{r^*}\!\rho_{\ell m} \, ,
\end{align}
where the second term is fixed by $J^+_{\ell m}$ through $\Box_\eta \rho_{\ell m} = - f\left(r\right) J^+_{\ell m}$. The charge $Q_{\text{cut}}$ is not conserved by the scattering near the horizon, and the reason is already contained in \eqref{eqn:endpointContributions}. That equation is a balance law, which equates the change of the cut charge between the two ends of $\Sigma$ to the flux through it. At null infinity the corresponding law does lead to a conserved charge \cite{He:2014cra, Strominger:2017zoo}: the cut charge is taken to vanish at the far end of $\mathcal{I}^+$ and of $\mathcal{I}^-$, so that minus its change is the cut charge at the other end, and the Ward identity equates the two cut charges that meet at spatial infinity. Neither step is available on $\Sigma$. It is a timelike surface at finite distance which charged matter crosses in both directions, so the cut charge does not vanish at either of its ends, and every charge that crosses $\Sigma$ changes it.

The charge we consider instead is what remains of $Q_{\text{cut}}$ once the matter flux is removed. By \eqref{eqn:eomLongitudinalCombination} this is
\begin{align}\label{eqn:softCharge}
    Q_{\text{soft}}\left[\lambda_{\ell m}\right] ~ &\coloneqq ~ Q_{\text{cut}}\left[\lambda_{\ell m}\right] - Q_{\text{hard}}\left[\lambda_{\ell m}\right] \nonumber \\
    &= ~ - \, \ell\left(\ell+1\right) \lambda_{\ell m} \int \mathrm{d}w \, \partial_w \Psi_{\ell m} ~ = ~ - \ell\left(\ell+1\right) \lambda_{\ell m} \, \Delta \Psi_{\ell m} \, ,
\end{align}
which we call the soft charge. It is built from the $\Psi_{\ell m}$ of \eqref{eqn:chiaSoln}, whereas $Q_{\text{cut}}$ is built from $\mathcal{E}_{\ell m}$, and the two fields are sourced differently. The first equation below is an identity, and the second is the divergence of \eqref{eqn:MoncriefMaster}:
\begin{subequations}\label{eqn:sourcedEquations}
\begin{align}
	\left(\Box_\eta - \dfrac{\ell\left(\ell+1\right) f\left(r\right)}{r^2}\right)\Psi_{\ell m} ~ &= ~ \dfrac{f\left(r\right)}{r^2}\Big[\mathcal{E}_{\ell m} - \ell\left(\ell+1\right)\Psi_{\ell m}\Big] \, , \\
        \left(\Box_\eta - \dfrac{\ell\left(\ell+1\right) f\left(r\right)}{r^2}\right)\mathcal{E}_{\ell m} ~ &= ~ \partial^a\left({\epsilon_a}^{b} \, r^2 J^{\ell m}_b\right) \, .
\end{align}
\end{subequations}
The bracket in the first is the piecewise constant of \eqref{eqn:MoncriefMaster}, so the source of $\Psi_{\ell m}$ is suppressed by $f\left(r\right)$ near the horizon, while $\mathcal{E}_{\ell m}$ is sourced at order one by the curl of the current. We show in \secref{sec:wardidentity} that $Q_{\text{soft}}$ commutes with the $S$-matrix. In all of these charges, we are yet to take the limit to the horizon before evaluating their action on external states. We will discuss this limit after a short detour into the magnetic charge. 

\paragraph{Magnetic charge} The expression \eqref{eqn:softCharge} is of course the electric charge. The corresponding magnetic charge is obtained by inserting the Hodge dual of the field strength in \eqref{eqn:electricCharge} as:
\begin{align}\label{eqn:magneticCharge}
    Q\left[\lambda\right] ~ &= ~ \int \left(\mathrm{d}^2 x\right)_{\mu\nu} \, \lambda \, \left(\star F^{\mu \nu}\right) \nonumber \\
    &= ~ \dfrac{1}{2} \int \left(\mathrm{d}^2 x\right)_{\mu\nu} \, \dfrac{\lambda}{\sqrt{-g}} \, g^{\rho \tau} g^{\sigma \kappa} {\epsilon^{\mu\nu}}_{\rho\sigma} F_{\tau \kappa} \, .
\end{align}
Restricting the integral to be on the transverse coordinates as before, to find the magnetic flux on the sphere, we have:
\begin{align}
    Q\left[\lambda\right] ~ &= ~ \dfrac{1}{8} \int \mathrm{d} x^A \wedge \mathrm{d} x^B \epsilon_{\mu\nu A B} \, \lambda \, g^{\rho \tau} g^{\sigma \kappa} {\epsilon^{\mu\nu}}_{\rho\sigma} F_{\tau \kappa} \nonumber \\
    &= ~ - \dfrac{1}{2} \int \mathrm{d} x^A \wedge \mathrm{d} x^B \, \lambda \, \left(\delta^\tau_A \delta^\kappa_B - \delta^\tau_B \delta^\kappa_A\right) F_{\tau \kappa} \nonumber \\
    &= ~ - \int \mathrm{d} x^A \wedge \mathrm{d} x^B \, \lambda \, F_{A B} \nonumber \\
    &= ~ \sum_{\ell m} \sum_{\ell' m'} \int \mathrm{d} x^A \wedge \mathrm{d} x^B \, \lambda_{\ell' m'} \,  Y_{\ell' m'} \, A^-_{\ell m} \left({\epsilon_B}^C \hat{\nabla}_A - {\epsilon_A}^C \hat{\nabla}_B \right) \partial_C Y_{\ell m} \nonumber \\
    &= ~ 2 \sum_{\ell m} \sum_{\ell' m'} \int \mathrm{d} x^A \wedge \mathrm{d} x^B \, \lambda_{\ell' m'} \,  Y_{\ell' m'} \, A^-_{\ell m} {\epsilon_B}^C \hat{\nabla}_A \partial_C Y_{\ell m} \, ,
\end{align}
where in the last line, we used the anti-symmetry of the wedge product. Now, we recognise that $\mathrm{d} x^A \wedge \mathrm{d} x^B = \epsilon^{A B} \mathrm{d} \theta \mathrm{d} \phi = \gamma^{A D} {\epsilon_D}^B \mathrm{d} \theta \mathrm{d} \phi $ to rewrite the above equation as
\begin{align}
    Q\left[\lambda\right] ~ &= ~ 2 \sum_{\ell m} \sum_{\ell' m'} \int \gamma^{A D} {\epsilon_D}^B \mathrm{d} \theta \mathrm{d} \phi \sqrt{\gamma} \, \lambda_{\ell' m'} \,  Y_{\ell' m'} \, A^-_{\ell m} {\epsilon_B}^C \hat{\nabla}_A \partial_C Y_{\ell m} \nonumber \\
    &= ~ - 2 \sum_{\ell m} \sum_{\ell' m'} \int \mathrm{d} \Omega \, \lambda_{\ell' m'} \,  Y_{\ell' m'} \, A^-_{\ell m}  \gamma^{A C} \hat{\nabla}_A \partial_C Y_{\ell m} \, ,
\end{align}
where we made use of the identity ${\epsilon_D}^B {\epsilon_B}^C = - \delta^C_D$ and the definition $\int \mathrm{d} \theta \mathrm{d} \phi \sqrt{\gamma} \eqqcolon \int \mathrm{d}\Omega$. We may now simply pick up the eigenvalue of the spherical harmonic under the action of the sphere Laplacian and integrate over the sphere to find
\begin{align}\label{eqn:magneticChargeNearHor}
    Q\left[\lambda_{\ell m}\right] ~ &= ~ 2 \ell \left(\ell + 1\right) \lambda_{\ell m} \, A^-_{\ell m} 
    \, .
\end{align}
Evidently, the charge carries no hard component. As it turns out, this is consistent with the fact that the odd-mode of the photon does not contribute to the soft theorem in the near-horizon region. We demonstrate this in \appref{app:magneticsofttheorem}. 

Note that this charge is not the generator of \eqref{eqn:gaugetransfo}, under which $A^-_{\ell m}$ is invariant; it is the charge of the large gauge transformations of the dual potential \cite{Strominger:2015bla}, which leave $F_{\mu\nu}$ and hence $A^-_{\ell m}$ unchanged. In a phase space built on $A_{\mu}$, it has no boundary mode to act on \cite{Freidel:2018fsk, Hosseinzadeh:2018dkh}. Its endpoint difference $-2\ell(\ell+1)\lambda_{\ell m}\Delta A^-_{\ell m}$ is the zero mode of the odd radiative field, the magnetic counterpart of \eqref{eqn:softCharge}.

\subsection{The cut charge and its action on external states}
The radiative mode obeys \eqref{eqn:MoncriefEoM}, a two-dimensional Klein--Gordon equation whose mass depends on position. In the static chart $\Box_\eta = -\partial^2_t + \partial^2_{r^*}$, and near the horizon $f\left(r\right) \simeq e^{r^*/R}$ up to a constant that we absorb into the origin of $r^*$, so that

\begin{equation}
\left(-\partial^2_t + \partial^2_{r^*}\right)\Psi_{\ell m} ~ = ~ \mu^2_\ell\left(r^*\right)\Psi_{\ell m} \, , \qquad \mu^2_\ell\left(r^*\right) ~ = ~ \dfrac{f\left(r\right)\,\ell\left(\ell+1\right)}{r^2} ~ \simeq ~ \dfrac{\ell\left(\ell+1\right)}{R^2}\, e^{r^*/R} \, .
\end{equation}
At fixed Killing frequency, $\Psi_{\ell m} = e^{-i\omega t}\psi_\omega\left(r^*\right)$, the radial function satisfies
\begin{equation}
\psi_\omega'' + \left[\omega^2 - \mu^2_\ell\left(r^*\right)\right]\psi_\omega ~ = ~ 0 \, .
\end{equation}
To solve it, change variable to $x = 2R\mu_\ell\left(r^*\right)$, which grows exponentially in $r^*$ and so satisfies $\mathrm{d}x/\mathrm{d}r^* = x/2R$. Then $\mathrm{d}^2/\mathrm{d}r^{*2} = \left(x^2\,\mathrm{d}^2/\mathrm{d}x^2 + x\,\mathrm{d}/\mathrm{d}x\right)/4R^2$ and $\mu^2_\ell = x^2/4R^2$, and the equation becomes
\begin{equation}
x^2\psi_{xx} + x\psi_x - \left(x^2 - 4R^2\omega^2\right)\psi ~ = ~ 0 \, ,
\end{equation}
the modified Bessel equation of imaginary order $\nu = \pm 2i\omega R$. This equation has two solutions $I_{\pm 2i\omega R}\left(2R\mu_\ell\right)$, and their series, $I_\nu\left(x\right) = \left(x/2\right)^\nu\left[1 + \left(x/2\right)^2/\left(\nu+1\right) + \ldots\right]/\Gamma\left(\nu+1\right)$, gives with $x/2 = R\mu_\ell \propto e^{r^*/2R}$
\begin{equation}\label{eqn:photonRadialExact}
I_{\pm 2i\omega R}\left(2R\mu_\ell\right) ~ \propto ~ e^{\pm i\omega r^*}\left[1 + \dfrac{R^2\mu^2_\ell\left(r^*\right)}{1 \pm 2i\omega R} + \mathcal{O}\left(R^4\mu^4_\ell\right)\right] \, . 
\end{equation}
The prefactor is $\left(R\mu_\ell\right)^{\pm 2i\omega R}$, a plane wave of wavenumber $\omega$ because $\mu_\ell$ is exponential in $r^*$; the bracket is the effect of the mass, and its size is $R^2\mu^2_\ell/\left|1 \pm 2i\omega R\right|$. There are two regimes, according to the size of $\omega R$.

For $\omega R \gg 1$ the correction is the phase $\mp iR\mu^2_\ell/2\omega$, and it is the first term of the WKB solution. Writing $\psi = e^{iS}$ in the radial equation and dropping $S''$ gives $S' = \pm\tilde\omega$ with $\tilde\omega\left(r^*\right) = \sqrt{\omega^2 - \mu^2_\ell\left(r^*\right)}$ the local radial wavenumber, so that $\psi \simeq e^{\pm i\int\tilde\omega\,\mathrm{d}r^*}$. This is a wave whose wavenumber adjusts to the mass at each radius. The neglected second derivative of $S$ is small when $\left|\tilde\omega'\right| \ll \tilde\omega^2$, which for $\omega \gg \mu_\ell$ holds whenever $\omega R \gg 1$. Expanding $\tilde\omega - \omega \simeq -\mu^2_\ell/2\omega$ and using $\int^{r^*}\mu^2_\ell = R\mu^2_\ell\left(r^*\right)$ gives $\psi \simeq e^{\pm i\omega r^*}\left(1 \mp iR\mu^2_\ell/2\omega\right)$, which is \eqref{eqn:photonRadialExact} in this regime. Over any stretch of $r^*$ that is short compared with $R$, on which $\mu_\ell$ does not change appreciably, the WKB wave is the plane wave $e^{\pm i\tilde\omega r^*}$ of a two-dimensional field of constant mass $\mu_\ell$, and this is the sense in which the mode has a mass shell. In this regime the mode is oscillatory where $\mu_\ell\left(r^*\right) < \omega$ and evanescent where $\mu_\ell\left(r^*\right) > \omega$, exactly as for a Schrödinger wave at energy $\omega^2$ in the rising potential $\mu^2_\ell\left(r^*\right)$. Since $\mu_\ell$ falls exponentially towards the horizon, every mode is oscillatory for $r^*$ below the turning point $r^*_\omega$ defined by $\mu_\ell\left(r^*_\omega\right) = \omega$ and evanescent above it, where the exact solution has the large-argument behaviour $I_\nu\left(x\right) \sim e^{x} = e^{2R\mu_\ell}$. At a fixed radius, $\mu_\ell\left(r^*\right)$ is the frequency below which a mode does not propagate there: a photon emitted at $r^*_0$ with $\omega < \mu_\ell\left(r^*_0\right)$ propagates only at radii below $r^*_\omega$, while between $r^*_\omega$ and the peak of the potential barrier, where the emission takes place, the field it produces falls off exponentially away from the source, and the amplitude that emerges as a propagating wave at $r^*_\omega$ carries the factor $e^{-\int_{r^*_\omega}^{r^*_0}\sqrt{\mu^2_\ell - \omega^2}}$, which differs appreciably from one only when $R\mu_\ell\left(r^*_0\right) \gg 1$.

For $\omega R \ll 1$ the correction in \eqref{eqn:photonRadialExact} is real, $R^2\mu^2_\ell \simeq f\left(r\right)\ell\left(\ell+1\right)$,\footnote{This combination is the only dimensionless parameter the two scales form, and it is small exactly when the partial wave cannot resolve the variation of its own mass over a wavelength; $R^2\mu^2_\ell\ll1$, that is $f(r)\ll1/\ell(\ell+1)$, is what we mean by partial wave $\ell$ being in the near-horizon region at radius $r$.} so it changes the amplitude of the wave and not its wavenumber, and it does not involve $\omega$ at all. Wherever $R^2\mu^2_\ell \ll 1$ both solutions are plane waves of wavenumber $\omega$, whether $\omega$ lies above $\mu_\ell\left(r^*\right)$ or below it: there is no threshold, no turning point, and the spectrum is continuous down to $\omega = 0$. As mentioned above, the WKB wave would be evanescent at the radii where $\omega < \mu_\ell\left(r^*\right)$, and is not the right description of a mode whose wavelength $1/\omega$ exceeds the scale $R$ on which the mass varies.

Where $\mu_\ell \ll \omega$ the two descriptions agree, both giving a plane wave of wavenumber close to $\omega$ with a small correction. We shall write the mode expansion in the WKB form, with the wavenumber $\tilde\omega$ and a frequency integral starting at the threshold $\mu_\ell$, because that is the form in which the amplitude of \secref{sec:softtheorem} is computed. For $\omega R \ll 1$ it is to be read with $\tilde\omega = \omega$ and the integral starting at zero.

We build the mode expansion from the plane wave $e^{-i\omega t + i\sigma\tilde\omega r^*}$, with $\sigma = \pm1$ the sign of the radial wavenumber. In the chart adapted to a horizon, $w = t + \eta^{wr^*}r^*$, we write the plane wave as $e^{-i\omega w + ik_{(\sigma)}r^*}$ with
\begin{equation}\label{eqn:ka}
k_a ~ = ~ \left(-\omega \, , \, k_{(\sigma)}\right) ~ = ~ \left(-\omega \, , \, \sigma\tilde\omega + \eta^{wr^*}\omega\right) \, , \qquad \tilde\omega ~ = ~ \sqrt{\omega^2 - \mu^2_\ell} \, , 
\end{equation}
and $k_{(\sigma)}$ are the two roots of the mass-shell condition $\eta^{ab}k_ak_b = -\mu^2_\ell$ at fixed $\omega$, since $\eta^{ab}k_ak_b = k_{(\sigma)}^2 - 2\eta^{wr^*}\omega\,k_{(\sigma)} = \tilde\omega^2 - \omega^2$. The label $\sigma$ does not depend on which horizon the chart is adapted to as we discuss below. 

The radiative mode $\Psi_{\ell m}$ can be quantised and normalised to satisfy the canonical symplectic brackets; we present the detailed calculation in \appref{app:radiativeModeQuantisation}. The result is \eqref{appeqn:PsiQuantisation}
\begin{align}\label{eqn:PsiQuantisation}
    \Psi_{\ell m} ~ = ~ \dfrac{1}{\sqrt{\ell\left(\ell+1\right)}} \sum_{\sigma = \pm} \int_{\mu_\ell}^\infty \dfrac{\mathrm{d}\omega}{2\pi} \, \dfrac{i}{\sqrt{2\tilde{\omega}}} \left[b^{(\sigma)}_{\ell m}\left(\omega\right) e^{- i \omega w + i k_{(\sigma)} r^*} - \text{h.c.}\right] \, ,
\end{align}
where the modes satisfy the algebra
\begin{align}\label{eqn:PsiAlgebra}
    \left[b^{(\sigma)}_{\ell m}\left(\omega\right) \, , \,
    b^{(\sigma')\dagger}_{\ell' m'}\left(\omega'\right)\right] ~ = ~
    \delta_{\sigma \sigma'} \, \delta_{\ell \ell'} \, \delta_{m m'} \, 2\pi \,
    \delta\left(\omega - \omega'\right) \, .
\end{align}
The factor $1/\sqrt{2\tilde\omega}$ is the canonical normalisation of a two-dimensional field of mass $\mu_\ell$, fixed by the symplectic form in \appref{app:radiativeModeQuantisation}, and the integral starts at the WKB threshold. The radiative mode is physical only for $\ell>0$. 

The polarisation vector is not an independent datum: it is generated by $\epsilon_{ab}\partial^b$ acting on the scalar mode, as in \eqref{eqn:polarisation}, giving
\begin{align}
    \varepsilon^{(\sigma)}_a ~ = ~ - \, \epsilon_{ab}k^b_{(\sigma)} ~ = ~ \left(- \sigma \tilde{\omega} \, , \, \eta^{wr^*} \sigma \tilde{\omega} + \omega\right) \, ,
\end{align}
which is automatically transverse, $\varepsilon^{(\sigma)}\!\cdot k = 0$, and of norm $\varepsilon^{(\sigma)}\!\cdot\varepsilon^{(\sigma)} = \mu^2_\ell$ on shell.
 
The additional label $\sigma$ distinguishes the two branches of the radial solution, and it has a chart-independent meaning: in the massless limit the phase is $e^{-i\omega\left(t - \sigma r^*\right)}$, so that $\sigma = +1$ is outgoing and a function of $u$, while $\sigma = -1$ is ingoing and a function of $v$, whichever null coordinate $w$ is used. What depends on the chart is which branch has a profile on the horizon the chart is adapted to. The future horizon is the surface $u = +\infty$. An ingoing wave moves along rays of constant $v$, which reach it, so its values on the horizon are its profile there, and it is the wave that falls in; an outgoing wave moves along rays of constant $u$, which near $\mathcal{H}^+$ run parallel to it, so it never reaches the horizon, and since $e^{-i\omega u} = e^{-i\omega v}e^{2i\omega r^*}$ oscillates infinitely fast as $r^* \rightarrow -\infty$ it leaves no finite profile there in the strict limit. On the past horizon, $v = -\infty$, the roles are exchanged: the outgoing wave emerges through it, the ingoing wave runs parallel to it. We say that a branch \emph{crosses} the horizon in the first case and runs \emph{along} it in the second. In the chart adapted to the horizon in question, $w$ is the null coordinate that varies along it, the crossing branch is $\eta^{wr^*}\sigma = -1$ and the branch along it is $\eta^{wr^*}\sigma = +1$, on both horizons. The hard modes have the same two branches as we discuss in \secref{sec:externalstates}.

The field appearing in $Q_{\text{cut}}$ is $\mathcal{E}_{\ell m}$, by \eqref{eqn:totalCharge}. Since $\mathcal{E}_{\ell m}/\ell\left(\ell+1\right)$ solves \eqref{eqn:MoncriefEoM} wherever the current vanishes, it has the expansion \eqref{eqn:PsiQuantisation} at the early and late ends of $\Sigma$, whether or not charges are present. The mode function of $\partial_w\mathcal{E}_{\ell m}$ is therefore $\left(\omega\sqrt{\ell\left(\ell+1\right)/2\tilde\omega}\right)e^{-i\omega w + ik_{(\sigma)}r^*}$, and two conditions govern its reduction to a boundary mode. The first is the soft limit
\begin{equation}\label{eqn:softlimit}
    \mu_\ell ~ \ll ~ \omega ~ \ll ~ E_p \, ,
\end{equation}
with $E_p$ the energies of the hard scalars. The two inequalities do different work. The first places the photon above the threshold $\mu_\ell\left(r\right)$ at the radius at which the process sits: there its radial wavenumber is real and $\tilde\omega \simeq \omega$, so that the mass-shell form \eqref{eqn:ka} describes the mode to relative accuracy $\mu^2_\ell/\omega^2$ or $R^2\mu^2_\ell$, whichever regime applies. The second is the soft limit proper, the photon frequency small compared with the energies of the legs that emit it; it does not enter the charges of this section, which contain no hard momentum, and it is the hierarchy in which the amplitude of \secref{sec:softtheorem} is computed. Both $\omega$ and $E_p$ are Killing frequencies and carry no radial dependence, whereas $\mu^2_\ell$ does, so that at fixed Killing energy the window widens as the horizon is approached. Dividing by $\sqrt{f\left(r\right)}$ shows what the hierarchy says for a static observer, since $\mu_\ell$ is a mass with respect to $\eta_{ab}$ while the block metric of \eqref{eqn:metric} is $f\left(r\right)\eta_{ab}$: it reads $\sqrt{\ell\left(\ell+1\right)}/r \ll \omega/\sqrt{f} \ll E_p/\sqrt{f}$, a window between the transverse mass of the partial wave, which is finite on the horizon, and the local energy of the legs. At fixed local energy its width is finite, independent of the stretching and $\ell$-dependent, and it closes at $\ell\left(\ell+1\right) \sim \left(E_pR/\sqrt{f}\right)^2$, the largest angular momentum a quantum of that energy can carry on a sphere of radius $R$.

The second condition is of a different character. The two radial branches of \eqref{eqn:ka} behave differently under the near-horizon expansion. On the crossing branch, $\eta^{w r^*}\sigma = -1$, the two terms cancel in the massless limit,
\begin{equation}
    k_{(\sigma)} ~ = ~ \eta^{w r^*} \left( \omega - \widetilde{\omega} \right) ~ \longrightarrow ~ 0 \, ,
\end{equation}
and what remains of the radial dependence of the mode function is the correction of \eqref{eqn:photonRadialExact}: the phase $e^{\mp i R \mu^2_\ell\left(r^*\right)/2\omega}$ for $\omega R \gg 1$, the amplitude $1 + R^2 \mu^2_\ell$ for $\omega R \ll 1$. Both are negligible, since $\mu^2_\ell$ vanishes exponentially in $r^*$ while $\left|r^*\right|$ grows only linearly. On the branch along the horizon, $\eta^{w r^*}\sigma = +1$, no such cancellation occurs, $k_{(\sigma)} \simeq 2 \eta^{w r^*}\omega$, and the phase $e^{2 i \eta^{w r^*} \omega r^*}$ survives; this is the statement that the branch along the horizon carries no data on it in the strict limit, and retaining this data is what the correlated limit below is for. For both branches to contribute to the boundary mode --- that is, for the complete radial phase space to be retained --- the boundary value must be read off where
\begin{equation}\label{eqn:boundaryModeCondition}
    r^{*} ~ \to ~ -\infty \quad \text{and} \quad \omega ~ \to ~ 0 \quad \text{subject to} \quad \dfrac{\mu_{\ell}}{\omega} ~ \ll ~ 1 \quad \text{and} \quad \omega \left| r^* \right| ~ \ll ~ 1 \, .
\end{equation}
This second condition is not a dynamical hierarchy. It is a prescription for the limiting path, chosen so that both radial branches survive. Of its two requirements, $\omega\left|r^*\right| \ll 1$ is the one that keeps the branch along the horizon. It also places the boundary mode in the regime $\omega R \ll 1$, since $\left|r^*\right| \gg R$; the branches there are the plane waves of \eqref{eqn:photonRadialExact}. The requirement $\mu_\ell/\omega \ll 1$ keeps the mode above the threshold at which the WKB form of the frequency integral starts. For $\omega R \ll 1$ the integral starts at zero, and the requirement becomes $R^2\mu^2_\ell \ll 1$, which holds automatically as $r^* \rightarrow -\infty$. The two requirements are compatible along the path because $\mu_\ell \propto e^{r^*/2R}$ falls exponentially while $\omega$ need only fall as a power of $\left|r^*\right|$. This condition is used only in the construction of the boundary mode and plays no role in the amplitude of \secref{sec:softtheorem}.

In this limit the mode function of $\partial_w\mathcal{E}_{\ell m}$ reduces to
\begin{align}\label{eqn:modeApprox}
        \omega \sqrt{\dfrac{\ell(\ell+1)}{2\tilde{\omega}}} \, e^{-i\omega w+i k_{(\sigma)}r^*} ~ \approx ~ \sqrt{\dfrac{\ell \left(\ell + 1 \right) \omega}{2}} \, e^{-i\omega w}  \, ,
\end{align}
and the mode expansion becomes the quantised boundary mode
\begin{align}\label{eqn:PsiModeExpansion}
	\partial_w \mathcal{E}^{(0)}_{\ell m} ~ = ~ \sqrt{\ell(\ell+1)}\sum_{\sigma=\pm} \int_0^{\infty}\frac{\mathrm{d}\omega}{2\pi} \sqrt{\frac{\omega}{2}} \left[ b_{\ell m}^{(\sigma)}\left(\omega\right) e^{-i\omega w} + b_{\ell m}^{(\sigma)\dagger}\left(\omega\right) e^{i\omega w} \right] \, .
\end{align}
Here, we introduced an extra superscript $(0)$ to refer to the leading boundary mode in the limit discussed above. This allows us to write the cut charge $Q_{\text{cut}}$ on the horizon as
\begin{align}\label{eqn:softChargeOnHorizons}
        Q_{\text{cut}}\!\left[\lambda^{(0)}_{\ell m}\right] ~ & = ~ - \, \lambda^{(0)}_{\ell m}\int_{-\infty}^{\infty} \mathrm{d}w \, \partial_w \mathcal{E}^{(0)}_{\ell m}\left(w\right) \nonumber \\
    &= ~ - \sqrt{\ell(\ell+1)}\, \lambda^{(0)}_{\ell m} \sum_{\sigma = \pm} \int_0^\infty \frac{\mathrm{d}\omega}{2\pi} \sqrt{\frac{\omega}{2}} \Bigg[ \int_{-\infty}^{\infty} \mathrm{d} w \,e^{-i\omega w}\, b^{(\sigma)}_{\ell m}\left(\omega\right) \nonumber \\
    &\hspace{7cm} + \int_{-\infty}^{\infty} \mathrm{d} w \, e^{i\omega w} \, b^{(\sigma)\dagger}_{\ell m}\left(\omega\right) \Bigg] \nonumber\\
    &= ~ - \sqrt{\ell(\ell+1)} \, \lambda^{(0)}_{\ell m} \sum_{\sigma = \pm} \int_0^\infty \mathrm{d} \omega \, \sqrt{\frac{\omega}{2}} \, \delta(\omega)\, \Big( b^{(\sigma)}_{\ell m}\left(\omega\right) + b^{(\sigma)\dagger}_{\ell m}\left(\omega\right) \Big) \nonumber\\
    &= ~ - \dfrac{1}{2} \sqrt{\ell(\ell+1)} \, \lambda^{(0)}_{\ell m} \sum_{\sigma = \pm} \lim_{\omega \to 0} \sqrt{\dfrac{\omega}{2}} \, \Big( b^{(\sigma)}_{\ell m}\left(\omega\right) + b^{(\sigma)\dagger}_{\ell m}\left(\omega\right)\Big) \, .
\end{align}
In the second equality we inserted the quantisation \eqref{eqn:PsiModeExpansion}, identified the $\delta$-function in the third, and integrated the frequency out in the last.

\subsubsection{External states}\label{sec:externalstates}

We define all external states to be near the horizon and stay agnostic of their global embedding. These hard modes consist of two branches labelled by $\sigma$, as was the case for the photon. The null coordinate is related to Schwarzschild time by $w = t + \eta^{wr^*} r^*$, and we parametrise a scalar momentum two-vector as $\left(- E_p , p^{(\sigma)}_{r^*}\right)$ with $p^{(\sigma)}_{r^*} = \sigma \Omega_p + \eta^{wr^*} E_p$ and $\Omega_p = \sqrt{E^2_p - m^2_{\ell}}$, entirely as for the photon below \eqref{eqn:ka}. A plane wave is then
\begin{equation}
    e^{- i E_p w + i p^{(\sigma)}_{r^*} r^*} ~ = ~ e^{- i E_p \left(t + \eta^{wr^*} r^*\right) + i \left(\sigma \Omega_p + \eta^{wr^*} E_p\right) r^*} ~ = ~ e^{- i E_p t + i \sigma \Omega_p r^*} \, ,
\end{equation}
in either chart. The label is therefore chart independent as mentioned earlier: $\sigma = +1$ modes move radially outward and $\sigma = -1$ modes fall radially inward, and in the massless limit the two are functions of $u = t - r^*$ and of $v = t + r^*$ respectively.

The two branches are similar to those of the photon: in the chart adapted to a given horizon, the branch with $\eta^{wr^*}\sigma = -1$ crosses the horizon and the branch with $\eta^{wr^*}\sigma = +1$ runs along it. In momentum space this is read off from the radial momentum. In the massless limit $p^{(\sigma)}_{r^*} = \left(\sigma + \eta^{wr^*}\right)E_p$, which vanishes for the crossing branch, a function of $w$ alone, and equals $2\eta^{wr^*}E_p$ for the branch along the horizon. The phase $e^{2i\eta^{wr^*}E_p r^*}$ of that branch oscillates without limit as $r^* \rightarrow -\infty$. Note that $p_{r^*} = 0$ means the mode is constant along $\partial_{r^*}\big|_w$, which is the null direction through the horizon, not the direction along it. On the past horizon, $w = u$ and $\eta^{ur^*} = -1$, the outgoing modes cross, emerging through $\mathcal{H}^-$ on their way to future null infinity, and the ingoing modes run along it, having come through the potential barrier from past null infinity on their way into the hole. On the future horizon, $w = v$ and $\eta^{vr^*} = +1$, the ingoing modes cross and fall in, and the outgoing modes run along it and either reflect off the barrier or escape to future null infinity. Both branches are present near both horizons as free waves; what changes between $\mathcal{H}^-$ and $\mathcal{H}^+$ is which of the two crosses and which runs along.

At early Killing time the near-horizon region lies along $\mathcal{H}^-$ and at late Killing time along $\mathcal{H}^+$. It is therefore natural to think of the in-states as being defined near the past horizon, where the outgoing branch has just crossed and the ingoing branch is on its way in along it, and of the out-states as being defined near the future horizon, where the ingoing branch is about to cross and the outgoing branch is on its way out along it.

The states so defined describe the field for a limited Killing time. The mass $\mu_\ell\left(r^*\right)$ of the radiative mode grows away from the horizon, and an outgoing wave emitted near the horizon is reflected by it and returns as an ingoing wave. For a process at $r^*_0$ this happens after a Killing time of about $2\left|r^*_0\right|$. The out-state is the free out-expansion, which represents the field before the reflected wave returns, and the charges on $\mathcal{H}^+$ are built from it. The integrals over $w$ in these charges, and the limit $\omega \rightarrow 0$ in \eqref{eqn:softChargeOnHorizons}, refer to this free expansion continued to all $w$, and not to the field at Killing times after the return. The reflection is free propagation in the full potential and is not part of the near-horizon $S$-matrix. The same applies to the in-state, which represents the field for a Killing time of the same length before the process. This time grows without bound as the process is taken to the horizon.

The vacuum on which these external states are built is the one annihilated by the annihilation operators (such as $b^{(\sigma)}_{\ell m}\left(\omega\right)$ of \eqref{eqn:PsiQuantisation}), which are positive frequency in $w$ at fixed $r^*$, that is with respect to Killing time. It is the state in which the correlated limit \eqref{eqn:boundaryModeCondition} is taken. Near the horizon that is the Boulware state. It is not the state of a physical black hole: its renormalised stress tensor diverges on the horizon in a freely falling frame, so it is compatible with a fixed background only at finite $f\left(r\right)$, and it is in that sense the state of a cold stretched horizon. We will discuss the Hartle--Hawking state in detail in \secref{sec:irstate}.

\paragraph{Commutator of the cut charge with the $S$-matrix} The commutator of the cut charge with the $S$-matrix, between states of hard scalars can now be written as:
\begin{align}\label{eqn:softCommutator}
    \langle\text{out}| Q^+_{\text{cut}} \, S - S Q^-_{\text{cut}} | \text{in} \rangle ~ &= ~ - \dfrac{ \lambda^{(0)}_{\ell m}}{2} \sqrt{\dfrac{\ell \left(\ell + 1\right)}{2}} \nonumber \\
    &\hspace{1.5cm} \times \sum_{\sigma = \pm} \lim_{\omega \to 0} \sqrt{\omega} \, \langle \text{out} | \, b^{+,(\sigma)}_{\ell m}\left(\omega\right) S - S \, b^{-,(\sigma)\dagger}_{\ell m}\left(\omega\right) | \text{in} \rangle ,
\end{align}
with $b^{\pm}$ the soft operators on $\mathcal H^\pm$. The $\sqrt{\omega}$ appearing in the above commutator is a remnant of the canonical normalisation of the radiative mode.

\subsection{Hard charge and its action on external states}
The hard sector involves sources coupled minimally to Maxwell theory \eqref{eqn:MaxwellAction}. We consider massive charged scalars for this matter sector
\begin{equation}\label{eqn:MatterAction}
    S_{\text{matter}} ~ = ~ \int \mathrm{d}^4 x \, \mathcal{L}_{\text{matter}} ~ = ~ - \int \mathrm{d}^4 x \sqrt{-g} \, \left[\left(D_\mu \tilde{\phi} \right)^* \left(D^\mu \tilde{\phi} \right) + m^2 \left|\tilde{\phi}\right|^2\right] \, ,
\end{equation}
where $D_\mu \tilde{\phi} = \left(\nabla_\mu - i q A_\mu\right) \tilde{\phi}$ is the gauge and gravity covariant derivative written in terms of the gravity covariant derivative $\nabla$, the charge of the field $q$, and the gauge field, and $\tilde{\phi}$ is the scalar prior to the rescaling \eqref{eqn:scalarRescaling}. The conserved current associated with the global $U(1)$ symmetry (generated by $\tilde{\phi} \rightarrow e^{i \alpha} \tilde{\phi}$) is given by
\begin{align}\label{eqn:current}
    J_\mu ~ &= ~ - i q \left[\tilde{\phi} \left(D_\mu \tilde{\phi}\right)^* - \tilde{\phi}^* D_\mu \tilde{\phi}\right] \nonumber \\
    &= ~ - i q \left[\tilde{\phi} \, \partial_\mu \tilde{\phi}^* - \tilde{\phi}^* \partial_\mu \tilde{\phi}\right] + 2 q^2 A_\mu \left|\tilde{\phi}\right|^2 \, ,
\end{align}
where $\nabla_\mu$ reduces to $\partial_\mu$ on a scalar. We restrict our attention to the $A_\mu$-independent scalar-bilinear contribution to the hard current, which acts within the scalar-only hard sector. We therefore omit the gauge-field-dependent contribution associated with the quartic seagull interaction; it changes the photon number and contributes to Ward identities involving additional photon states or at higher perturbative order.

The combination of the current that appears in the hard charge \eqref{eqn:hardCharge} is ${\epsilon_w}^{b} J_b$. Writing $\widetilde{\partial} \coloneqq \partial_w + \eta^{w r^*}\partial_{r^*}$, and using ${\epsilon_w}^{b}\partial_b = \partial^{r^*} = \eta^{w r^*}\widetilde{\partial}$ together with the linearity of the current in the derivative,
\begin{align}\label{eqn:currentEpsilonCombination}
    {\epsilon_w}^{b} J_b ~ = ~ \eta^{w r^*} J \, , \qquad \text{with} \qquad
    J ~ \coloneqq ~ J_w + \eta^{w r^*} J_{r^*} ~ = ~ - i q \left(\tilde{\phi} \, \widetilde{\partial} \tilde{\phi}^* - \tilde{\phi}^* \widetilde{\partial} \tilde{\phi}\right) \, .
\end{align}
We now pass to the rescaled field $\phi_{\ell m} = r \, \tilde{\phi}_{\ell m}$ of \eqref{eqn:scalarRescaling}, in which the hard modes are naturally quantised. In the Wronskian combination the terms involving $\partial_a r$ cancel between the two halves,
\begin{align}\label{eqn:currentRescaling}
    \tilde{\phi}_{\ell_1 m_1} \partial_a \tilde{\phi}^*_{\ell_2 m_2} - \tilde{\phi}^*_{\ell_2 m_2}\partial_a \tilde{\phi}_{\ell_1 m_1} ~ = ~ \dfrac{1}{r^2}\left(\phi_{\ell_1 m_1} \partial_a \phi^*_{\ell_2 m_2} - \phi^*_{\ell_2 m_2}\partial_a \phi_{\ell_1 m_1}\right) \, ,
\end{align}
so that $r^2 J^{\ell m}_a$ is independent of $r$: the explicit factor of $r^2$ in \eqref{eqn:hardCharge} is exactly absorbed by the rescaling. Expanding the scalars in partial waves and projecting onto $Y_{\ell m}$ we find
\begin{align}\label{eqn:r2CurrentInPWs}
    r^2 \, {\epsilon_w}^{b} J^{\ell m}_b ~ = ~ - i q \, \eta^{w r^*} \sum_{\substack{\ell_1 m_1 \\ \ell_2 m_2}} C^{(3)} \left(\phi_{\ell_1 m_1} \widetilde{\partial}\phi^*_{\ell_2 m_2} - \phi^*_{\ell_2 m_2}\widetilde{\partial}\phi_{\ell_1 m_1}\right) \, ,
\end{align}
where we defined the integral of three spherical harmonics as in \eqref{eqn:Yintegrals}:
\begin{equation}\label{eqn:Y3integral}
	C^{(3)} ~ \coloneqq ~ \int \mathrm{d} \Omega \, Y_{\ell m} Y_{\ell_1 m_1} Y_{\ell_2 m_2}  ~ \eqqcolon ~ C\left[\ell m; \ell_{1} m_{1}; \ell_{2} m_{2}\right] \, .
\end{equation}
In what follows, we will use $C^{(3)}$ and $C\left[\ell m; \ell_{n} m_{n}; \ell_{j} m_{j}\right]$ interchangeably; we specify the arguments where there is potential for confusion when other quantities carry the same summed indices. 

The hard charge \eqref{eqn:hardCharge} of course contains a second term dependent on $\rho_{\ell m}$. It is negligible on the horizon. Since $J^+_{\ell m}$ carries only angular derivatives, $\Box_\eta \rho_{\ell m} = - f\left(r\right) J^+_{\ell m}$ gives $\partial^{r^*}\!\rho_{\ell m} \sim f\left(r\right) J^+_{\ell m}/E$ for modes of null energy $E$, whereas $r^2 {\epsilon_w}^{b}J^{\ell m}_b \sim E \, r^2 J^+_{\ell m}$ carries one null derivative. The two therefore stand in the ratio $\ell\left(\ell+1\right)f\left(r\right)/r^2E^2 \sim m^2_\ell/E^2$, which vanishes in the near-horizon limit by \eqref{eqn:scalarMass}. 

The hard charge \eqref{eqn:hardCharge} therefore reads
\begin{align}\label{eqn:hardChargeCurrentExpansion}
    Q_{\text{hard}}\left[\lambda^{(0)}_{\ell m}\right] ~ &= ~ i q \, \eta^{w r^*} \lambda^{(0)}_{\ell m} \sum_{\substack{\ell_1 m_1 \\ \ell_2 m_2}} C^{(3)} \lim_{r^* \to -\infty} \int \mathrm{d} w \, \Bigg( \phi_{\ell_1 m_1} \widetilde{\partial} \, \phi^*_{\ell_2 m_2} - \phi^*_{\ell_2 m_2} \widetilde{\partial} \, \phi_{\ell_1 m_1} \Bigg) \, ,
\end{align}
where we have also inserted the near-horizon expansion for the gauge parameter and kept the leading term.

In order to then evaluate this charge in the near-horizon limit, we need the appropriate quantisations of the hard scalars which we detail in \appref{app:hardModeQuantisation}. Defining the momentum two vector as
\begin{align}
    p_a ~ = ~ \left(- E_p , p^{(\sigma)}_{r^*}\right) \quad \text{where} \quad p^{(\sigma)}_{r^*}\left(E_p\right) ~ \coloneqq ~ \left(\sigma \Omega_p + \eta^{w r^*} E_p\right) \, , 
\end{align}
and
\begin{align}
    \Omega_p ~ \coloneqq ~ \sqrt{E_p^2 - m^2_{\ell}} \quad \text{where} \quad m^2_{\ell} ~ = ~ f\left(r\right) \left(\dfrac{\ell \left(\ell + 1\right) + 1}{R^2} + m^2\right) \, ,
\end{align}
we find the hard mode quantisation in \appref{app:hardModeQuantisation} to be
\begin{align}
\phi_{\ell_1 m_1} ~ &= ~ \sum_{\sigma_1 = \pm} \int_{m_{\ell_p}}^\infty \dfrac{\mathrm{d} E_p}{4\pi \Omega_p} \left(c^{(\sigma_1)}_{\ell_1 m_1}\left(E_p\right) V^{(\sigma_1)}_{E_p} + d^{(-\sigma_1)\dagger}_{\ell_1 m_1}\left(E_p\right) V^{(-\sigma_1)*}_{E_p} \right) \, ,
\end{align}
where
\begin{align}
    V^{(\sigma)}_{E_p} ~ = ~ e^{ - i w E_p + i r^* p^{(\sigma)}_{r^*}\left(E_p\right)} \, .
\end{align}
Using this mode expansion, it is straightforward to work out the hard charge. We detail this calculation in \appref{app:hardChargeCalculation} resulting in \eqref{appeqn:hardChargeResult} which we repeat here
\begin{align}\label{eqn:hardChargeResult}
    Q_{\mathrm{hard}}\left[\lambda^{(0)}_{\ell m}\right] ~ &= ~ q \lambda^{(0)}_{\ell m} \sum_{\substack{\ell_1m_1\\\ell_2m_2}} \sum_{\sigma_1=\pm} \int_0^\infty \frac{\mathrm{d}E_p}{4\pi E_p} \, \sigma_1 \nonumber \\
    &\hspace{2.5cm} \times \Big[ c^{(\sigma_1)\dagger}_{\ell_2m_2}(E_p) c^{(\sigma_1)}_{\ell_1m_1}(E_p) - d^{(\sigma_1)\dagger}_{\ell_1m_1}(E_p) d^{(\sigma_1)}_{\ell_2m_2}(E_p) \Big] C^{(3)} \, .
\end{align}

\paragraph{Action on external states} We now define external states in the following standard way
\begin{align}
    | \text{in}\rangle ~ = ~ \prod_{n' \in \text{in}} c^{q_{n'}, \left(\sigma_{n'}\right) \dagger}_{\ell_{n'} m_{n'}}\left(E_{p_{n'}}\right) | 0 \rangle \, ,
\end{align}
where the product runs over all allowed in-states and the extra label $q_{n'}=\pm q$ keeps track of the charge (and therefore whether particles or anti-particles are being created). That is 
\begin{align}
    c^{+,\left(\sigma\right)\dagger}_{\ell_{n'} m_{n'}} \left(E_{p_{n'}}\right) ~ = ~ c^{\left(\sigma\right)\dagger}_{\ell_{n'} m_{n'}} \left(E_{p_{n'}}\right) \quad \text{and} \quad  c^{-,\left(\sigma\right)\dagger}_{\ell_{n'} m_{n'}}\left(E_{p_{n'}}\right) ~ = ~ d^{\left(\sigma\right)\dagger}_{\ell_{n'} m_{n'}}\left(E_{p_{n'}}\right) \, .
\end{align}
Then, the action of the hard charge on such a generic multi-particle external state can be worked out to be
\begin{align}
    Q_{\mathrm{hard}}\left[\lambda^{(0)}_{\ell m}\right] | \text{in}\rangle ~ &= ~ \lambda^{(0)}_{\ell m} \sum_{\ell_j m_j} \sum_{n' \in \text{in}} \, \sigma_{n'} q_{n'} C\left[\ell m , \ell_{n'} m_{n'} , \ell_j m_j \right] | \mathcal{M}^{\text{in},\left(\sigma_{n'}\right)}_{n' j}\rangle \, ,
\end{align}
where we defined the quantity 
\begin{align}
    | \mathcal{M}^{\text{in},\left(\sigma_{n'}\right)}_{n' j}\rangle ~ \coloneqq ~ c^{q_1 , \left(\sigma_{1}\right)\dagger}_{\ell_1 m_1} \left(E_{p_1}\right) \dots c^{q_{n'} , \left(\sigma_{n'}\right) \dagger}_{\ell_j m_j} \left(E_{p_{n'}}\right) \dots |0\rangle \, ,
\end{align}
and made use of the commutators of the operators \eqref{eqn:scalarAlgebra}. 

This allows us to write the commutator of the hard charge with the scattering matrix in the following way 
\begin{align}\label{eqn:hardCommutator}
    \langle \text{out} | Q^+_{\text{hard}} S - S Q^-_{\text{hard}} | \text{in}\rangle &= \lambda^{(0)}_{\ell m} \sum_{\ell_j m_j} \Bigg( \sum_{n \in \text{out}} \, \sigma_{n} q_{n} C\left[\ell m ; \ell_n m_n ; \ell_j m_j \right] \mathcal{M}^{\text{out},\left(\sigma_{n}\right)}_{n j}\nonumber \\
    &\hspace{2.25cm} - \sum_{n' \in \text{in}} \, \sigma_{n'} q_{n'} C\left[\ell m ; \ell_{n'} m_{n'} ; \ell_j m_j \right] \mathcal{M}^{\text{in},\left(\sigma_{n'}\right)}_{n' j}\Bigg) ,
\end{align}
where we now defined the scattering matrix elements
\begin{align}
    \mathcal{M}^{\text{in},\left(\sigma_{n'}\right)}_{n' j} ~ \coloneqq ~ \langle \text{out}| S | \mathcal{M}^{\text{in},\left(\sigma_{n'}\right)}_{n' j}\rangle \quad \text{and} \quad \mathcal{M}^{\text{out},\left(\sigma_{n}\right)}_{n j} ~ \coloneqq ~ \langle \mathcal{M}^{\text{out},\left(\sigma_{n}\right)}_{n j}| S | \text{in} \rangle \, .
\end{align}
The hard commutator carries no power of $\omega$.

\subsection{The Ward--Takahashi identity near the horizon}\label{sec:wardidentity}

The Ward--Takahashi identity near the horizon is the statement that the soft charge \eqref{eqn:softCharge} commutes with the $S$-matrix,
\begin{align}\label{eqn:WardIdentitySoft}
    \langle \text{out}| Q^+_{\text{soft}} S - S Q^-_{\text{soft}} | \text{in} \rangle ~ &= ~ 0 \, .
\end{align}
Since $Q_{\text{soft}} = Q_{\text{cut}} - Q_{\text{hard}}$, this is the same as
\begin{align}\label{eqn:WardIdentityCharges}
    \langle \text{out}| Q^+_{\text{cut}} S - S Q^-_{\text{cut}} | \text{in} \rangle ~ &= ~ \langle \text{out}| Q^+_{\text{hard}} S - S Q^-_{\text{hard}} | \text{in} \rangle \, .
\end{align}
Here $Q^-$ and $Q^+$ are the charges of \secref{sec:covariantCharge} with the field replaced by the free in- and out-expansions of \secref{sec:externalstates}, which hold at the early and late ends of $\Sigma$ respectively.

The symmetry behind \eqref{eqn:WardIdentitySoft} is the large gauge transformation: by \appref{app:softGenerates} the soft charge generates the shift of the boundary coordinate. We show that the Ward identity \eqref{eqn:WardIdentitySoft} is equivalent to a soft theorem, which we establish in \secref{sec:softtheorem}, and the invariance of the $S$-matrix under the shift follows from that. The charge $Q_{\text{cut}}$ is the generator of the residual transformation with respect to the symplectic flux through $\Sigma$, in both sectors, by \appref{app:hardGenerates}. On the external states it produces the same shift, and in addition it acts on the scalars through $Q_{\text{hard}}$, which weights the two radial branches with opposite signs, as the eigenvalues in \eqref{eqn:hardCommutator} show. The scattering turns charges around, so $Q_{\text{hard}}$ does not commute with the $S$-matrix, and \eqref{eqn:WardIdentityCharges} states that $Q_{\text{cut}}$ fails to commute with it by exactly the amount by which $Q_{\text{hard}}$ does.

The simplest case is $\ell = 0$. There the soft charge \eqref{eqn:softCharge} vanishes identically, as does the term in $\rho_{\ell m}$ of \eqref{eqn:hardCharge}, so that $Q_{\text{cut}} = Q_{\text{hard}}$. The cut charge \eqref{eqn:electricCharge} is the monopole flux of the radial electric field through the cut, which is the charge enclosed by $\Sigma$. Take an in-state with a charge $q$ moving inward and an out-state in which it moves outward. The hard charge of the in-state is $-q\lambda_{00}/\sqrt{4\pi}$ and that of the out-state is $+q\lambda_{00}/\sqrt{4\pi}$, so both sides of \eqref{eqn:WardIdentityCharges} are $2q\lambda_{00}/\sqrt{4\pi}$ times the amplitude. Neither charge is conserved, and the identity holds because the two are the same operator. For $\ell > 0$ they differ by the soft charge and the identity is nontrivial: the change in the hard charge is equal to the change in $Q_{\text{cut}}$, which by \eqref{eqn:softChargeOnHorizons} is carried by soft photons. We verify this at tree level in \secref{sec:softtheorem}. For free propagation, which preserves the branch label of every leg, both sides vanish.

Inserting \eqref{eqn:softCommutator} and \eqref{eqn:hardCommutator} into \eqref{eqn:WardIdentityCharges}, we may write the identity as
\begin{align}\label{eqn:WardIdentity}
    &\sum_{\sigma = \pm} \langle \text{out} | b^{+,(\sigma)}_{\ell m} S - S b^{-,(\sigma) \dagger}_{\ell m} | \text{in} \rangle = - \, 2 \sqrt{\dfrac{2}{\ell \left(\ell + 1\right)}} \nonumber \\
    &\hspace{5.5cm} \times ~ \sum_{\ell_j m_j} \Bigg( \sum_{n \in \text{out}} \, \dfrac{\sigma_{n} q_{n}}{\sqrt{\omega}} C\left[\ell m ; \ell_n m_n ; \ell_j m_j \right] \mathcal{M}^{\text{out},\left(\sigma_{n}\right)}_{n j} \nonumber \\
    &\hspace{6.5cm}- \sum_{n' \in \text{in}} \, \dfrac{\sigma_{n'} q_{n'}}{\sqrt{\omega}} C\left[\ell m ; \ell_{n'} m_{n'} ; \ell_j m_j \right] \mathcal{M}^{\text{in},\left(\sigma_{n'}\right)}_{n' j} \Bigg) .
\end{align}
As mentioned below \eqref{eqn:softCommutator}, the apparent $1/\sqrt{\omega}$ pole is an artefact of the normalisation of the state and not owed to the dynamics.

\section{A new soft photon theorem near the horizon}\label{sec:softtheorem}

We are interested in the leading single soft photon theorem near the horizon. The corresponding effective theory governing the near-horizon scattering was first studied in \cite{Feleppa:2023eoi} building on the framework in \cite{Gaddam:2020mwe, Gaddam:2020rxb, Gaddam:2021zka, Gaddam:2022pnb}. While the photons there were virtual participants in ladder diagrams, we are now interested in on-shell external photons. Therefore, it is convenient to work with an effective theory for the radiative mode described in \secref{sec:radMode}. We review the construction of the effective theory for this on-shell photon mode in \appref{app:MaxwellinPWs}. The relevant Feynman rule follows from the interaction \eqref{eqn:Psivertex} and is given by \eqref{eqn:vertex},

\vspace{0.5cm}
\begin{fmffile}{feyn_bhvertex}
\begin{align}
  \begin{fmfgraph*}(70,70)
    \fmfleft{i8,i7}
    \fmfright{o2}
    \fmf{wiggly}{i7,v1}
    \fmf{fermion}{i8,v1}
    \fmf{fermion}{v1,o2}
    \fmflabel{$\Psi_{\ell m}\left(k\right)$}{i7}
    \fmflabel{$\phi_{\ell_j m_j} \left(p_n + k\right)$}{i8}
    \fmflabel{$\phi^{*}_{\ell_n m_n} \left(p_n\right)$}{o2}
    \fmfdot{v1}
    \fmfv{label=$ \dfrac{2 i q }{\sqrt{2 \tilde{\omega} \ell \left(\ell + 1\right)}} \varepsilon_{(\sigma)}\left(\omega\right) \cdot p_{n} C^{(3)}$,label.angle=60,label.dist=30}{v1}
    \marrow{a}{right}{rt}{$k$}{v1,i7}
    \marrow{b}{right}{rt}{$p_j$}{i8,v1}
    \marrow{c}{up}{top}{$p_n$}{v1,o2}
  \end{fmfgraph*}  \qquad \qquad \qquad \label{eqn:bhvertex}
\end{align}
\vspace{0cm}
\end{fmffile}

\noindent Here, we parametrise the photon momentum on the horizon as $k_a = \left(-\omega, k_{(\sigma)}\right)$ with $k_{(\sigma)}$ fixed by the mass-shell condition \eqref{eqn:ka}, defined $\tilde{\omega} = \sqrt{\omega^2 - \mu^2_\ell}$, and $\varepsilon^{(\sigma)}_a\left(\omega\right)$ is the polarisation vector \eqref{eqn:polarisation} generated by $\epsilon_{ab}\partial^b$ acting on the Moncrief mode. The scalar propagator is given by \eqref{eqn:bhscalarprop}
\begin{align}
\mathcal{P}_{\phi_{\ell m}} ~ &= ~ \dfrac{- i}{p^{a} p_{a} + m^2_{\ell} - i \epsilon} \, . 
\end{align}
Then, the soft-emission amplitude is obtained in the standard way: attach the soft $\Psi_{\ell m}$ leg to each external hard scalar leg, include the scalar propagator on the shifted leg, and sum over the intermediate partial wave $(\ell_j,m_j)$:
\begin{align}\label{eqn:softAmplitude}
&\mathcal{M}_{b^{(\sigma)}}
\left(
k \ell m;\{p^{\mathrm{out}}\ell^{\mathrm{out}}m^{\mathrm{out}}\};
\{p^{\mathrm{in}}\ell^{\mathrm{in}}m^{\mathrm{in}}\}
\right)
\nonumber\\[0.3em]
&\qquad
=
\sum_{\ell_j m_j}
\Biggl(
\sum_{n=1}^{n_{\mathrm{out}}}
\mathcal{M}
\left(
p^{\mathrm{out}}_1\ell^{\mathrm{out}}_1m^{\mathrm{out}}_1,\dots,
\left(p^{\mathrm{out}}_n + k\right)\ell_j m_j,\dots,
p^{\mathrm{out}}_{n_{\text{out}}} \ell^{\mathrm{out}}_{n_{\text{out}}} m^{\mathrm{out}}_{n_{\text{out}}};
\{p^{\mathrm{in}},\ell^{\mathrm{in}}m^{\mathrm{in}}\}
\right)
\nonumber\\
&\qquad\qquad\qquad\qquad\times
\frac{-i}{\bigl((p_n^{\mathrm{out}} + k)^2+m^2_{\ell_j}-i\epsilon\bigr)}
\left[
\dfrac{2 \, i \, q_n \left(\varepsilon_{(\sigma)}\left(\omega\right) \cdot p^{\mathrm{out}}_{n} \right)}{\sqrt{2 \tilde{\omega} \ell \left(\ell + 1\right)}} \, C^{(3)} 
\right]
\nonumber\\
&\qquad\qquad
+
\sum_{n'=1}^{n_{\mathrm{in}}}
\mathcal{M}
\left(
\{p^{\mathrm{out}},\ell^{\mathrm{out}}m^{\mathrm{out}}\};
p^{\mathrm{in}}_1\ell^{\mathrm{in}}_1m^{\mathrm{in}}_1,\dots,
\left(p^{\mathrm{in}}_{n'} - k\right)\ell_j m_j,\dots,
p^{\mathrm{in}}_{n_{\text{in}}}\ell^{\mathrm{in}}_{n_{\text{in}}} m^{\mathrm{in}}_{n_{{\text{in}}}}
\right)
\nonumber\\
&\qquad\qquad\qquad\qquad\times
\frac{-i}{\bigl((p_{n'}^{\mathrm{in}} - k)^2+m^2_{\ell_j}-i\epsilon\bigr)}
\left[
\dfrac{2 \, i \, q_{n'} \left(\varepsilon_{(\sigma)}\left(\omega\right) \cdot p^{\mathrm{in}}_{n'} \right)}{\sqrt{2 \tilde{\omega} \ell \left(\ell + 1\right)}} \, C^{(3)}
\right]
\Biggl) \, .
\end{align}
In writing this expression, we have already inserted the momentum conserving delta functions to perform the $p_j^{\mathrm{out/in}}$ integrals and consequently set:
\begin{equation}
    p_j^{\mathrm{out}} ~ = ~ p_n^{\mathrm{out}} + k \quad \text{and} \quad p_j^{\mathrm{in}} ~ = ~ p_{n'}^{\mathrm{in}} - k \, .
\end{equation}
To avoid clutter of notation, we now define the quantities
\begin{align}
    \mathcal{M}^{\text{out}}_{n}\left(k\right) ~ &\coloneqq ~ \mathcal{M}
    \left(
    p^{\mathrm{out}}_1\ell^{\mathrm{out}}_1m^{\mathrm{out}}_1,\dots,
    \left(p^{\mathrm{out}}_n + k\right) \ell_j m_j,\dots,
    p^{\mathrm{out}}_{n_{\mathrm{out}}} \ell^{\mathrm{out}}_{n_{\mathrm{out}}}m^{\mathrm{out}}_{n_{\mathrm{out}}};
    \{p^{\mathrm{in}},\ell^{\mathrm{in}}m^{\mathrm{in}}\}
    \right) , \\
    \mathcal{M}^{\text{in}}_{n'}\left(k\right) ~ &\coloneqq ~ \mathcal{M}
    \left(
    \{p^{\mathrm{out}},\ell^{\mathrm{out}}m^{\mathrm{out}}\};
    p^{\mathrm{in}}_1\ell^{\mathrm{in}}_1m^{\mathrm{in}}_1,\dots,
    \left(p^{\mathrm{in}}_{n'} - k\right) \ell_j m_j,\dots,
    p^{\mathrm{in}}_{n_{\mathrm{in}}} \ell^{\mathrm{in}}_{n_{\mathrm{in}}}m^{\mathrm{in}}_{n_{\mathrm{in}}}
    \right) \, .
\end{align}
The denominators simplify on shell. Imposing $\left(p^{\mathrm{out}}_n\right)^2 + m^2_{\ell_n} = 0$, $\left(p^{\mathrm{in}}_{n'}\right)^2 + m^2_{\ell_{n'}} = 0$ and $k^2 + \mu^2_\ell = 0$ for the hard scalars and the photon,
\begin{align}
    \left(p^{\mathrm{out}}_n + k\right)^2 + m^2_{\ell_j} ~ &= ~ \phantom{-} 2 \left[ k \cdot p^{\mathrm{out}}_{n} + \tfrac{1}{2}\left(m^2_{\ell_j} - m^2_{\ell_n} - \mu^2_\ell\right)\right] \, , \nonumber \\
    \left(p^{\mathrm{in}}_{n'} - k\right)^2 + m^2_{\ell_j} ~ &= ~ - 2 \left[ k \cdot p^{\mathrm{in}}_{n'} - \tfrac{1}{2}\left(m^2_{\ell_j} - m^2_{\ell_{n'}} - \mu^2_\ell\right)\right] \, ,
\end{align}
the relative sign being the origin of the relative minus between the outgoing and incoming sums below. The amplitude then becomes
\begin{align}
&\mathcal{M}_{b^{(\sigma)}}
\left(
k \ell m;\{p^{\mathrm{out}}\ell^{\mathrm{out}}m^{\mathrm{out}}\};
\{p^{\mathrm{in}}\ell^{\mathrm{in}}m^{\mathrm{in}}\}
\right)
\nonumber\\[0.3em]
&\quad
=
\sum_{\ell_j m_j}
\Biggl(
\sum_{n=1}^{n_{\mathrm{out}}}
\mathcal{M}^{\text{out}}_{n}\left(k\right)
\frac{1}{\sqrt{2 \tilde{\omega} \ell \left(\ell + 1\right)}}
\left[
q_n \dfrac{ \left(\varepsilon_{(\sigma)}\left(\omega\right) \cdot p^{\mathrm{out}}_{n} \right)}{\bigl(k \cdot p_n^{\mathrm{out}} \bigr) + \frac{1}{2} \left(m^2_{\ell_j} - m^2_{\ell_n} - \mu^2_\ell\right)} \, C^{(3)} 
\right]
\nonumber\\
&\hspace{1.5cm}
-
\sum_{n'=1}^{n_{\mathrm{in}}}
\mathcal{M}^{\text{in}}_{n'}\left(k\right)
\frac{1}{\sqrt{2 \tilde{\omega} \ell \left(\ell + 1\right)}}
\left[
q_{n'} \dfrac{ \left(\varepsilon_{(\sigma)}\left(\omega\right) \cdot p^{\mathrm{in}}_{n'} \right)}{\bigl( k \cdot p_{n'}^{\mathrm{in}} \bigr) - \frac{1}{2}\left(m^2_{\ell_j} - m^2_{\ell_{n'}} - \mu^2_\ell\right)} \, C^{(3)}
\right]
\Biggl) \, .
\end{align}

To simplify this expression, we resort to the near-horizon soft limit described in \eqref{eqn:softlimit}. All squared masses are proportional to $f\left(r\right)$ and are therefore exponentially suppressed in the near-horizon limit $r^* \to -\infty$,
\begin{align}\label{eqn:masses}
    \mu^2_{\ell} ~ = ~ \dfrac{f\left(r\right) \ell \left(\ell+1\right)}{r^2} ~ \sim ~ \dfrac{e^{r^*/R} \, \ell \left(\ell+1\right)}{r^2} \, , \qquad m^2_{\ell} ~ \propto ~ f\left(r\right) \, .
\end{align}
In the same limit $\tilde{\omega} \rightarrow \omega$. In momentum space the masses cannot be dropped at the outset, since on the half of the phase space where photon and hard leg move on the same branch they fix an otherwise undefined ratio, and the value they fix it to does not depend on them; \appref{app:interactionTerms} states the regime in which holding them fixed is justified, and what fixes the soft factor outside it. The ratio of vertex numerator to propagator denominator is computed there for both branches, collinear and non-collinear, and is the same in the two cases \eqref{appeqn:vertexRatio}. We repeat the result here
\begin{align}\label{eqn:vertexRatio}
    \dfrac{\varepsilon_{(\sigma)}\left(\omega\right)\cdot p_{n}}{k \cdot p_n \pm \frac{1}{2}\left(m^2_{\ell_j} - m^2_{\ell_n} - \mu^2_\ell\right)} ~ \approx ~ - \, \sigma_{p_n} \, ,
\end{align}
independently of the photon branch label $\sigma$ and of the sign of the mass term, which is subleading in the limit \eqref{eqn:softlimit}. In the plane-wave basis the corresponding factor $\varepsilon\cdot p_n/k\cdot p_n$, with a unit polarisation, carries the $1/\omega$ pole. In the partial wave basis, it is independent of $\omega$: the polarisation vector \eqref{eqn:polarisation} is built from the photon momentum and carries a power of $\omega$ itself, so that numerator and denominator scale alike on the half of the phase space where photon and scalar leg move on opposite branches, while on the collinear half their leading terms cancel and the masses fix the ratio, again with no power of $\omega$. The soft limit is therefore finite. Substituting \eqref{eqn:vertexRatio} into the amplitude above,
\begin{align}\label{eqn:softAmplitudeReduced}
&\mathcal{M}_{b^{(\sigma)}}
\left(
k \ell m;\{p^{\mathrm{out}}\ell^{\mathrm{out}}m^{\mathrm{out}}\};
\{p^{\mathrm{in}}\ell^{\mathrm{in}}m^{\mathrm{in}}\}
\right) 
\nonumber\\[0.3em]
&\qquad
=
\frac{-1}{\sqrt{2 \omega \ell \left(\ell + 1\right)}} \sum_{\ell_j m_j}
\Biggl(
\sum_{n=1}^{n_{\mathrm{out}}}
q_n \sigma_{n} C\left[\ell m , \ell_n m_n ; \ell_j m_j\right] \mathcal{M}^{\text{out}}_{n}\left(k\right) \nonumber \\
&\hspace{5cm} -
\sum_{n'=1}^{n_{\mathrm{in}}}
q_{n'} \sigma_{{n'}} C\left[\ell m , \ell_{n'} m_{n'} ; \ell_j m_j\right] \mathcal{M}^{\text{in}}_{n'}\left(k\right)
\Biggl)  \, .
\end{align}
This is the result for a single branch of the photon. Defining the matrix elements appearing in the Ward identity \eqref{eqn:WardIdentity},
\begin{align}
    \mathcal M_{b^{(\sigma)}} ~ \coloneqq ~ \langle\mathrm{out}| b^{+,(\sigma)}_{\ell m} S |\mathrm{in}\rangle \qquad \text{and} \qquad \mathcal M_{b^{(\sigma)\dagger}} ~ \coloneqq ~ \langle\mathrm{out}| S \, b^{-,(\sigma)\dagger}_{\ell m} |\mathrm{in}\rangle \, ,
\end{align}
we observe that the right-hand side of \eqref{eqn:softAmplitudeReduced} is independent of $\sigma$, since the ratio quoted above is. As noted in \appref{app:interactionTerms}, this holds leg by leg in the gauge used there and, in the gauge $A_w = 0$ in which the emission of each leg sits on its own branch, for the sum over legs, by charge conservation; the amplitude is the same. Writing $\mathcal{M}_{b}$ for this common value,
\begin{align}\label{eqn:branchSum}
    \sum_{\sigma} \mathcal{M}_{b^{(\sigma)}} ~ = ~ 2 \, \mathcal{M}_{b} \qquad \text{and} \qquad \sum_{\sigma} \mathcal M_{b^{(\sigma)\dagger}} ~ = ~ - \, 2 \, \mathcal M_{b} \, .
\end{align}
The second equality arises from the observation that the crossed incoming photon contribution from the creation operator reverses the sign of the photon momentum $k_a$ in \eqref{eqn:softAmplitude}. This results in the propagator appearing with the opposite sign and consequently leads to the overall minus sign. Finally, combining all insertions as in the Ward identity \eqref{eqn:WardIdentity}, we find
\begin{align}\label{eqn:finalAmp}
\sum_{\sigma} \Big(\mathcal{M}_{b^{(\sigma)}} - \mathcal{M}_{b^{(\sigma)\dagger}}\Big) &= - \, 2 \sqrt{\frac{2}{\omega \ell \left(\ell + 1\right)}} \nonumber \\
&\hspace{1.5cm} \times \sum_{\ell_j m_j}
\Biggl(
\sum_{n=1}^{n_{\mathrm{out}}}
q_n \sigma_{n} C\left[\ell m , \ell_n m_n ; \ell_j m_j\right] \mathcal{M}^{\text{out}}_{n}\left(k\right) \nonumber \\
&\hspace{3cm} -
\sum_{n'=1}^{n_{\mathrm{in}}}
q_{n'} \sigma_{{n'}} C\left[\ell m , \ell_{n'} m_{n'} ; \ell_j m_j\right] \mathcal{M}^{\text{in}}_{n'}\left(k\right)
\Biggl) .
\end{align}
This result is valid for all physical external radiative photon modes $\ell > 0$, and agrees with the Ward identity \eqref{eqn:WardIdentity}. It establishes the identity at tree level on the half of the phase space where photon and hard leg move on opposite branches, for every frequency in \eqref{eqn:softlimit}, and on the collinear half in the regime \eqref{eqn:frozenMassRegime}. Outside that regime the collinear emission is not completed within the near-horizon region, and the momentum-space computation does not control the collinear half. There the radiated field is computed in \appref{app:interactionTerms} from the field equation \eqref{eqn:sourcedEquations} for $\mathcal{E}_{\ell m}$, at leading order near the process. The result \eqref{eqn:classicalStep} is the same step on each branch, with the value that \eqref{eqn:softAmplitudeReduced} gives and with no reference to the masses. A derivation in the full potential is left to future work. With this, the Ward identity \eqref{eqn:WardIdentity} holds, and the $S$-matrix commutes with the generator of the large gauge transformation.

As mentioned below \eqref{eqn:WardIdentity}, the soft factor is finite in the soft limit $\omega \to 0$. The apparent $1/\sqrt{\omega}$ pole is a mere artefact of the canonical phase space normalisation of the radiative mode \eqref{appeqn:PsiQuantisation}. This is a property of the partial-wave basis and of the polarisation it generates, not of the horizon. Writing $\varepsilon_{(\sigma)} = \omega\,\hat\varepsilon_{(\sigma)}$ with $\hat\varepsilon_{(\sigma)}$ finite as $\omega \rightarrow 0$, \eqref{eqn:vertexRatio} reads $\hat\varepsilon_{(\sigma)}\cdot p_n/k\cdot p_n = -\sigma_{p_n}/\omega$ on both halves of the phase space, Weinberg’s simple pole with residue $-\sigma_{p_n}q_n$; the factor $\omega$ of the polarisation and the normalisation $\left(2\tilde\omega\right)^{-1/2}$ of the external photon multiply every diagram alike and turn the pole into the finite value above. Universality is therefore guaranteed by Weinberg’s argument: attachments to internal lines and vertices are off shell by hard amounts and contribute at relative order $\omega/E_n$, as does the dependence of the hard amplitude on $k$, and \eqref{eqn:finalAmp} is the exact residue at tree level, corrected by $\mathcal{O}\left(\omega/E_n\right)$ and $\mathcal{O}\left(\mu^2_\ell/\omega^2\right)$. The same applies to the partial waves of the flat-space photon, including at null infinity. In the plane-wave basis the counting is different because of the normalisation of the oscillators. With the covariant normalisation the mode expansion of the radiative field at $\mathcal{I}^+$ carries $\mathrm{d}\omega$ with no power of $\omega$, and the pole sits entirely in the amplitude. With oscillators normalised to a unit commutator in frequency, in plane waves or in partial waves, the field carries a factor of $\omega^{-1/2}$ per mode and the amplitude carries the other factor of $\omega^{-1/2}$, which enters through the field in the vertex. The symplectic form fixes only the product. What does not depend on these choices is the emission probability of \secref{sec:ircounting} and the step in the radiated field, which we compute in \eqref{eqn:radiatedStep}.

\subsection{Gauge invariance and charge conservation}\label{sec:gaugeinvariance}

By \eqref{eqn:gaugetransfo} the transverse even-sector mode $A^{+}_{\ell m}$ is pure gauge and carries no dynamical degree of freedom in the bulk, yet it couples to the matter through the vertex \eqref{eqn:A+vertex}. In what follows, we will show that this pure-gauge configuration can be removed by a field redefinition. A configuration in which the gauge-invariant one-form vanishes, $\chi^{\ell m}_a = 0$ with $A^{+}_{\ell m} = \lambda_{\ell m}$, is by \eqref{eqn:AaSoln} and \eqref{eqn:gaugetransfo} the four-dimensional configuration $A_\mu = \partial_\mu \lambda$. Its $\ell = 0$ part is $A^{00}_a = \partial_a \lambda_{00}$, with no transverse component. Writing the charged scalar of \eqref{eqn:MatterAction} as
\begin{equation}\label{eqn:fieldRedefinition}
    \tilde{\phi} \eqqcolon e^{i q \lambda} \, \tilde{\phi}' \, ,
    \quad \text{so that} \quad
    D_\mu \tilde{\phi} = \left(\nabla_\mu - i q \, \partial_\mu \lambda\right)
    e^{i q \lambda}\tilde{\phi}' = e^{i q \lambda} \nabla_\mu \tilde{\phi}'
    \quad \text{and} \quad
    \left|\tilde{\phi}\right|^2 = \left|\tilde{\phi}'\right|^2 \, ,
\end{equation}
the matter action becomes the free action for $\tilde{\phi}'$ and $\lambda$ has disappeared. This is exact and to all orders in $q$.

The redefinition \eqref{eqn:fieldRedefinition} is not the identity on the asymptotic states. By \eqref{lambda} the residual gauge parameter is independent of the null coordinate, so on each horizon its value is a function on the sphere alone, with leading near-horizon term $\lambda^{(0)}_{\ell m}$ as in \eqref{eqn:AplusNearHor}. Each external leg of charge $q_n$ localised on the sphere at $\Omega_n$ is multiplied by $e^{i q_n \lambda}$ evaluated there, and therefore
\begin{equation}\label{eqn:SmatrixPhase}
    S \Big|_{A_{\mu} = \partial_{\mu} \lambda} ~ = ~
    \exp\left[ i \sum_{\ell m} \lambda^{(0)}_{\ell m}
    \left( \sum_{n \in \mathrm{out}} q_n Y_{\ell m}\left(\Omega_n\right)
    - \sum_{n' \in \mathrm{in}} q_{n'} Y_{\ell m}\left(\Omega_{n'}\right)\right)\right] S \, .
\end{equation}
The entire effect of the pure-gauge configuration on the $S$-matrix is this phase, and it depends on $\lambda_{\ell m}$ only through its boundary value $\lambda^{(0)}_{\ell m}$. Three statements follow immediately.

\begin{itemize}
    \item \emph{Small gauge transformations decouple.} If $\lambda^{(0)}_{\ell m} = 0$ the phase is unity for every $\left(\ell , m\right)$, whatever the profile of $\lambda_{\ell m}\left(r^*\right)$ in the bulk. Decoupling is a condition on the boundary value alone.
    \item \emph{The $s$-wave is charge conservation.} For $\ell = 0$ the harmonic is constant, the exponent is proportional to $\sum_{n \in \mathrm{out}} q_n - \sum_{n' \in \mathrm{in}} q_{n'}$, and the phase is unity by charge conservation. This has no radiative counterpart. The gauge parameter $\lambda_{\ell m}$ exists for every $\ell \geq 0$, whereas $A^{+}_{\ell m}$ and the Moncrief mode $\Psi_{\ell m}$ exist only for $\ell > 0$. At $\ell = 0$ the parameter is constant on the sphere, and the transformation is the global phase rotation.
    \item \emph{Large transformations act on the boundary, not on the hard phase.} For $\ell > 0$ with $\lambda^{(0)}_{\ell m} \neq 0$ the exponent of \eqref{eqn:SmatrixPhase} is the difference between the outgoing and incoming charge distributions on the sphere, angle by angle. At leading order in the near-horizon limit the charge at each angle is conserved, as we show in \eqref{eqn:angularLocality} below, and the phase is unity for every $\ell$: the pure-gauge mode decouples from the hard $S$-matrix for large transformations as it does for small ones. What a large transformation does change is the boundary. The corner term of the presymplectic potential \eqref{eqn:MaxwellThetaPolarised} evaluated on $\delta_\lambda A^{+}_{\ell m}$ is $-Q_{\mathrm{soft}}\left[\lambda_{\ell m}\right]$, which by \appref{app:softGenerates} generates the shift of the boundary coordinate, vanishes for a small parameter and does not for a non-vanishing constant one. How this relates to the Ward identity \eqref{eqn:WardIdentitySoft} is discussed below.
\end{itemize}

\paragraph{Conservation of the charge at each angle.} The current \eqref{eqn:current}, with its seagull term, is conserved, $\nabla_\mu J^\mu = 0$. With $\sqrt{-g} = f\left(r\right) r^2 \sqrt{\gamma}$, $g^{ab} = \eta^{ab}/f\left(r\right)$ and $g^{AB} = \gamma^{AB}/r^2$ this reads
\begin{equation}\label{eqn:angularLocality}
    \eta^{ab} \partial_a \left(r^2 J_b\right) + f\left(r\right) \hat{\nabla}^A J_A ~ = ~ 0 \, .
\end{equation}
The transverse current enters with a factor of $f\left(r\right)$. Relative to the first term it is of order $m^2_\ell/E^2_p$, the ratio that was neglected above \eqref{eqn:hardChargeCurrentExpansion}. At leading order in the near-horizon limit the two-dimensional current $r^2 J_a$ is therefore conserved at each angle separately. This uses only the global $U(1)$ symmetry and the form of the metric \eqref{eqn:metric}. It holds for the partial waves with $\ell\left(\ell+1\right) \ll E^2_p R^2/f\left(r\right)$, below the value at which the window \eqref{eqn:softlimit} closes, and for processes confined to the near-horizon region. On in- and out-states that contain every charge, including those that have crossed the surface, the charge at each angle is then the same, and the exponent of \eqref{eqn:SmatrixPhase} vanishes for every such $\left(\ell, m\right)$.

\paragraph{The phase and the Ward identity.} The boundary coordinate $\varphi_{\ell m}$ of \eqref{eqn:AplusBranches} enters the dynamics only through the configuration $A^{+}_{\ell m} = \varphi_{\ell m}$ with $\chi^{\ell m}_a = 0$, which is the pure-gauge configuration $A_\mu = \partial_\mu \lambda$ with $\lambda_{\ell m} = \varphi_{\ell m}$. By \eqref{eqn:SmatrixPhase} the $S$-matrix depends on it only through the phase, with $\lambda^{(0)}_{\ell m}$ replaced by $\varphi^{(0)}_{\ell m}$, and by \eqref{eqn:angularLocality} the phase is unity. The scattering of the hard scalars therefore does not depend on $\varphi^{(0)}_{\ell m}$. This is consistent with the Ward identity \eqref{eqn:WardIdentitySoft}, whose charge generates the shift of $\varphi^{(0)}_{\ell m}$, but it does not by itself imply it. By \eqref{eqn:softCharge} the soft charge is the difference of $Q_{\text{cut}}$, which is linear in the radiative zero mode, and $Q_{\text{hard}}$, and the identity is the statement that their changes cancel. Deriving that from the symmetry would require the brackets on the phase space of both sectors with the corner pair included, which we have not constructed.

As at null infinity, the identity is established by the soft theorem. The amplitude of \secref{sec:softtheorem} gives it where the momentum-space computation is controlled, and elsewhere the field equations do. By \eqref{eqn:classicalStep} and \eqref{eqn:GaussStep} the change of $\mathcal{E}_{\ell m}$ between the in- and out-expansions equals that of the bracket of \eqref{eqn:MoncriefMaster}, so that the two expansions of $\Psi_{\ell m}$ coincide,
\begin{equation}\label{eqn:DeltaPsiConserved}
    \langle \text{out} | \, \Delta\Psi^{+}_{\ell m} \, S - S \, \Delta\Psi^{-}_{\ell m} \, | \text{in} \rangle ~ = ~ 0 \, ,
\end{equation}
which is \eqref{eqn:WardIdentitySoft} by \eqref{eqn:softCharge}. This holds for Killing times before the reflected wave of \secref{sec:externalstates} returns. The source of $\Psi_{\ell m}$ in \eqref{eqn:sourcedEquations} is of order one far from the horizon, and its response reaches the near-horizon region only then. In the form \eqref{eqn:WardIdentityCharges} the two radial branches enter with opposite signs and the difference does not cancel.

\paragraph{Diagrammatics.} It is useful to see the infinitesimal form of \eqref{eqn:SmatrixPhase} emerge from the vertex \eqref{eqn:A+vertex}, both as a check on the partial-wave reduction and for comparison with \eqref{eqn:finalAmp}. Attaching a single external $A^{+}_{\ell m}$ leg to each hard scalar leg, exactly as in \eqref{eqn:softAmplitude},
\begin{align}
&\mathcal{M}_{A^+_{\ell m}}
\left(
k \ell m;\{p^{\mathrm{out}}\ell^{\mathrm{out}}m^{\mathrm{out}}\};
\{p^{\mathrm{in}}\ell^{\mathrm{in}}m^{\mathrm{in}}\}
\right) ~
\nonumber\\[0.3em]
&\quad
=
\sum_{\ell_j m_j}
\Biggl(
\sum_{n=1}^{n_{\mathrm{out}}}
\mathcal{M}
\left(
p^{\mathrm{out}}_1\ell^{\mathrm{out}}_1m^{\mathrm{out}}_1,\dots,
\left(p^{\mathrm{out}}_n + k\right)\ell_j m_j,\dots,
p^{\mathrm{out}}_{n_{\text{out}}} \ell^{\mathrm{out}}_{n_{\text{out}}} m^{\mathrm{out}}_{n_{\text{out}}};
\{p^{\mathrm{in}},\ell^{\mathrm{in}}m^{\mathrm{in}}\}
\right)
\nonumber\\
&\qquad\qquad\times
\frac{-i}{\bigl((p_n^{\mathrm{out}} + k)^2+m^2_{\ell_j}-i\epsilon\bigr)}
\left[
- i q_n \Big( k \cdot \left(2 p^{\mathrm{out}}_{n} + k\right) + m^2_{\ell_j} - m^2_{\ell_n} \Big) C^{(3)}
\right]
\nonumber\\
&\qquad\qquad
+
\sum_{n'=1}^{n_{\mathrm{in}}}
\mathcal{M}
\left(
\{p^{\mathrm{out}},\ell^{\mathrm{out}}m^{\mathrm{out}}\};
p^{\mathrm{in}}_1\ell^{\mathrm{in}}_1m^{\mathrm{in}}_1,\dots,
\left(p^{\mathrm{in}}_{n'} - k\right)\ell_j m_j,\dots,
p^{\mathrm{in}}_{n_{\text{in}}}\ell^{\mathrm{in}}_{n_{\text{in}}} m^{\mathrm{in}}_{n_{{\text{in}}}}
\right)
\nonumber\\
&\qquad\qquad\times
\frac{-i}{\bigl((p_{n'}^{\mathrm{in}} - k)^2+m^2_{\ell_j}-i\epsilon\bigr)}
\left[
- i q_{n'} \Big( k \cdot \left(2 p^{\mathrm{in}}_{n'} - k\right) + m^2_{\ell_{n'}} - m^2_{\ell_j} \Big) C^{(3)}
\right]
\Biggl) \, ,
\end{align}
where the vertex is quoted directly as the Feynman rule, without the additional factor of $-i$ that accompanies \eqref{eqn:vertex}, because the external $A^{+}_{\ell m}$ leg is a c-number gauge parameter and not a quantised mode. On shell, $\left(p^{\mathrm{out}}_n\right)^2 + m^2_{\ell_n} = \left(p^{\mathrm{in}}_{n'}\right)^2 + m^2_{\ell_{n'}} = 0$, each vertex bracket is exactly the inverse propagator of the internal line, up to a sign fixed by the direction of charge flow,
\begin{align}\label{eqn:A+vertexCancels}
    k \cdot \left(2 p^{\mathrm{out}}_{n} + k\right) + m^2_{\ell_j} - m^2_{\ell_n} ~ &= ~ \phantom{-} \left(p^{\mathrm{out}}_n + k\right)^2 + m^2_{\ell_j} \, , \nonumber \\
    k \cdot \left(2 p^{\mathrm{in}}_{n'} - k\right) + m^2_{\ell_{n'}} - m^2_{\ell_j} ~ &= ~ - \left[\left(p^{\mathrm{in}}_{n'} - k\right)^2 + m^2_{\ell_j}\right] \, ,
\end{align}
so the propagator cancels identically against the vertex on every leg, with no soft pole left over, and
\begin{align}\label{eqn:A+amplitude}
&\mathcal{M}_{A^+_{\ell m}}
\left(
k \ell m;\{p^{\mathrm{out}}\ell^{\mathrm{out}}m^{\mathrm{out}}\};
\{p^{\mathrm{in}}\ell^{\mathrm{in}}m^{\mathrm{in}}\}
\right)
\nonumber\\[0.3em]
&\qquad
=
- \sum_{\ell_j m_j}
\Biggl(
\sum_{n=1}^{n_{\mathrm{out}}}
\mathcal{M}^{\text{out}}_n\left(k\right) q_n C\left[\ell m, \ell_n m_n; \ell_j m_j\right] -
\sum_{n'=1}^{n_{\mathrm{in}}}
\mathcal{M}^{\text{in}}_{n'}\left(k\right) q_{n'} C\left[\ell m, \ell_{n'} m_{n'}; \ell_j m_j\right]
\Biggl) \, .
\end{align}
The cancellation \eqref{eqn:A+vertexCancels} is algebraic and required no soft limit, which is the diagrammatic statement that a pure-gauge mode has no pole to produce. For $\ell = 0$ there is no $A^{+}_{00}$, and the external leg is the gauge parameter $\lambda_{00}$ itself, entering through $A^{00}_a = \partial_a \lambda_{00}$. Only the longitudinal part of \eqref{eqn:A+vertex} is then present, $C^{(3)}$ sets $\ell_j = \ell_n$, and \eqref{eqn:A+vertexCancels} and \eqref{eqn:A+amplitude} hold unchanged.

Only emissions from external legs have been written in \eqref{eqn:A+amplitude}. That this is the whole answer is the content of \eqref{eqn:SmatrixPhase}, and the diagrammatic reason is that \eqref{eqn:A+vertexCancels} makes the insertions telescope along each charged line: successive insertions cancel in pairs and only the two ends of each line survive, which are the external legs. The telescoping is complete, however, only once the quartic vertex is included. At a point where the charged line also emits a photon internal to $\mathcal{M}$, the cubic identity \eqref{eqn:A+vertexCancels} alone does not cancel both adjacent propagators, and the seagull term of \eqref{eqn:totinteraction} supplies the missing contribution; this is the standard Ward--Takahashi structure of scalar QED. We do not compute it --- the same term is omitted from the hard current below \eqref{eqn:current} --- and we do not need to: \eqref{eqn:fieldRedefinition} disposes of both vertices at once, so \eqref{eqn:SmatrixPhase} already contains the seagull, and \eqref{eqn:A+amplitude} is its infinitesimal form.

That the two agree is a short calculation. An angle-dependent phase rotation acts on the
scalar as $\delta_{\lambda_{\ell m}} \tilde{\phi}\left(\Omega\right) = i q \lambda_{\ell m}
Y_{\ell m}\left(\Omega\right) \tilde{\phi}\left(\Omega\right)$, which in the partial-wave
basis reads
\begin{align}\label{eqn:angularPhaseRotation}
    \delta_{\lambda_{\ell m}} \phi_{\ell_j m_j} ~ = ~ i q \lambda_{\ell m}
    \sum_{\ell_n m_n} C\left[\ell m , \ell_n m_n ; \ell_j m_j\right] \phi_{\ell_n m_n} \, .
\end{align}
Since the amplitude is multilinear in the external scalars, summing \eqref{eqn:angularPhaseRotation} over the external legs with the relative sign between in and out states reproduces the right-hand side of \eqref{eqn:A+amplitude}, which is therefore the partial-wave transcription of the exponent of \eqref{eqn:SmatrixPhase}, up to the overall normalisation of the external $A^{+}_{\ell m}$ leg fixed by the convention above.

\paragraph{Charge and flux.} Comparing \eqref{eqn:A+amplitude} with the soft theorem \eqref{eqn:finalAmp}, the two right-hand sides are the same object but for the branch labels,
\begin{align}\label{eqn:chargeVsFlux}
    \text{pure gauge:} ~ &\sum_{n \in \text{out}} q_n \, C \, \mathcal{M}^{\text{out}}_n ~ - \sum_{n' \in \text{in}} q_{n'} \, C \, \mathcal{M}^{\text{in}}_{n'} \, , \nonumber \\
    \text{radiative:} ~ &\sum_{n \in \text{out}} \sigma_n q_n \, C \, \mathcal{M}^{\text{out}}_n - \sum_{n' \in \text{in}} \sigma_{n'} q_{n'} \, C \, \mathcal{M}^{\text{in}}_{n'} \, .
\end{align}
The first vanishes on shell, by the angular locality just discussed; the second does not. The difference has a clear origin. The charge $Q_{\text{cut}} = Q_{\mathrm{soft}} + Q_{\mathrm{hard}}$ contains the hard charge, and by \eqref{eqn:hardCharge} this piece is the radial charge \emph{flux} $r^2\,{\epsilon_w}^{b}J^{\ell m}_b$ integrated along the generators, not the charge on a spatial slice. Its mode expansion \eqref{eqn:hardChargeResult} is weighted by the branch label of the hard scalar, so that
\begin{align}\label{eqn:hardChargeOnModes}
    \left[Q_{\mathrm{hard}}\left[\lambda^{(0)}_{\ell m}\right] \, , \, c^{(\sigma)\dagger}_{\ell_n m_n}\right] ~ \propto ~ \sigma \, q_n \, \lambda^{(0)}_{\ell m} \, C\left[\ell m, \ell_n m_n ; \ell_j m_j\right] \, c^{(\sigma)\dagger}_{\ell_j m_j} \, ,
\end{align}
and the two radial branches are rotated by opposite phases. The bulk rotation \eqref{eqn:angularPhaseRotation} rotates them by the same phase, which is why \eqref{eqn:A+amplitude} counts every hard leg once while \eqref{eqn:finalAmp} counts it with $\sigma_n$. The two statements also differ in which partial waves they cover: \eqref{eqn:SmatrixPhase} holds for every $\ell \geq 0$, its $\ell = 0$ member being ordinary charge conservation, whereas the radiative mode exists only for $\ell > 0$, and at $\ell = 0$ the identity \eqref{eqn:WardIdentityCharges} holds trivially, as discussed in \secref{sec:wardidentity}. The charge that enters the balance along the horizon is therefore not the restriction of the bulk charge. It is the flux of the current through the horizon: since ${\epsilon_w}^{b} J_b = J^{r^*}$, the hard charge is $r^2 J^{r^*}$ integrated over all $w$, the net charge carried across $\Sigma$ counted with the sign of the direction of crossing, independent of where $\Sigma$ sits by conservation of the current, and on the states of \secref{sec:externalstates} the charge-weighted sum of radial directions, $\sum_n q_n \sigma_n$. Free propagation near the horizon preserves the branch of every leg and leaves this sum unchanged; the interaction turns legs around and changes it. The two modes of the photon also couple to different parts of the current: the pure-gauge mode to $\partial^a J_a$, which vanishes, and the radiative mode, by \eqref{eqn:Psivertex}, to $\epsilon^{ab}\partial_a J_b$, the curl of the current, which does not. The Ward identity \eqref{eqn:WardIdentity} is the statement that the change in $\sum_n q_n \sigma_n$ equals the change in $Q_{\text{cut}}$, which is carried by soft photons, and it is this flux that pairs with the memory of the boundary mode in \secref{sec:memory}. By \appref{app:hardGenerates}, $Q_{\text{cut}}$ is the Noether flux of the residual gauge transformation through $\Sigma$, with both sectors included. It is not conserved, for the reason given in \secref{sec:covariantCharge}; the charge that commutes with the $S$-matrix is $Q_{\mathrm{soft}}$, and the amplitude verifies this at tree level.

This should not be confused with the charge--flux split familiar from the covariant phase space treatment of null boundaries in general relativity
\cite{Wald:1999wa, Chandrasekaran:2018aop, Odak:2023pga}. There the split originates in an obstruction: at a boundary through which symplectic flux leaks, $\iota_{X_\xi}\Omega$ fails to be an exact variation, the symmetry admits no Hamiltonian, and a prescription is required before a charge can be assigned to a cross section at all --- the flux being what is left over. No such obstruction arises here. The residual gauge parameter of \eqref{lambda} is field independent and the background is fixed, so \eqref{eqn:fullGenerator} holds as an exact variation, off shell and for the Maxwell and the matter sectors alike, and both the charge on a cut of the horizon and the flux through it are unambiguously defined. What is at issue is only which combination commutes with the $S$-matrix, and it is the flux through $\Sigma$ with its matter part removed, $Q_{\text{cut}} - Q_{\mathrm{hard}}$.

At null infinity the distinction is invisible, for the reason we spell out in \secref{sec:memoryAtScri}: by \eqref{eqn:scriBranches} the branch crossing $\mathcal{I}^+$ is the outgoing one and the branch along it carries no data there, so the analogue of $\sigma_n$ takes a single value on every leg and the two weightings of \eqref{eqn:chargeVsFlux} cannot be told apart. A horizon does not have this property: it admits modes that cross it and modes that propagate along its generators, both signs of radial motion occur among the external states, and the two notions differ. We add that the horizon in question is one through which charge is flowing; for a horizon in equilibrium the fluxes associated with the boundary symmetries vanish identically \cite{Ashtekar:2021wld, Ashtekar:2024bpi}, so the regime in which the present distinction has content is precisely the one those treatments set aside.

\section{Boundary mode on the horizon and the memory effect}\label{sec:memory}

We now show that the soft charge is the memory conjugate to the boundary mode on the horizon, identify which fields a detector must measure to see it, and separate two contributions to it. The first is the zero mode of the field radiated by the scattering, which the soft theorem computes and which enters the horizon through the magnetic observable alone. The second is the Coulomb memory left by every charge that crosses the horizon, scattered or not, which enters both observables. The Ward identity is the statement that their sum is conserved, the first changing by minus the change in the second. We will show that the two radial branches enter the two boundary observables in the symmetric and antisymmetric combinations respectively. Only the symmetric combination is a charge, and the radiated field does not contribute to the antisymmetric combination while a crossing charge does.

\subsection{The soft charge is a memory}\label{sec:memoryIsSoftCharge}

The soft charge \eqref{eqn:softCharge} is already a total null derivative,
\begin{equation}\label{eqn:QsoftMemory}
    Q_{\mathrm{soft}}\left[\lambda_{\ell m}\right] ~ = ~ - \, \ell\left(\ell+1\right) \lambda_{\ell m} \int \mathrm{d}w \, \partial_w \Psi_{\ell m} ~ = ~ - \ell\left(\ell+1\right) \lambda_{\ell m} \, \Delta \Psi_{\ell m} \, .
\end{equation}
This is a direct consequence of the Moncrief variable being the potential for the gauge-invariant one-form in \eqref{eqn:chiaSoln}: the integration along the horizon generator has already been carried out in defining $\Psi_{\ell m}$, and is not something that has to be arranged by a choice of gauge. We therefore define the horizon memory operator gauge invariantly as
\begin{equation}\label{eqn:memoryOperator}
    N_{\ell m} ~ \coloneqq ~ \Delta \Psi_{\ell m} ~ = ~
    \Psi_{\ell m}\big|_{w \rightarrow +\infty} - \Psi_{\ell m}\big|_{w \rightarrow -\infty} \, ,
\end{equation}
with the orientation \eqref{eqn:orientation}, so that on $\mathcal{H}^+$ it is the memory accumulated from the bifurcation sphere to the future endpoint and on $\mathcal{H}^-$ from the past endpoint to the bifurcation sphere, and
\begin{equation}\label{eqn:QsoftIsMemory}
    Q_{\mathrm{soft}}\left[\lambda_{\ell m}\right] ~ = ~ - \ell\left(\ell+1\right) \lambda_{\ell m} \, N_{\ell m} \, .
\end{equation}

\paragraph{The memory as a vacuum transition.} In \appref{app:symplecticform} we derive the symplectic form and canonical brackets of the radiative phase space. In the text surrounding \eqref{eqn:largegaugetrafo} we show that the degenerate horizon vacua are labelled by the boundary coordinate $\varphi^{(0)}_{\ell m}$, which shifts by $\lambda_{\ell m}$ under a large gauge transformation and therefore labels distinct states. The canonical bracket on the boundary is \eqref{appeqn:MemoryGoldstoneBracket}, which we repeat for convenience,
\begin{align}\label{eqn:MemoryGoldstoneBracket}
    \left[\Delta\Psi_{\ell' m'} \, , \, \varphi_{\ell m}\right] ~ = ~ \dfrac{i}{\ell\left(\ell+1\right)} \, \delta_{\ell\ell'}\delta_{mm'} \, .
\end{align}
Since $N_{\ell m}$ is conjugate to $\varphi^{(0)}_{\ell m}$, the operator $\exp\left[-i\ell\left(\ell+1\right)\lambda_{\ell m}N_{\ell m}\right]$ translates $\varphi^{(0)}_{\ell m}$ by $\lambda_{\ell m}$, and a change in the zero mode leaves the horizon in a different member of the degenerate family from the one it started in.

\paragraph{Gauss-law form of the memory.} The memory is not merely a formal endpoint difference; it is fixed by the flux through the two ends of the horizon and by the charge that crossed in between. To make this transparent, we use the variable $\mathcal{E}_{\ell m}$ of \eqref{eqn:radialFlux} and introduce a second one,
\begin{equation}\label{eqn:fluxVariable}
    \mathcal{E}_{\ell m} ~ = ~ - \, \dfrac{r^2}{f\left(r\right)} \, F^{\ell m}_{w r^*} \qquad \text{and} \qquad
    G_{\ell m} ~ \coloneqq ~ \mathcal{E}_{\ell m} - \ell\left(\ell+1\right) \Psi_{\ell m} \, ,
\end{equation}
the first being the radial electric flux through a cut, which is what the covariant charge \eqref{eqn:electricCharge} measures, and the second the bracket that appears in \eqref{eqn:MoncriefMaster}. Dropping the term in $\rho_{\ell m}$, which is sourced at order $f\left(r\right)$ and plays no role on the horizon, that equation says
\begin{equation}\label{eqn:GaussFunction}
    \partial_a G_{\ell m} ~ = ~ {\epsilon_a}^{b} \, r^2 J^{\ell m}_b \, ,
\end{equation}
so that $G_{\ell m}$ is constant wherever there are no charges and jumps across the worldline of each one; it is the two-dimensional Gauss function, defined up to an additive constant that drops out of every difference below. Its endpoint difference along the horizon is the hard charge,
\begin{equation}\label{eqn:DeltaG}
    \mathcal{Q}_{\ell m} ~ \coloneqq ~ - \Delta G_{\ell m} ~ = ~ - \int \mathrm{d}w \, r^2 \, {\epsilon_w}^{b} J^{\ell m}_b ~ = ~ \dfrac{Q_{\mathrm{hard}}\left[\lambda_{\ell m}\right]}{\lambda_{\ell m}} \, ,
\end{equation}
by \eqref{eqn:hardCharge}, and by \eqref{eqn:hardChargeResult} it has the eigenvalue $\sum_n \sigma_n q_n Y_{\ell m}\left(\Omega_n\right)$ on a state of charges $q_n$ at angles $\Omega_n$ moving on the radial branches $\sigma_n$: it is the net flux-weighted charge that crossed the surface. Taking the endpoint difference of \eqref{eqn:fluxVariable} then gives
\begin{equation}\label{eqn:GaussLawMemory}
    \ell\left(\ell+1\right) N_{\ell m} ~ = ~ \Delta \mathcal{E}_{\ell m} ~ + ~ \mathcal{Q}_{\ell m} \, ,
\end{equation}
which is nothing but the soft/hard split of the covariant charge \eqref{eqn:softCharge} read backwards. We write its two terms as
\begin{equation}\label{eqn:memorySplit}
    N_{\ell m} ~ = ~ N^{\text{rad}}_{\ell m} + N^{\mathrm{C}}_{\ell m} \, , \qquad N^{\text{rad}}_{\ell m} ~ \coloneqq ~ \dfrac{\Delta \mathcal{E}_{\ell m}}{\ell\left(\ell+1\right)} \, , \qquad N^{\mathrm{C}}_{\ell m} ~ \coloneqq ~ \dfrac{\mathcal{Q}_{\ell m}}{\ell\left(\ell+1\right)} \, .
\end{equation}
The first is the zero mode of the radiative oscillators, by \eqref{eqn:softChargeOnHorizons}, and the second is the Coulomb memory of the charges that crossed, to which we return in \secref{sec:twoObservables}. In these terms $Q_{\text{cut}} = - \ell\left(\ell+1\right)\lambda_{\ell m} N^{\text{rad}}_{\ell m}$ and $Q_{\mathrm{hard}} = \ell\left(\ell+1\right)\lambda_{\ell m} N^{\mathrm{C}}_{\ell m}$, and the Ward identity \eqref{eqn:WardIdentitySoft} is the conservation of their sum. Note that \eqref{eqn:GaussLawMemory} is a measurement of the radial electric flux on a \emph{cut} of the horizon, taken at the two ends; it is not an impulse on a detector. The remainder of this section is concerned with what an impulse measures, which is where the branch structure enters.

\subsection{The two boundary observables}\label{sec:twoObservables}

\paragraph{Symmetric and antisymmetric branch combinations.} In the gauge $A_w = 0$, \eqref{eqn:AaSoln} gives $\chi^{\ell m}_w = - \partial_w A^+_{\ell m}$ exactly, so by the orientation \eqref{eqn:orientation},
\begin{equation}\label{eqn:DeltaAplus}
    \int \mathrm{d}w \, \chi^{\ell m}_w ~ = ~ - \Delta A^{+}_{\ell m} ~ = ~ - \sum_{\sigma = \pm} \sigma \, \Delta \Psi^{(\sigma)}_{\ell m} + \Delta \rho_{\ell m} \, ,
\end{equation}
the second equality following from \eqref{eqn:AplusBranches}, in which the $w$-independent piece $\varphi_{\ell m}$ drops out of the endpoint difference, and holding for any field of the form $\Psi_{\ell m} = \Psi^{(+)}_{\ell m}\left(u\right) + \Psi^{(-)}_{\ell m}\left(v\right)$ in the near-horizon region, radiative or not. Comparing with \eqref{eqn:memoryOperator}, the horizon carries two boundary observables built from the endpoint data, and they are the two branch parities:
\begin{equation}\label{eqn:MemoryBranchSplit}
    N_{\ell m} ~ = ~ \sum_{\sigma = \pm} \Delta \Psi^{(\sigma)}_{\ell m}
    \qquad \text{and} \qquad
    \Delta A^{+}_{\ell m} ~ = ~ \sum_{\sigma = \pm} \sigma \,
    \Delta \Psi^{(\sigma)}_{\ell m} \, .
\end{equation}
Equivalently, $\Delta\Psi^{(\sigma)}_{\ell m}$ on the two branches is recovered as
\begin{equation}\label{eqn:BranchesFromObservables}
    \Delta \Psi^{(\sigma)}_{\ell m}\Big|_{\eta^{wr^*}\sigma = \pm 1} ~ = ~
    \dfrac{1}{2}\left(N_{\ell m} \pm \eta^{wr^*} \Delta A^{+}_{\ell m}\right) \, ,
\end{equation}
so that the branch along the horizon is visible only through the combination $N_{\ell m} + \eta^{wr^*}\Delta A^{+}_{\ell m}$.

The distinction between the two is important because two different branch labels are in play. By \eqref{eqn:drstarOnBranches}, a derivative \emph{normal} to the boundary weights the branches of whatever field it acts on by that field's own $\sigma$,
\begin{equation}\label{eqn:drstarLimitRepeat}
    \chi^{\ell m}_w ~ = ~ \partial^{r^*}\!\Psi_{\ell m} ~ = ~
    - \sum_{\sigma = \pm} \sigma \, \partial_w \Psi^{(\sigma)}_{\ell m} \, ,
\end{equation}
whereas a derivative \emph{along} the boundary does not. In the constraint \eqref{eqn:eomLongitudinalCombination} from which the charge is read off, the radiative term appears with the derivative $\partial_w$ along the generators and the current with the normal derivative $\partial^{r^*}$, as \eqref{eqn:currentEpsilonCombination} and the text above it make explicit. The three objects therefore carry three different weightings:
\begin{itemize}
    \item $Q_{\mathrm{soft}}$ carries \emph{no} photon-branch weight. It is the branch-symmetric combination \eqref{eqn:memoryOperator}, and it descends from the corner term \eqref{eqn:CornerSplit}.
    \item $Q_{\mathrm{hard}}$ carries the branch weight $\sigma_n$ of the \emph{hard scalars}, \eqref{eqn:hardChargeResult}, and is a radial flux. This weighting is what gives the identity content, since the unweighted sum over hard legs is the total charge and vanishes by conservation, whereas the flux-weighted combination \eqref{eqn:softAmplitudeReduced} does not. It is physical in the sense made precise below \eqref{eqn:scalarAlgebra}: a charge crossing outward contributes with the opposite sign to one crossing inward.
    \item $\Delta A^{+}_{\ell m}$ carries the branch weight $\sigma$ of the \emph{photon}. No term in \eqref{eqn:eomLongitudinalCombination} carries that weighting.
\end{itemize}
Being a flux is therefore not what disqualifies $\Delta A^{+}_{\ell m}$ from being a charge, since the hard charge is a flux and is a charge. What disqualifies it is that it is not produced by contracting a large gauge transformation into the symplectic form. Since $\delta_\lambda \Psi_{\ell m} = 0$ by \eqref{eqn:largegaugetrafo}, the bulk term \eqref{eqn:BulkSymplecticBranches}, which is the only place the photon-branch weighting appears in the phase space, drops out of $i_{X_\lambda}\Omega$ identically,
\begin{equation}\label{eqn:BulkTermDropsOut}
    i_{X_\lambda} \left[\sum_{\ell m}\,\ell\left(\ell+1\right)\int \mathrm{d}w \;
    \delta\!\left(\partial^{r^*}\!\Psi_{\ell m}\right) \wedge \delta \Psi_{\ell m}\right]
    ~ = ~ 0 \, ,
\end{equation}
and the entire charge comes from the corner. Both $N_{\ell m}$ and $\Delta A^{+}_{\ell m}$ are endpoint data of the same field; only the first of them is a charge.

\paragraph{Transverse electric and transverse magnetic fields.} To say which field each of these two objects is, we pass to the static chart. The two charts $\left(w, r^*\right)$ and $\left(t, r^*\right)$ share their second coordinate, so that in the chain rule for $\partial_w$ the radial term is absent identically while in the chain rule for $\partial_{r^*}$ the time term is not:
\begin{equation}\label{eqn:chartChainRule}
    \partial_w\big|_{r^*} ~ = ~ \partial_t\big|_{r^*} \, ,
    \qquad \qquad
    \partial_{r^*}\big|_{w} ~ = ~ \partial_{r^*}\big|_{t}
    - \eta^{wr^*} \partial_t \, ,
\end{equation}
since $w = t + \eta^{wr^*}r^*$ gives $\partial r^*/\partial w\big|_{r^*} = 0$ and $\partial t/\partial r^*\big|_{w} = - \eta^{wr^*}$. This is the usual asymmetry of Eddington--Finkelstein coordinates: shifting the time coordinate by a function of $r^*$ alone leaves the time basis vector untouched and tilts the radial one. It is also what makes $w$ null, since it returns $\eta^{ww} = \eta^{tt} + \left(\eta^{wr^*}\right)^2 \eta^{r^* r^*} = 0$ from $\eta^{tt} = -1$ and $\eta^{r^*r^*} = +1$, and it turns $-\partial^2_t + \partial^2_{r^*}\big|_t$ into the operator $\Box_\eta$ defined below \eqref{eqn:epsilonIdentities}. Using \eqref{eqn:chartChainRule},
\begin{equation}\label{eqn:tildePartialIsStaticRadial}
    \partial^{r^*} ~ = ~ \eta^{w r^*}\widetilde{\partial}
    ~ = ~ \eta^{w r^*}\left(\partial_w + \eta^{w r^*}\partial_{r^*}\big|_w\right)
    ~ = ~ \partial_{r^*}\big|_{t} \, ,
\end{equation}
and the Jacobian of $\left(w, r^*\right)\rightarrow \left(t, r^*\right)$ is unimodular, so $\epsilon_{t r^*} = \epsilon_{w r^*} = +1$ and \eqref{eqn:chiaSoln} reads, in the static chart,
\begin{equation}\label{eqn:chiStaticChart}
    \chi^{\ell m}_{t} ~ = ~ \partial^{r^*}\!\Psi_{\ell m} + \partial_t \rho_{\ell m}
    \qquad \text{and} \qquad
    \chi^{\ell m}_{r^*}\big|_{t} ~ = ~ \partial_w \Psi_{\ell m}
    + \partial_{r^*}\rho_{\ell m}\big|_t \, .
\end{equation}
Note that $\chi^{\ell m}_{t} = \chi^{\ell m}_{w}$, because $\partial_w$ and $\partial_t$ are the same basis vector by \eqref{eqn:chartChainRule}; it is only the radial component that differs between the two charts, and it differs by exactly that component,
\begin{equation}\label{eqn:radialComponentsRelated}
    \chi^{\ell m}_{r^*}\big|_{t} ~ = ~ \chi^{\ell m}_{r^*}\big|_{w}
    ~ + ~ \eta^{wr^*} \chi^{\ell m}_{w}
    ~ = ~ - \eta^{wr^*}\partial_{r^*}\Psi_{\ell m}\big|_w
    + \eta^{wr^*} \partial^{r^*}\!\Psi_{\ell m}
    ~ = ~ \partial_w \Psi_{\ell m} \, ,
\end{equation}
which is the reason $\chi^{\ell m}_{r^*}\big|_w$ is purely the branch along the horizon, as in \eqref{eqn:chirstarCrossing} below, while $\chi^{\ell m}_{r^*}\big|_t$ is branch-symmetric. Substituting into the transverse part of \eqref{eqn:FinTermsofChi_a} and suppressing the odd-parity term,
\begin{subequations}\label{eqn:EandBcomponents}
    \begin{align}
    F_{A t} ~ &= ~ \sum_{\ell m} \left(\partial^{r^*}\!\Psi_{\ell m}
      + \partial_t \rho_{\ell m}\right) \partial_A Y_{\ell m}
      && \text{transverse electric,} \\
    F_{A r^*}\big|_{t} ~ &= ~ \sum_{\ell m} \left(\partial_w \Psi_{\ell m}
      + \partial_{r^*}\rho_{\ell m}\big|_t\right) \partial_A Y_{\ell m}
      && \text{transverse magnetic,}
    \end{align}
\end{subequations}
the labels being those of a static observer, whose four-velocity is proportional to $\partial_t$, so that $E_A \propto F_{At}$ while $F_{A r^*}$ is purely spatial. The $\rho_{\ell m}$ terms are sourced by $J^{+}_{\ell m}$ at order $f\left(r\right)$, by \eqref{eqn:chiaSoln}, and are dropped below; we return to them in the note below.

In the orthonormal frame of the static observer, $e_{\hat t} = f\left(r\right)^{-1/2}\partial_t$, $e_{\hat r} = f\left(r\right)^{-1/2}\partial_{r^*}$ and $e_{\hat A} = r^{-1}\hat{e}_A$, the two components of \eqref{eqn:EandBcomponents} carry the same redshift factor,
\begin{equation}\label{eqn:orthonormalFrameFields}
    E_{\hat A} ~ = ~ \dfrac{F_{A t}}{r \sqrt{f\left(r\right)}}
    \qquad \text{and} \qquad
    F_{\hat A \hat r} ~ = ~ \dfrac{F_{A r^*}\big|_{t}}{r \sqrt{f\left(r\right)}} \, ,
\end{equation}
and it cancels against the proper time $\mathrm{d}\tau = \sqrt{f\left(r\right)} \, \mathrm{d}t = \sqrt{f\left(r\right)}\,\mathrm{d}w$ of the same observer, so that $\int \mathrm{d}\tau \, E_{\hat A} = r^{-1}\!\int \mathrm{d}w \, F_{At}$ and likewise for the magnetic component. Both integrals are therefore finite along the family of static observers as it is taken to the horizon, and the statements below are about locally measured fields.

Integrating \eqref{eqn:EandBcomponents} along the generator with the orientation \eqref{eqn:orientation}, which runs with the proper time of the observer on both horizons, and using \eqref{eqn:DeltaAplus} and \eqref{eqn:memoryOperator},
\begin{equation}\label{eqn:EandBmemory}
    \int \mathrm{d}w \, F_{At} ~ = ~ - \sum_{\ell m} \Delta A^{+}_{\ell m} \,
    \partial_A Y_{\ell m}
    \qquad \text{and} \qquad
    \int \mathrm{d}w \, F_{A r^*}\big|_{t} ~ = ~ \sum_{\ell m} N_{\ell m} \,
    \partial_A Y_{\ell m} \, .
\end{equation}
This is the sharp form of the statement that the two branch parities are two different fields: the branch-antisymmetric combination is the time-integrated transverse electric field, and the branch-symmetric combination, the one the soft charge measures, is the time-integrated transverse magnetic field.

\paragraph{What a detector samples.} A test particle of charge $q'$ obeys $\mathrm{d}p_\mu/\mathrm{d}\tau = q' F_{\mu\nu}u^\nu$, so along any worldline $\gamma$ with negligible angular motion,
\begin{equation}\label{eqn:kickLineIntegral}
    \Delta p_A ~ = ~ q' \int_\gamma F_{A a} \, \mathrm{d}x^{a}
    ~ = ~ q' \sum_{\ell m} \partial_A Y_{\ell m} \int_\gamma \chi^{\ell m} \, ,
    \qquad \chi^{\ell m} ~ \coloneqq ~ \chi^{\ell m}_a \, \mathrm{d}x^a \, ,
\end{equation}
which holds without approximation and is gauge invariant, with $\chi^{\ell m}$ the full one-form of \eqref{eqn:chiaSoln} and no restriction to the radiative sector. Its exterior derivative is the two-dimensional field strength,
\begin{equation}\label{eqn:dchi}
    \mathrm{d}\chi^{\ell m} ~ = ~ F^{\ell m}_{w r^*} \, \mathrm{d}w \wedge \mathrm{d}r^*
    ~ = ~ - \, \dfrac{f\left(r\right)}{r^2} \, \mathcal{E}_{\ell m} \, \mathrm{d}w \wedge \mathrm{d}r^*
\end{equation}
by \eqref{eqn:fluxVariable}, so that by Stokes' theorem the impulses on two worldlines with common endpoints differ by the flux of the radial electric field through the strip between them. In the near-horizon region that flux is of order $f\left(r\right)$ per unit Killing time, since $\mathcal{E}_{\ell m}$ is finite on the horizon and the coordinate components of the radial field carry the redshift. Over the duration of the field, therefore, $\chi^{\ell m}$ is closed to leading order, and the impulse depends on the worldline only through its endpoints. For a field of the form below \eqref{eqn:DeltaAplus} the primitive is explicit. In the radiative sector $\chi^{\ell m} = \star \, \mathrm{d}\Psi_{\ell m}$, with $\star$ the Hodge dual of $\eta_{ab}$ normalised by \eqref{eqn:chiStaticChart}, $\star \mathrm{d}t = \mathrm{d}r^*$ and $\star \mathrm{d}r^* = \mathrm{d}t$, and the null one-forms are its eigenforms, $\star \mathrm{d}u = - \mathrm{d}u$ and $\star \mathrm{d}v = \mathrm{d}v$. Hence
\begin{equation}\label{eqn:chiExactDifferential}
    \chi^{\ell m} ~ = ~ \star \, \mathrm{d} \Big( \Psi^{(+)}_{\ell m}\left(u\right) + \Psi^{(-)}_{\ell m}\left(v\right) \Big)
    ~ = ~ \mathrm{d}\Psi^{(-)}_{\ell m} - \mathrm{d}\Psi^{(+)}_{\ell m}
    ~ = ~ - \sum_{\sigma = \pm} \sigma \, \mathrm{d}\Psi^{(\sigma)}_{\ell m} \, ,
\end{equation}
the differential of the branch-antisymmetric combination. Its integral along $\gamma$ is the endpoint difference of that combination. On a worldline that traverses the complete support of both pulses within the near-horizon region, so that $u$ and $v$ each run through the range on which the corresponding branch varies, this is $- \sum_\sigma \sigma \, \Delta\Psi^{(\sigma)}_{\ell m} = - \Delta A^{+}_{\ell m}$ with the orientation \eqref{eqn:orientation}. Therefore
\begin{equation}\label{eqn:memoryKick}
    \Delta p_A ~ = ~ - q' \sum_{\ell m} \Delta A^{+}_{\ell m} \, \partial_A Y_{\ell m}
\end{equation}
on every such worldline, at fixed $r^*$ or moving radially with any velocity profile: the magnetic Lorentz force that the radial velocity multiplies is compensated by the time the detector spends inside each pulse, as it must be for a boundary term. The static observer at fixed $r^*$ is the case in which both pulses pass in full; a worldline that crosses the horizon, or leaves the near-horizon region, before one of them has passed samples only part of that endpoint difference, with the finite-$f\left(r\right)$ corrections of \eqref{eqn:dchi} integrated over the portion it does traverse. In components, the $\mathrm{d}r^*$ part of \eqref{eqn:chiExactDifferential} in the $\left(w, r^*\right)$ chart comes from the branch along the surface alone, which is a function of $w - 2\eta^{wr^*}r^*$ while the crossing branch is a function of $w$,
\begin{equation}\label{eqn:chirstarCrossing}
    \chi^{\ell m}_{r^*}\big|_w ~ = ~ - \eta^{w r^*} \partial_{r^*} \Psi_{\ell m}\big|_w
    ~ = ~ \sum_{\sigma = \pm} \left(1 + \eta^{w r^*}\sigma\right)
    \partial_w \Psi^{(\sigma)}_{\ell m}
    ~ = ~ 2 \, \partial_w \Psi^{(\sigma)}_{\ell m}
    \Big|_{\eta^{w r^*}\sigma = + 1} \, ,
\end{equation}
in agreement with \eqref{eqn:BranchesFromObservables}; this is the component the velocity multiplies, and the time the detector spends in that branch is stretched by exactly the factor that cancels it. The memory $N_{\ell m}$ is therefore not sampled by any impulse. It is sampled, by \eqref{eqn:EandBmemory}, as the time integral of the transverse magnetic field at fixed radius, and by \eqref{eqn:GaussLawMemory} as the change in the radial electric flux through a cut together with the flux of charge through the surface; both are field measurements. The null worldline along a generator of the horizon is the singular limit of the timelike family: the differential of the branch along the surface pulls back to zero on it, so that its impulse comes from the crossing branch alone and is $\tfrac{1}{2}\left(\eta^{wr^*} N_{\ell m} - \Delta A^{+}_{\ell m} \right)$ by \eqref{eqn:BranchesFromObservables}. No massive charge realises it, and the detectors of this section are the static observers at fixed $r^*$ of \eqref{eqn:orthonormalFrameFields} and the timelike worldlines moving radially among them. Everything in \eqref{eqn:memoryKick} is a one-point function on the state of the detector's surroundings. It has two contributions which we compute in what follows.

\paragraph{Note:} Restoring the longitudinal piece of \eqref{eqn:chiaSoln} does not change the above. In the gauge $A_w = 0$, equation \eqref{eqn:DeltaAplus} is exact with $\chi^{\ell m}_w$ the full one-form, so $\Delta A^{+}_{\ell m}$ as defined there already contains it; what \eqref{eqn:AplusBranches} omits is a term $-\rho_{\ell m}$,
\begin{equation}\label{eqn:AplusWithRho}
    A^{+}_{\ell m} ~ = ~ \sum_{\sigma = \pm} \sigma \, \Psi^{(\sigma)}_{\ell m}
    ~ - ~ \rho_{\ell m} ~ + ~ \varphi_{\ell m}\left(r^*\right) \, ,
\end{equation}
whose endpoint difference is sourced by $J^{+}_{\ell m}$ at order $f\left(r\right)$ and vanishes on the horizon. The sources enter the memory elsewhere, through $\Psi_{\ell m}$ itself, by \eqref{eqn:GaussLawMemory}, at order one.

\paragraph{The shock of a crossing charge.} The kick \eqref{eqn:memoryKick} is a one-point function, and a physical state carries the field of every charge in it. So, before asking what the scattering contributes, it is useful to note what a charge contributes by crossing. Here a crossing charge is one that passes through the surface on which the memory is measured, on whichever branch it moves; a stretched horizon is timelike and every charge passes through it. To leading order in $f\left(r\right)$ the near-horizon geometry is the product of two-dimensional Rindler space with a sphere of radius $R$. Moreover, every hard particle is null by \eqref{eqn:scalarMass}. And the field of a charge $q$ at angle $\Omega_q$ moving along $v = v_q$ or $u = u_q$ is the electromagnetic shock wave on that null plane, a function of that one null coordinate with a step across the charge. For $\ell > 0$ it carries no radial flux, so $\Delta\mathcal{E}_{\ell m} = 0$ in \eqref{eqn:GaussLawMemory}. The memory of a crossing charge is therefore its Coulomb memory alone,
\begin{equation}\label{eqn:coulombMemory}
    N^{\mathrm{C}}_{\ell m} ~ = ~ \dfrac{\mathcal{Q}_{\ell m}}{\ell\left(\ell+1\right)} ~ = ~ \dfrac{\sigma_q \, q \, Y_{\ell m}\left(\Omega_q\right)}{\ell\left(\ell+1\right)} \, ,
\end{equation}
independent of the energy of the charge and of where the surface is stretched.\footnote{A memory at Coulombic order fixed by the flux of charged particles through a surface at finite distance is also what the Maxwell constraint gives on a null plane at large but finite radius in flat space, for massless and for massive sources \cite{Raclariu:2026mfs}; see \secref{sec:memoryAtScri}.} This is the memory of the charge's own field rather than of anything the charge did, and it is the memory that the dressing of \secref{sec:irdressing} cancels against the zero mode of the radiative field. Since the shock sits on the single branch $\sigma_q$, by \eqref{eqn:MemoryBranchSplit} it enters both boundary observables with equal magnitude: $N^{\mathrm{C}}_{\ell m}$ weighted by the direction of crossing, and $\Delta A^{+ \, \mathrm{C}}_{\ell m} = q \, Y_{\ell m}\left(\Omega_q\right)/\ell\left(\ell+1\right)$ by the charge alone, which are the two weightings of \eqref{eqn:chargeVsFlux}. Any timelike detector near the horizon is therefore kicked by every charge that falls past it, by the transverse kick of the shock with the Green's function of the sphere. On that detector, \eqref{eqn:memoryKick} summed over partial waves gives, for one crossing charge,
\begin{equation}\label{eqn:coulombKick}
    \Delta p_A ~ = ~ - \, q \, q' \, \partial_A \mathcal{G}\left(\Omega , \Omega_q\right) \, .
\end{equation}
Here,
\begin{equation}
    \mathcal{G}\left(\Omega , \Omega'\right) ~ \coloneqq ~ \sum_{\ell > 0} \sum_{m} \dfrac{Y_{\ell m}\left(\Omega\right) Y_{\ell m}\left(\Omega'\right)}{\ell\left(\ell+1\right)} ~ = ~ - \, \dfrac{1}{4\pi} \left[ \ln \dfrac{1 - \cos\gamma}{2} + 1 \right] \, ,
\end{equation}
with $\gamma$ the angle between $\Omega$ and $\Omega_q$ and $\mathcal{G}$ the Green's function of the sphere Laplacian on the $\ell > 0$ modes. At small separation $\mathcal{G} \simeq - \left(2\pi\right)^{-1}\ln\gamma$, so the kick is $\left(q q'/2\pi\right)\partial_A \ln \gamma$, directed away from the crossing charge when the charges have the same sign: it is the transverse kick of the electromagnetic shock wave, with the circumference of the horizon in place of the transverse plane. A timelike detector near a horizon through which charges are falling is kicked by every one of them and all the kicks add. What the next subsection shows is that the field radiated by the scattering contributes to $N^{\text{rad}}_{\ell m}$ and to nothing else.

\subsection{The memory of a scattering process}\label{sec:vanishingKick}

A scattering process adds to the horizon the zero mode of the field it radiates. The soft theorem of \secref{sec:softtheorem} computes it, and two things need to be settled about it: how large it is, and on which branch it sits. The answer to the second is that its zero modes on the two branches are equal, so that the electric observable receives nothing from the scattering at leading order in the soft limit. This is not visible from the Ward identity, because both sides of \eqref{eqn:WardIdentity} involve the photon branches only through the unweighted sum $\sum_\sigma$: the photon side because \eqref{eqn:softChargeOnHorizons} is the integral along $\Sigma$ of a derivative along it, with no normal derivative, and the hard side because the $\sigma_1$ appearing in \eqref{eqn:hardChargeResult} is the branch label of the \emph{scalar}, not of the photon. The photon-branch-antisymmetric combination $\sum_\sigma \sigma\left(\cdot\right)$ never appears in the identity, and its vanishing cannot be a consequence of it. It follows instead from the amplitude together with the conservation of the bulk charge, \eqref{eqn:A+amplitude}.

\paragraph{The two observables as operators.} In the window \eqref{eqn:boundaryModeCondition} both branches carry the same mode function $e^{-i\omega w}$, which is what produces the branch sum in \eqref{eqn:PsiModeExpansion}. Repeating the steps of \eqref{eqn:softChargeOnHorizons} once with weight $1$ and once with weight $\sigma$ gives the radiative parts of the two boundary observables in identical form,
\begin{subequations}\label{eqn:NandAplusModes}
    \begin{align}
        N^{\text{rad}}_{\ell m} ~ &= ~ \dfrac{1}{2 \sqrt{\ell\left(\ell+1\right)}}
      \sum_{\sigma = \pm} \lim_{\omega \rightarrow 0} \sqrt{\dfrac{\omega}{2}} 
      \left( b^{(\sigma)}_{\ell m}\left(\omega\right)
      + b^{(\sigma)\dagger}_{\ell m}\left(\omega\right)\right) \, , \\
          \Delta A^{+ \, \text{rad}}_{\ell m} ~ &= ~ \dfrac{1}{2 \sqrt{\ell\left(\ell+1\right)}}
      \sum_{\sigma = \pm} \sigma \lim_{\omega \rightarrow 0} \sqrt{\dfrac{\omega}{2}}
      \left( b^{(\sigma)}_{\ell m}\left(\omega\right)
      + b^{(\sigma)\dagger}_{\ell m}\left(\omega\right)\right) \, ,
    \end{align}
\end{subequations}
the endpoint prescription $\int_0^\infty \mathrm{d}\omega \, \delta\left(\omega\right) = 1/2$ entering both alike. These are the zero modes of the free radiative field, which by \eqref{eqn:softChargeOnHorizons} is $\mathcal{E}_{\ell m}/\ell\left(\ell+1\right)$ on the external states, and they are the operators whose matrix elements the amplitude computes. Taking the same matrix element as in \eqref{eqn:softCommutator}, and using the definitions below \eqref{eqn:softAmplitudeReduced},
\begin{subequations}\label{eqn:NandAplusMatrixElements}
    \begin{align}
    \langle \text{out} | N^{\text{rad} +}_{\ell m} S - S N^{\text{rad} -}_{\ell m} | \text{in}\rangle ~ &= ~ 
      \dfrac{1}{2\sqrt{2 \ell\left(\ell+1\right)}}
      \lim_{\omega \rightarrow 0} \sqrt{\omega} \sum_{\sigma = \pm}
      \left(\mathcal{M}_{b^{(\sigma)}} - \mathcal{M}_{b^{(\sigma)\dagger}}\right) \, , \\
          \langle \text{out} | \Delta A^{+ \, \text{rad}}_{\ell m} S - S \, \Delta A^{+ \, \text{rad}}_{\ell m}
      | \text{in}\rangle ~ &= ~ 
      \dfrac{1}{2\sqrt{2 \ell\left(\ell+1\right)}}
      \lim_{\omega \rightarrow 0} \sqrt{\omega} \sum_{\sigma = \pm} \sigma
      \left(\mathcal{M}_{b^{(\sigma)}} - \mathcal{M}_{b^{(\sigma)\dagger}}\right) \, .
    \end{align}
\end{subequations}
Write $\mathcal{N}_{\ell m}$ for the bracket on the right-hand side of \eqref{eqn:softAmplitudeReduced}
\begin{align}\label{eqn:softCoefficient}
\mathcal{N}_{\ell m} ~ \coloneqq ~ \sum_{\ell_j m_j} \Biggl( \sum_{n = 1}^{n_{\text{out}}} q_n \sigma_n \, C\left[\ell m , \ell_n m_n ; \ell_j m_j\right] \mathcal{M}^{\text{out}}_n ~ - ~ \sum_{n' = 1}^{n_{\text{in}}} q_{n'} \sigma_{n'} \, C\left[\ell m , \ell_{n'} m_{n'} ; \ell_j m_j\right] \mathcal{M}^{\text{in}}_{n'} \Biggr) \, ,
\end{align}
so that $\mathcal{M}_{b^{(\sigma)}} = - \mathcal{N}_{\ell m}/\sqrt{2\omega\ell\left(\ell+1\right)}$. By \eqref{eqn:branchSum} and \eqref{eqn:finalAmp} the first is
\begin{equation}\label{eqn:radiatedMemory}
	\langle \text{out} | N^{\text{rad} +}_{\ell m} S - S N^{\text{rad} -}_{\ell m} | \text{in}\rangle ~ = ~ - \, \dfrac{\mathcal{N}_{\ell m}}{\ell\left(\ell+1\right)} ~ = ~ - \, \dfrac{\mathcal{Q}^{\text{out}}_{\ell m} - \mathcal{Q}^{\text{in}}_{\ell m}}{\ell\left(\ell+1\right)} \; \langle \text{out} | S | \text{in} \rangle \, ,
\end{equation}
the second equality holding on states localised on the sphere, where $\mathcal{Q}^{\text{out}}_{\ell m}$ and $\mathcal{Q}^{\text{in}}_{\ell m}$ are the eigenvalues $\sum_n \sigma_n q_n Y_{\ell m}\left(\Omega_n\right)$ of the hard charge on the out- and in-states. It is non-vanishing, equal to the insertion of $Q_{\text{cut}}$ in \eqref{eqn:softCommutator} up to the factor $-\ell\left(\ell+1\right)\lambda^{(0)}_{\ell m}$, as it must be by \eqref{eqn:memorySplit}. Comparing with \eqref{eqn:coulombMemory}, it is minus the difference between the Coulomb memory of the outgoing charges and that of the incoming charges, so that
\begin{equation}\label{eqn:memoryConserved}
    \langle \text{out} | N^+_{\ell m} S - S N^-_{\ell m} | \text{in}\rangle ~ = ~ 0 \, ,
\end{equation}
which is the Ward identity \eqref{eqn:WardIdentitySoft}: the zero mode of the radiation cancels the change in the Coulomb memory. The second matrix element vanishes because $\mathcal{M}_{b^{(\sigma)}}$ is independent of $\sigma$, which holds leg by leg in the gauge of \appref{app:interactionTerms} and, in the gauge $A_w = 0$, for the sum over legs by charge conservation:
\begin{equation}\label{eqn:branchAntisymmetric}
    \sum_{\sigma = \pm} \sigma \, \mathcal{M}_{b^{(\sigma)}} ~ = ~ 0
    \qquad \text{and} \qquad
    \sum_{\sigma = \pm} \sigma \, \mathcal{M}_{b^{(\sigma)\dagger}} ~ = ~ 0 \, ,
\end{equation}
to be contrasted with \eqref{eqn:branchSum}. Hence
\begin{equation}\label{eqn:noKick}
	    \langle \text{out} | \Delta A^{+ \, \text{rad}}_{\ell m} S
    - S \, \Delta A^{+ \, \text{rad}}_{\ell m} | \text{in}\rangle ~ = ~ 0 \, :
\end{equation}
the radiated field has equal zero modes on the two branches, and adds nothing to the electric observable at leading order. The Coulomb part of $\Delta A^{+}_{\ell m}$ is weighted by the charge alone, by \eqref{eqn:coulombMemory}, and is unchanged by the scattering as well. On a physical state the impulse \eqref{eqn:memoryKick}, on every timelike radial worldline, is therefore the shock of what crossed, \eqref{eqn:coulombMemory}, and not of what scattered.

\paragraph{Where the cancellation comes from.} Equation \eqref{eqn:branchAntisymmetric} is a consequence of \eqref{appeqn:vertexRatio}. The numerator and denominator of \eqref{eqn:vertexRatio} depend on the photon branch only through the product $\sigma \sigma_{p_n}$. When $\sigma \sigma_{p_n} = -1$ the leading terms of the expansions above \eqref{eqn:kdotp} survive and are proportional, giving $+\sigma = -\sigma_{p_n}$; when $\sigma \sigma_{p_n} = +1$ they cancel and the mass terms give $-\sigma = -\sigma_{p_n}$, the explicit $\sigma$-dependence flipping sign between the two configurations in exactly the way needed for the ratio to be $\sigma$-independent. Keeping the first subleading terms of those expansions,
\begin{subequations}\label{eqn:vertexRatioCorrections}
    \begin{align}
    \left.\dfrac{\varepsilon_{(\sigma)}\left(\omega\right)\cdot p_n}
      {k\cdot p_n + \frac{1}{2}\left(m^2_{\ell_j}-m^2_{\ell_n}-\mu^2_\ell\right)}
      \right|_{\sigma\sigma_{p_n} = -1} &= - \sigma_{p_n}
      \left[ 1 + \dfrac{m^2_{\ell_j}-m^2_{\ell_n}-\mu^2_\ell}{4 \, \omega E_n}
      + \ldots \right] , \\
    \left.\dfrac{\varepsilon_{(\sigma)}\left(\omega\right)\cdot p_n}
      {k\cdot p_n + \frac{1}{2}\left(m^2_{\ell_j}-m^2_{\ell_n}-\mu^2_\ell\right)}
      \right|_{\sigma\sigma_{p_n} = +1} &= - \sigma_{p_n}
      \left[ 1 + \dfrac{\omega\left(m^2_{\ell_j}-m^2_{\ell_n}-\mu^2_\ell\right)}
      {E_n \, \mu^2_\ell}
      - \dfrac{2\,\omega^2 m^2_{\ell_n}}{E^2_n \, \mu^2_\ell} + \ldots \right] .
    \end{align}
\end{subequations}
Both corrections are $\sigma$-independent once written in terms of $\sigma_{p_n}$, so the branch-antisymmetric sum receives no contribution from the leading terms and is given by the difference of the two brackets. The second dominates the first by $\left(\omega/\mu_\ell\right)^2 \gg 1$, and $m^2_{\ell_n}/\mu^2_\ell$ is finite on the horizon, both squared masses being proportional to $f\left(r\right)$ by \eqref{eqn:masses}, so the difference is $\mathcal{O}\left(\omega/E_n\right)$:
\begin{equation}\label{eqn:antisymmetricSuppression}
    \sum_{\sigma = \pm} \sigma \, \mathcal{M}_{b^{(\sigma)}}
    ~ = ~ \mathcal{O}\!\left(\dfrac{\omega}{E_n}\right) \mathcal{M}_{b}
    \qquad \text{against} \qquad
    \sum_{\sigma = \pm} \mathcal{M}_{b^{(\sigma)}} ~ = ~ 2 \, \mathcal{M}_{b} \, .
\end{equation}
The remainder is suppressed by the hierarchy $\omega \ll E_n$ that defines the soft limit and is annihilated outright by the factor $\lim_{\omega\rightarrow 0}\sqrt{\omega}$ in \eqref{eqn:NandAplusMatrixElements}, since $\mathcal{M}_{b} \propto \omega^{-1/2}$ by \eqref{eqn:ampFactorised}. Equation \eqref{eqn:noKick} is therefore exact in the soft limit and does not refer to the masses. Outside the regime \eqref{eqn:frozenMassRegime} the same result follows from \eqref{eqn:classicalStep}. This is the computation in the gauge $A^{+}_{\ell m} = 0$ of \appref{app:interactionTerms}, in which each leg radiates the two branches equally. 

In the gauge $A_w = 0$ the emission of each leg sits on its own branch, as the shock \eqref{eqn:coulombMemory} does, the two branch totals differ by the charge-weighted combination of \eqref{eqn:chargeVsFlux}, and \eqref{eqn:noKick} is charge conservation. The per-leg distribution between the branches is not gauge invariant; the totals are, and they agree.

Equation \eqref{eqn:vertexRatio} carries both halves of the branch bookkeeping at once, and they might otherwise look like competing claims. That the ratio is \emph{odd} in $\sigma_{p_n}$ is what makes the hard side of \eqref{eqn:WardIdentity} a non-trivial flux: each hard leg enters \eqref{eqn:softAmplitudeReduced} weighted by $q_n \sigma_n$, and it is that weighted combination, rather than the total charge, which vanishes by conservation, that survives. That the same ratio is \emph{even} in $\sigma$, indeed independent of it, is what makes the branch-symmetric photon combination survive with multiplicity two in \eqref{eqn:branchSum} while the antisymmetric one vanishes in \eqref{eqn:branchAntisymmetric}. The two boundary parities therefore survive on opposite sides of the identity: the matter enters through its branch-antisymmetric flux, the photon through its branch-symmetric corner. Both facts are consequences of \eqref{eqn:kdotp}: numerator and denominator depend on the photon branch only through $\sigma \sigma_{p_n}$, so that the explicit $\sigma$ cancels against $\sigma_{p_n}$ in both configurations and what survives is odd in the branch of the leg alone.

\paragraph{The standing-wave interpretation.} The content of \eqref{eqn:branchAntisymmetric} is that the radiated field, summed over the legs, has equal zero modes on the two branches. 

As boundary data on the horizon its steps on the two branches are then equal, so that the antisymmetric combination in \eqref{eqn:NandAplusModes} cancels while the symmetric one doubles. By \eqref{eqn:EandBmemory} this is the familiar statement that the transverse electric field of a standing wave has a node where the transverse magnetic field has an antinode: for the zero mode of the field it radiates, the horizon behaves like the surface at which that node sits, and the whole of the radiated even-parity data reappears in the magnetic channel, where $N^{\text{rad}}_{\ell m}$ collects it. The statement is about the radiated field summed over the legs. The shock of a single crossing charge sits on one branch, and by \eqref{eqn:coulombMemory} it has no node at the horizon; the zero mode of the radiated field is minus the difference between the Coulomb memories of the out- and in-configurations, \eqref{eqn:radiatedMemory}, and it is divided equally between the two branches.

\paragraph{The radiated field in position space.} Both results can be read off from the field the process radiates. Let $\mathcal{E}^{\text{out}}_{\ell m}$ and $\mathcal{E}^{\text{in}}_{\ell m}$ be $\ell\left(\ell+1\right)$ times the expansion \eqref{eqn:PsiQuantisation} built from $b^{+,(\sigma)}_{\ell m}$ and $b^{-,(\sigma)}_{\ell m}$, and take the matrix element of \eqref{eqn:NandAplusMatrixElements} with the field in place of its zero mode. Only the annihilation part contributes to the first term and only the creation part to the second. For $\mu_\ell \ll \omega$ the mode function of the branch $\sigma$ is $e^{-i\omega\left(t - \sigma r^*\right)}$, a function of $u$ for $\sigma = +1$ and of $v$ for $\sigma = -1$. The mode function of \eqref{eqn:PsiQuantisation} carries $1/\sqrt{2\omega\ell\left(\ell+1\right)}$. It enters here explicitly, through the expansion of the field at the point where it is evaluated. It has already entered $\mathcal{M}_{b^{(\sigma)}} = - \mathcal{M}_{b^{(\sigma)\dagger}} = - \mathcal{N}_{\ell m}/\sqrt{2\omega\ell\left(\ell+1\right)}$ through the field in the vertex \eqref{eqn:vertex}. The two factors multiply to the $1/2\omega$ of the two-point function of the radiative field:
\begin{align}\label{eqn:radiatedStep}
\langle \text{out} | \mathcal{E}^{\text{out}}_{\ell m} S - S \, \mathcal{E}^{\text{in}}_{\ell m} | \text{in} \rangle ~ &= ~ - \, \mathcal{N}_{\ell m} \sum_{\sigma = \pm} \int_0^\infty \dfrac{\mathrm{d}\omega}{2\pi} \, \dfrac{\sin \omega\left(t - \sigma r^*\right)}{\omega} \nonumber \\
&= ~ - \, \dfrac{\mathcal{N}_{\ell m}}{4} \Big[ \mathrm{sgn}\left(u\right) + \mathrm{sgn}\left(v\right) \Big] \, ,
\end{align}
with the origin of $u$ and $v$ at the process. The amplitude has this form only for $\omega \ll E_p$, so the result holds for $\left|u\right|$ and $\left|v\right|$ large compared with $1/E_p$. The radiated field is a step of the same height on each branch. Each step is $-\mathcal{N}_{\ell m}/2$. Their sum is \eqref{eqn:radiatedMemory} and their difference is \eqref{eqn:noKick}. For a single charge turned around at the origin, the step in $u$ cancels the shock of the outgoing leg in $\Psi_{\ell m}$, and the step in $v$ is the shock that the incoming leg would have left had it continued. The field $\Psi_{\ell m}$ is then what it would have been without the scattering, as it must be by \eqref{eqn:sourcedEquations}. On a surface of constant $r^*$ the two steps are separated by $2\left|r^*\right|$ in $w$, and the condition $\omega\left|r^*\right| \ll 1$ of \eqref{eqn:boundaryModeCondition} says that the zero mode does not resolve the separation. The step is to $\mathcal{E}_{\ell m}$ what $A_z \propto \Theta\left(u\right)$ is to the radiative field at $\mathcal{I}^+$, and $\partial_w \mathcal{E}_{\ell m}$ is the corresponding pulse. In the plane-wave basis at $\mathcal{I}^+$, with covariantly normalised oscillators, the $1/\omega$ that transforms into the step is the pole of the amplitude. With the unit normalisation used here, it is the mode normalisation taken twice, once at the vertex and once at the point of observation. How it is divided between the amplitude and the measure depends on the normalisation of the oscillators, at either boundary. In the covariant normalisation of \appref{app:radiativeModeQuantisation} the measure carries all of it and the amplitude none.

All of this refers to the out-states of \secref{sec:externalstates}, before the outgoing step has been reflected. A wave of vanishing frequency is reflected by the potential with a change of sign, so that once the reflected step has returned the two cancel, $\Delta\mathcal{E}_{\ell m}$ vanishes again, and $N_{\ell m}$ is the Coulomb memory of the outgoing charges alone. We do not study that regime here.

\subsection{Comparison with null infinity}\label{sec:memoryAtScri}

We begin with the observation that nothing in \secref{sec:twoObservables} used the horizon. The partial-wave reduction \eqref{eqn:AaSoln}--\eqref{eqn:MoncriefEoM}, the branch decomposition \eqref{eqn:ka}, the identifications \eqref{eqn:EandBcomponents} and the impulse \eqref{eqn:memoryKick} all hold verbatim in flat space, with $f\left(r\right) \rightarrow 1$, $r^* \rightarrow r$ and $\mu^2_\ell \rightarrow \ell\left(\ell+1\right)/r^2$. It is therefore possible to compare the two boundaries by tracking a single equation.

\paragraph{Which branch crosses $\mathcal{I}^+$.} The grading is fixed by \eqref{eqn:ka} alone, relative to whichever null surface carries the coordinate $w = t + \eta^{wr^*}r^*$. Future null infinity is reached at fixed $u$, so there $w = u$ and $\eta^{wr^*} = -1$, exactly as on $\mathcal{H}^-$, and
\begin{equation}\label{eqn:scriBranches}
    k_{(\sigma)} ~ \xrightarrow{\ \mu_\ell \rightarrow 0 \ } ~
    \eta^{w r^*} \omega \left(1 + \eta^{wr^*}\sigma\right) ~ = ~
    \begin{cases}
    0 \, , & \eta^{wr^*}\sigma = -1 \, , \text{ i.e. } \sigma = +1 \, , \\
    -\,2\omega \, , & \eta^{wr^*}\sigma = +1 \, , \text{ i.e. } \sigma = -1 \, .
    \end{cases}
\end{equation}
By the identification in \secref{sec:externalstates}, $\sigma = +1$ is the radially outgoing branch. It is the branch crossing $\mathcal{I}^+$, as it is the branch crossing $\mathcal{H}^-$: $\mathcal{I}^+$ is the surface $v \rightarrow +\infty$, its generators are parametrised by $u$, and outgoing radiation, propagating along the rays of constant $u$, meets it in its cuts and varies along its generators. The branch along $\mathcal{I}^+$ is the ingoing one, with phase $e^{-i\omega u - 2 i \omega r^*} = e^{-i \omega v}$; it is constant along the generators of $\mathcal{I}^+$, and its free data lives on $\mathcal{I}^-$.

\paragraph{One branch of data at $\mathcal{I}^+$.} The radiative datum at $\mathcal{I}^+$ is extracted as $r \rightarrow \infty$ at fixed $u$, which is what the standard falloff $A_A = \mathcal{O}\left(1\right)$, with $F_{rA} = \mathcal{O}\left(r^{-2}\right)$, encodes. The branch along $\mathcal{I}^+$ is a function of $v = u + 2r$ alone, so that limit evaluates it at $v \rightarrow +\infty$, where any wave packet in it vanishes, and it contributes nothing to the datum. In the $\left(w, r^*\right)$ chart this is $\chi^{\ell m}_{r^*} \rightarrow 0$, and by \eqref{eqn:chirstarCrossing} it says that the branch along $\mathcal{I}^+$ carries no news there:
\begin{equation}\label{eqn:scriNoCrossingNews}
    \partial_w \Psi^{(\sigma)}_{\ell m}\Big|_{\eta^{wr^*}\sigma = +1} ~ = ~ 0
    \qquad \Longrightarrow \qquad
    \Delta \Psi^{(\sigma)}_{\ell m}\Big|_{\eta^{wr^*}\sigma = +1} ~ = ~ 0 \, .
\end{equation}
This is not a condition imposed on the field. It is what the limit that defines the boundary does to the branch that runs along it, and it will hold on any null boundary approached in the same way. Feeding \eqref{eqn:scriNoCrossingNews} into \eqref{eqn:BranchesFromObservables} collapses the two boundary observables onto one,
\begin{equation}\label{eqn:scriDegeneration}
    \Delta A^{+}_{\ell m} ~ = ~ - \, \eta^{wr^*} N_{\ell m} ~ = ~ N_{\ell m}
    \qquad \text{at } \mathcal{I}^+ ,
\end{equation}
and equivalently, in \eqref{eqn:radialComponentsRelated}, $\chi^{\ell m}_{w} = \eta^{wr^*} \chi^{\ell m}_{r^*}\big|_t$, so that the time-integrated transverse electric and magnetic fields are the same function. At $\mathcal{I}^+$ there is one memory, not two.

\paragraph{The velocity kick.} With \eqref{eqn:scriDegeneration}, the impulse \eqref{eqn:memoryKick}, on any timelike radial detector, gives
\begin{equation}\label{eqn:scriKick}
    \Delta p_A = - q' \sum_{\ell m} \Delta A^{+}_{\ell m} \, \partial_A Y_{\ell m} = - q' \sum_{\ell m} N_{\ell m} \, \partial_A Y_{\ell m} = \sum_{\ell m} \dfrac{q'}{\ell\left(\ell+1\right)} \dfrac{Q_{\mathrm{soft}}\left[\lambda_{\ell m}\right]} {\lambda_{\ell m}} \, \partial_A Y_{\ell m} \, ,
\end{equation}
which is the statement that the large-gauge vacuum transition is physical because it is registered as a velocity kick on test charges \cite{Bieri:2013hqa, Pasterski:2015zua}. Every step of that argument survives on a stretched horizon except \eqref{eqn:scriNoCrossingNews}. There the kick measures $\Delta A^{+}_{\ell m}$ and the memory measures $N_{\ell m}$, and by \eqref{eqn:MemoryBranchSplit} the two differ by twice the data on the ingoing branch. The field radiated by a scattering process changes $N^{\text{rad}}_{\ell m}$ by \eqref{eqn:radiatedMemory} and adds nothing to $\Delta A^{+}_{\ell m}$, \eqref{eqn:noKick}. A crossing charge enters both, by \eqref{eqn:coulombMemory}, with the same sign if it crosses outward and with opposite signs if it crosses inward. The vacuum transition produced by a scattering process is still physical, still labelled by $\varphi^{(0)}_{\ell m}$, and still measurable. But on the horizon it is measured as the change in the radial electric flux through a cut, by \eqref{eqn:GaussLawMemory}, while any timelike detector is kicked by the shock of every charge that passed it, \eqref{eqn:coulombKick}. This is a memory of what crossed but not of what scattered.

\paragraph{The Coulomb memory at the two boundaries.} The second difference between the two boundaries is the geometric one behind \eqref{eqn:coulombMemory}. In the near-horizon region the metric \eqref{eqn:metric} is, to leading order in $f\left(r\right)$, the product of two-dimensional Rindler space with a sphere of radius $R$, and the null planes of the Rindler factor are the surfaces of constant $u$ and constant $v$. A charge crossing $\mathcal{H}^+$ at angle $\Omega_q$ moves along $v = v_q$, its shock is the null plane $v = v_q$, and that plane meets the horizon, and every stretched horizon, at the same value of $v$ at every angle: the step is deposited on the whole cross section at once, at finite transverse distance $R\gamma$ from every point of it, and its partial-wave moments are the order-one quantities of \eqref{eqn:coulombMemory}. In flat space the shock of a null charge is also a null plane, but a null plane is tilted with respect to every sphere about the origin. A radially outgoing charge at angle $\Omega_q$ has its shock on $t - r\cos\gamma = \text{const}$, which meets the sphere of radius $r$ at retarded time $u_\gamma = u_q - r\left(1 - \cos\gamma\right)$, and in the limit $r \rightarrow \infty$ at fixed $u$ that defines $\mathcal{I}^+$ this recedes to $u \rightarrow -\infty$ at every angle but $\gamma = 0$. The step has not disappeared; it has been pushed into the early boundary of $\mathcal{I}^+$, where it is the $u$-independent Coulomb datum $A_A \propto \partial_A \ln\left(1 - \cos\gamma\right)$ of the outgoing charge at $\mathcal{I}^+_-$, and the antipodal matching across $i^0$ identifies it with the datum that the corresponding incoming charge carries at $\mathcal{I}^-_+$. A charge that propagates freely through flat space therefore leaves no memory at $\mathcal{I}^+$, and the memory of a scattering process there is the difference between the out- and in-Coulomb data. On the horizon the zero mode of the radiated field is minus the corresponding difference, \eqref{eqn:radiatedMemory}, and the Coulomb memory of each crossing sits beside it, because the surface is at finite radius and the charges pass through it; the two cancel in $N_{\ell m}$. The horizon is not needed for this. Any finite distance suffices. On a sphere of finite radius in flat space, Gauss' law fixes the divergence of the tangential vector potential left behind by the total charge that has crossed the sphere \cite{Susskind:2015hpa}. On a null plane at large but finite distance from the sources, the Maxwell constraint likewise ties a memory at Coulombic order to the flux of charged particles through the plane, for massless and for massive sources \cite{Raclariu:2026mfs}. What the limit $r \rightarrow \infty$ at fixed $u$ does is push the step of a free charge off the surface. The branch along the horizon is treated alike by the two boundaries: a charge on it passes a stretched horizon at $r^*_0$ at $v = u_q + 2 r^*_0$, which recedes to the bifurcation sphere as $r^*_0 \rightarrow -\infty$, exactly as the flat-space step recedes to $\mathcal{I}^+_-$. The difference between the two boundaries lies entirely in the crossing branch, whose shock sits at finite $v$ at every angle on the horizon and at finite $u$ only at the charge’s own angle at $\mathcal{I}^+$. This is the same fact as the finite transverse mass of \secref{sec:irwindow}: a horizon is a null boundary with a sphere of fixed size attached to it, and a null charge crossing it is at finite distance from every point of that sphere.

\paragraph{Why the horizon keeps both branches.} On the horizon itself the same thing happens. The branch along $\mathcal{H}^+$ is $e^{-i\omega\left(v - 2 r^*\right)} = e^{-i\omega u}$, the outgoing modes that hug the horizon, and the limit $r^* \rightarrow -\infty$ at fixed $v$ that defines $\mathcal{H}^+$ evaluates it at $u \rightarrow +\infty$, where a wave packet in it vanishes just as the ingoing packet vanishes at $\mathcal{I}^+$; in Kruskal terms it is a function of $U$ alone and is constant on $U = 0$. A true horizon, like $\mathcal{I}^+$, carries one branch of characteristic data, and \eqref{eqn:scriNoCrossingNews} holds on it. The constructions of this paper are not placed on $\mathcal{H}^+$ but on a stretched horizon, by the correlated limit \eqref{eqn:boundaryModeCondition}, and the two boundaries differ.

What the strict limit would cost is half the phase space. In flat space the two branches are not independent data: the homogeneous radial solution regular at the origin is
\begin{subequations}\label{eqn:regularityAtOrigin}
\begin{align}
    \sin\!\left(\omega r - \dfrac{\ell \pi}{2}\right) ~ &= ~ \dfrac{1}{2 i}\left(e^{i \omega r} e^{-i\ell\pi/2} - e^{-i\omega r}e^{+i \ell \pi/2}\right) \quad \text{which implies} \\
    b^{(-)}_{\ell m}\left(\omega\right) ~ &= ~ \left(-1\right)^{\ell+1} b^{(+)}_{\ell m}\left(\omega\right) \, ,
\end{align}
\end{subequations}
at large $r$ and solves \eqref{eqn:MoncriefEoM} with $f\left(r\right)=1$. It is one function of $\omega$ per $\left(\ell, m\right)$ rather than two, so that discarding the branch along $\mathcal{I}^+$ discards no degrees of freedom: its content is carried by the crossing branch and recovered on $\mathcal{I}^-$, and $\mathcal{I}^+$ on its own is a complete boundary for the radiative field. Near the horizon there is no origin. The region behind $\mathcal{H}$ is causally excluded, the near-horizon expansion supplies no analogue of \eqref{eqn:regularityAtOrigin}, the two sets $b^{(\pm)}_{\ell m}$ are independent on the states of \secref{sec:externalstates}, which are defined before the outgoing branch has been reflected by the potential, and the hard legs of the scattering sit on both branches. Projecting onto the crossing branch alone would discard half the radial phase space and, as noted below \eqref{eqn:alongPhase}, would collapse the soft theorem \eqref{eqn:finalAmp} onto the pure-gauge statement \eqref{eqn:A+amplitude}.

What makes it possible to avoid the strict limit is the rate at which the phase of the branch along the surface accumulates. With $f\left(r\right) \simeq e^{r^*/R}$ near the horizon, the proper distance from a point to $\mathcal{H}$ is $2 R \sqrt{f\left(r\right)}$, and
\begin{equation}\label{eqn:phaseAccumulation}
    \left|k_{(\sigma)} r^*\right|_{\mathcal{I}^+} ~ = ~ 2 \omega r \, ,
    \qquad \qquad
    \left|k_{(\sigma)} r^*\right|_{\mathcal{H}} ~ = ~
    2 \omega R \left|\ln f\left(r\right)\right| ~ = ~
    4 \omega R \left|\ln \dfrac{2 R \sqrt{f\left(r\right)}}{2 R}\right| \, .
\end{equation}
At $\mathcal{I}^+$ the phase grows linearly in the proper distance to the boundary, which is infinite; the region $\omega r \lesssim 1$ in which the branch along $\mathcal{I}^+$ would still vary along the generators is the near zone, not an asymptotic region, and there is no surface that is both at large $r$ and carries both branches as data. Near the horizon the boundary is at finite proper distance and the phase grows only logarithmically in it, so that the condition $\omega \left|r^*\right| \ll 1$ is met throughout the near-horizon region out to $f\left(r\right) \sim e^{-1/\omega R}$, exponentially close to the horizon in $f$ and arbitrarily close to it in proper distance. On any such stretched horizon the branch along $\mathcal{H}^+$ still varies along the generators, both branches are data, and \eqref{eqn:scriNoCrossingNews} does not hold. This is the precise sense in which the correlated limit \eqref{eqn:boundaryModeCondition} is available on the horizon and not at null infinity, and it is the origin of the differences discussed above.

\paragraph{Summary of the comparison.} At $\mathcal{I}^+$: the crossing branch is the outgoing radiation; the branch along it is the ingoing one and carries no news; $N_{\ell m}$ and $\Delta A^{+}_{\ell m}$ degenerate by \eqref{eqn:scriDegeneration}; the transverse electric and magnetic memories coincide; a free charge leaves no memory, its shock having been pushed into the Coulomb datum at $\mathcal{I}^+_-$; and the memory of a scattering process is a velocity kick. On a stretched horizon, where the near-horizon scattering is placed: both branches are independent boundary data by \eqref{eqn:boundaryModeCondition}; $N_{\ell m}$ is the branch-symmetric combination and is the soft charge, $\Delta A^{+}_{\ell m}$ the branch-antisymmetric combination and not a charge at all; the first is the time-integrated transverse magnetic field, the second the time-integrated transverse electric field; every crossing charge leaves its shock on both, with equal magnitude, at the finite value of the generator coordinate at which it crossed; the field radiated by the scattering adds to the first alone, by \eqref{eqn:noKick}, minus the difference between the Coulomb memories of its out- and in-configurations, so that $N_{\ell m}$ is conserved; and the scattering is observed as the Gauss-law flux difference \eqref{eqn:GaussLawMemory} between the two ends of the horizon, which is not an impulse on a timelike detector.

\paragraph{Relation to other notions of horizon memory.} Several notions of memory have been defined on horizons and other null boundaries in the literature. In \cite{Donnay:2018ckb} the change in the horizon data produced by a transcient gravitational shock wave is shown to be the action of a horizon supertranslation composed with a superrotation on the stationary solution, and is measured by the change in the corresponding charges; in the Einstein--Maxwell extension the supertranslated Reissner--Nordstr\"om solution is obtained from a horizon superrotation together with a gauge transformation whose parameter depends on the electric charge. In \cite{Rahman:2019bmk} the memory tensor is the change in the induced metric of the horizon between early and late stationary eras, the time integral of the expansion and shear of the generators, and separately a supertranslation relates the preferred foliations of the two eras; the finding is that on a horizon the two are not related to each other, nor the memory tensor to the charges and fluxes of the horizon symmetries, because gravitational radiation changes the horizon geometry at zeroth order. In \cite{Adami:2021nnf} memory is the persistent change of the surface charges of a null boundary under a flux of news, read from the Raychaudhuri and Damour equations. In \cite{Danielson:2022tdw, Danielson:2022sga} the electromagnetic memory of a Killing horizon is the change $\Delta A_B$ in the transverse vector potential along the generators between early and late affine times, produced by a charge displaced in the exterior and held there; the constraint along the generators ties its exact part to the change in the radial electric field through the horizon, $D^A \Delta A_A = - \Delta E_U$, the analysis is stated for any dimension $d \geq 4$, and it is the record on the horizon of the position of the exterior charge whose superposition it decoheres. In \cite{Danielson:2025aji} this $\Delta A_B$ is promoted to an observable, central in the algebra of fields that decay along the generators and no longer central once the Weyl elements that shift it are adjoined, and the Killing-invariant state is extended to the enlarged algebra, with states of different memory orthogonal on the whole horizon while the algebra of the past of a cut does not contain the memory, and coherent states restricted to it have nonzero fidelity at finite affine time. The memory in this paper is the electromagnetic counterpart of the time-integrated shear of \cite{Rahman:2019bmk}, the zero mode $N_{\ell m}$ of the radiative field along the generators, and \eqref{eqn:GaussLawMemory} is the partial-wave form of the constraint of \cite{Danielson:2022sga}, with $\ell\left(\ell+1\right) N_{\ell m}$ in place of $D^A \Delta A_A$ and with the flux $\mathcal{Q}_{\ell m}$ of the charges that cross the surface added; what is new relative to \cite{Rahman:2019bmk} is the relation they found missing: $N_{\ell m}$ is canonically conjugate to the vacuum label by \eqref{eqn:MemoryGoldstoneBracket}, is the soft charge by \eqref{eqn:QsoftIsMemory}, and is fixed by the hard flux and the radial electric flux through \eqref{eqn:GaussLawMemory}, the latter being given by the amplitude through \eqref{eqn:radiatedMemory}, so that it stands to the horizon charges as the memory at $\mathcal{I}^+$ stands to the large gauge charges.

Where this paper and \cite{Danielson:2022sga} differ is in what the memory does to a charge on the horizon. There the data on $\mathcal{H}^+$ are the single function $A_B\left(V, x^A\right)$ along the generators, so that the change in radial flux and the transverse impulse $q \, \Delta A_B$ that a charge moving along a generator would receive are one statement, although no such impulse is discussed there; in the language of \secref{sec:memoryAtScri} this is the degeneration \eqref{eqn:scriDegeneration}, which holds on $\mathcal{H}^+$ as on $\mathcal{I}^+$, and their memory, the inward-propagating change in the Coulomb field of a charge held in the exterior, is a datum on the crossing branch alone, which by \eqref{eqn:coulombMemory} registers on both of our observables with equal magnitude. The kick that vanishes here, \eqref{eqn:noKick}, is the impulse \eqref{eqn:memoryKick} on any timelike detector near a stretched horizon that traverses both pulses, at fixed radius or moving radially, which is the antisymmetric combination of the two branches, a component that the data on $\mathcal{H}^+$ alone do not contain; a detector on a null generator is the singular limit of that family, with impulse $\tfrac{1}{2}\left(\eta^{wr^*} N_{\ell m} - \Delta A^{+}_{\ell m}\right)$ from the crossing branch alone, which is the only impulse their data determine and which no massive charge realises. Scattering near the horizon is not considered in \cite{Danielson:2022sga, Danielson:2025aji}, and the memory it radiates, \eqref{eqn:radiatedMemory}, has equal zero modes on the two branches. No timelike detector sees this contribution. The obstruction identified in \cite{Rahman:2019bmk} does not arise for a Maxwell field on a fixed background, which leaves the horizon geometry untouched; whether it returns in the gravitational case is a question we leave for future work.

\section{Infrared behaviour}\label{sec:irbehaviour}

The absence of a pole in \eqref{eqn:vertexRatio} is a statement about the vertex in the partial-wave basis. It is not, by itself, a statement about the infrared behaviour of the $S$-matrix, which is governed by the phase space of the soft mode rather than by the vertex. The logarithmic divergence of the flat-space $S$-matrix is present on the horizon as well. In the partial-wave basis it is the norm of a single zero mode per partial wave, and that zero mode is the radiative part $N^{\text{rad}}_{\ell m}$ of the memory of \secref{sec:memory}.

\subsection{Counting}\label{sec:ircounting}

In the plane wave basis in four dimensions the soft factor scales as $\omega^{-1}$ while the phase space measure for the emitted photon carries $\omega^2 \mathrm{d}\omega$, and the emission probability behaves as $\int \mathrm{d}\omega/\omega$. In the partial wave basis the angular directions have been traded for the labels $\left(\ell, m\right)$, and by \eqref{eqn:PsiAlgebra} the measure is $\mathrm{d}\omega$ with no accompanying power of the frequency. The two powers of $\omega$ that the squared soft factor has lost are precisely the two that the measure has lost. Naive counting therefore returns the same logarithm as in the plane wave basis in flat space, and whatever is different about the horizon must come from the range of the integration and from the interpretation of its endpoints, not from the integrand.

Throughout the window \eqref{eqn:softlimit} one has $\widetilde{\omega} \simeq \omega$, and by \eqref{eqn:softAmplitudeReduced} the amplitude factorises as
\begin{align}\label{eqn:ampFactorised}
\mathcal{M}_{b^{(\sigma)}}\left(\omega\right) ~ = ~ - \, \dfrac{\mathcal{N}_{\ell m}}{\sqrt{2 \, \omega \, \ell \left(\ell+1\right)}} \, , \qquad \text{independently of } \sigma \, ,
\end{align}
with the entire frequency dependence residing in the normalisation of the external state and the soft coefficient \eqref{eqn:softCoefficient} independent of $\omega$ to leading order in the soft limit. Using the algebra \eqref{eqn:PsiAlgebra}, the resolution of the identity on the one-photon sector of the mode $\left(\ell, m\right)$ is 
\begin{equation}
	\mathbb{1} ~ = ~ \sum_\sigma \int \dfrac{\mathrm{d}\omega}{2\pi} b^{(\sigma)\dagger}_{\ell m}\left(\omega\right)\left|0\right\rangle 			\left\langle 0 \right| b^{(\sigma)}_{\ell m}\left(\omega\right) \, , 
\end{equation}
so that the mean number of soft photons emitted into that mode is
\begin{align}\label{eqn:softProbability}
P_{\ell m} ~ = ~ \sum_{\sigma = \pm} \int \dfrac{\mathrm{d}\omega}{2\pi} \, \left|\mathcal{M}_{b^{(\sigma)}}\left(\omega\right)\right|^2 ~ = ~ \dfrac{\left|\mathcal{N}_{\ell m}\right|^2}{2\pi \, \ell \left(\ell+1\right)} \int \dfrac{\mathrm{d}\omega}{\omega} \, .
\end{align}
Everything that follows is a discussion of the two limits of this integral.

\subsection{The radiative window at a stretched horizon}\label{sec:irwindow}

Restricted to the window \eqref{eqn:softlimit}, with $c \gg 1$ a fixed number keeping the endpoints away from the two edges of the window, the integral is elementary. The infrared end is the transverse mass $\mu_\ell$, and although $\mu_\ell$ vanishes as the horizon is approached, what enters the logarithm is a ratio. Using $\mu^2_\ell = f\left(r\right)\ell\left(\ell+1\right)/r^2$,
\begin{equation}\label{eqn:windowWidth}
\ln \left( \dfrac{E_p}{\mu_\ell} \right) ~ = ~ \ln \left( \dfrac{E^{\text{loc}}_p \, r}{\sqrt{\ell \left(\ell + 1\right)}} \right) \, , \qquad \text{with} \qquad E^{\text{loc}}_p ~ \coloneqq ~ \dfrac{E_p}{\sqrt{f\left(r\right)}} \, ,
\end{equation}
identically. The right-hand side compares two quantities as measured by a static observer at $r$: the hard energy $E^{\text{loc}}_p$, and the transverse mass $\sqrt{\ell\left(\ell+1\right)}/r$, which is what $\mu_\ell$ becomes once the redshift is divided out, since $\mu_\ell$ is a mass with respect to $\eta_{ab}$ while the block metric is $f\left(r\right)\eta_{ab}$. Both are finite on the horizon, and
\begin{align}\label{eqn:softProbabilityResult}
P_{\ell m}\Big|_{\text{window}} ~ = ~ \dfrac{\left|\mathcal{N}_{\ell m}\right|^2}{2\pi \, \ell \left(\ell+1\right)} \left[ \, \ln \left( \dfrac{E^{\text{loc}}_p \, R}{\sqrt{\ell \left(\ell+1\right)}} \right) - 2 \ln c \, \right] \, ,
\end{align}
where the dependence on $c$ is an additive constant and does not affect the coefficient of the logarithm.

The Regge--Wheeler potential $V_\ell\left(r\right) = f\left(r\right)\ell\left(\ell+1\right)/r^2$ vanishes on the horizon, so the spectrum of Killing frequencies at every $\ell$ is continuous down to $\omega = 0$; $\mu_\ell$ is the value of $\sqrt{V_\ell}$ at the radius where the scattering has been placed. In the WKB description, a mode of Killing frequency $\omega$ propagates at that radius if $\omega > \mu_\ell\left(r\right)$; if $\omega < \mu_\ell\left(r\right)$ it does not propagate at $r$, but it propagates perfectly well closer in, at every radius inside the turning point $r_\omega$ defined by $\mu_\ell\left(r_\omega\right) = \omega$. At the radius $r$, therefore, a mode in the sector $\omega < \mu_\ell\left(r\right)$ is evanescent in the WKB sense. It is not thereby static: it keeps its time dependence and propagates closer in. Its $\omega \rightarrow 0$ member is the static $\ell$-pole profile solving $\left(r^2/f\right)\partial^2_{r^*}\Psi_{\ell m} - \ell\left(\ell+1\right)\Psi_{\ell m} = \text{const}$, which is \eqref{eqn:MoncriefMaster} with $\partial_w = 0$. The endpoint difference $\Delta\Psi_{\ell m}$ at fixed $r$ is the change of that static field between early and late times, which is what the Gauss-law form \eqref{eqn:GaussLawMemory} says.

In the WKB description, the scale $\mu_\ell\left(r\right)$ thus separates locally propagating from locally evanescent modes at a \emph{stretched} horizon at radius $r$, and \eqref{eqn:softProbabilityResult} is a statement about that surface: it counts the radiative soft photons resolved by a process placed on a stretched horizon, and it says that the count does not depend on where the horizon is stretched. The threshold is a statement about the local WKB wavenumber $\sqrt{\omega^2 - \mu^2_\ell\left(r\right)}$, real above $\mu_\ell\left(r\right)$ and imaginary below, and the exact radial functions \eqref{eqn:photonRadialExact} make no transition there: wherever $\mu_\ell R \ll 1$, which is everywhere a stretched horizon is placed, a mode of any Killing frequency $\omega \ll 1/R$ is a plane wave of wavenumber $\omega$ to relative accuracy $R^2\mu^2_\ell$, and neither it nor its WKB counterpart changes appreciably across a region of size $R$ on either side of the threshold. The scale $\mu_\ell\left(r\right)$ is therefore the resolution of the count rather than a threshold the field respects: it is the frequency below which a process placed at $r$ cannot tell a radiative mode from the static profile by its local dispersion. It is the cutoff that defines the count; we make no claim about what a detector of finite duration and bandwidth resolves. The count \eqref{eqn:softProbabilityResult} is the number of soft photons above that resolution, and its independence of the stretching is unaffected. This independence is evident in \eqref{eqn:windowWidth}. The process described by \eqref{eqn:finalAmp} is a local one, specified by the energies its participants carry in their own neighbourhood, so it is $E^{\text{loc}}_p$ that is held fixed as the scattering is placed at different radii, and the width of the radiative window is then set by the horizon radius alone. Holding the Killing energy $E_p$ fixed instead, while $r^* \rightarrow - \infty$, describes a different process at each radius, of unboundedly growing local energy, and it is that family which fails to have a finite width; we return to it in \secref{sec:irlimits}. The coefficient of the logarithm is built from \eqref{eqn:softCoefficient}, in which each hard leg enters weighted by $q_n \sigma_n$: the infrared logarithm is governed by the same flux-weighted combination that appears in the hard charge \eqref{eqn:hardChargeResult} and in the Ward identity \eqref{eqn:WardIdentity}, rather than by the total charge, which vanishes by conservation.

There is a structural difference between the horizon and null infinity here. The locally measured transverse mass of the $\ell$-th partial wave of a null boundary whose cross sections have radius $r_{\text{bdy}}$ is $\sqrt{\ell\left(\ell+1\right)}/r_{\text{bdy}}$. At $\mathcal{I}$ the cross sections have unbounded area, this scale goes to zero, and no invariant threshold separates the Coulomb sector from the radiation. At the horizon the cross sections have area $4\pi R^2$, transverse wavelengths are bounded by the circumference, and the threshold is finite. Said differently, the near-horizon limit $r^* \rightarrow -\infty$ does not enlarge the sphere, whereas the limit $r \rightarrow \infty$ at null infinity does. The same scale governs the Regge--Wheeler barrier separating the near zone from the exterior, whose peak sits at the photon sphere $r = 3R/2$ with
\begin{equation}\label{eqn:barrierPeak}
\sqrt{V^{\text{max}}_{\ell}} ~ = ~ \dfrac{2\sqrt{3}}{9} \, \dfrac{\sqrt{\ell \left(\ell + 1\right)}}{R} \, ,
\end{equation}
so that radiation below it reaches infinity only with probability $\mathcal{O}\big(\left(\omega R\right)^{2\ell+2}\big)$. The single channel for which this scale vanishes, $\ell = 0$, is exactly the one in which the radiative mode does not exist; there the Maxwell field has only the monopole Coulomb field, fixed by the enclosed charge, and the constant gauge parameter $\lambda_{00}$, which carries ordinary charge conservation and no radiation.

The barrier \eqref{eqn:barrierPeak} governs escape to infinity. It does not govern emission into the horizon, and one should not expect the sub-threshold modes $\omega < \mu_\ell\left(r\right)$ to be tunnelling-suppressed in that direction. The potential between the process at $r_0$ and the turning point $r_\omega$ is exponentially thin in the tortoise coordinate: with $\mu^2_\ell \propto e^{r^*/R}$ near the horizon, the WKB exponent across the forbidden region is
\begin{align}\label{eqn:wkbExponent}
\int_{r^*_\omega}^{r^*_0} \mathrm{d}r^* \sqrt{\mu^2_\ell\left(r^*\right) - \omega^2} ~ &= ~ 2R \left[ \sqrt{\mu^2_\ell\left(r_0\right) - \omega^2} - \omega \arccos \dfrac{\omega}{\mu_\ell\left(r_0\right)} \right] \nonumber \\
&\leq ~ 2 R \, \mu_\ell\left(r_0\right) \nonumber \\
&\simeq ~ 2 \sqrt{f\left(r_0\right) \ell\left(\ell+1\right)} \, ,
\end{align}
which vanishes as the process approaches the horizon. The near-horizon region is transparent to its own sub-threshold modes, which is the membrane-paradigm statement that a horizon reflects nothing sent towards it from close by. The sector below $\mu_\ell\left(r\right)$ is therefore excluded from \eqref{eqn:softProbabilityResult} not because it is dynamically suppressed but because it lies below the resolution of a count made at that radius. This excluded sector is the subject of \secref{sec:irzeromode}.

\subsection{The collinear sector}\label{sec:ircollinear}

Before turning to the zero mode, we isolate the collinear sector, because it is a separate and finite infrared phenomenon. For the configurations $\sigma\sigma_{p_n} = +1$, in which the soft photon and the hard leg it attaches to lie on the same radial branch, both the numerator and the denominator of \eqref{eqn:vertexRatio} vanish at zero mass, as the expansions above \eqref{eqn:kdotp} show. The near-horizon charges are effectively massless, $m^{2}_\ell \propto f\left(r\right)$ as in \eqref{eqn:masses}, and a massless charge radiating collinearly is a singular configuration in any dimension. In flat massless QED the collinear region is a measure-zero set of angles that contributes a logarithm to probabilities; in the partial wave basis the angular integral has become a sum over two branches, the collinear configurations are half of the branch combinations, and their contribution is finite and of order one. What makes it finite is shown in \appref{app:interactionTerms}. The masses fix the ratio of two vanishing quantities to $-\sigma_{p_n}$, a value that does not depend on them. Holding the masses fixed is justified in the regime \eqref{eqn:frozenMassRegime}, in which the collinear emission is completed over a distance short compared with $R$; outside that regime, which contains the whole of the zero-mode regime $\omega R \ll 1$ and, by \secref{sec:irwindow}, every partial wave with $\mu_\ell R \ll 1$ on a stretched horizon, the momentum-space computation does not control the collinear half, and the soft factor is fixed instead by the field equation for $\mathcal{E}_{\ell m}$ at leading order, as shown in \appref{app:interactionTerms}. The masses therefore set the width of the window \eqref{eqn:softlimit}, and they set the scale $\sqrt{\ell\left(\ell+1\right)}/R$ at which the sum over partial waves is cut off, which is where the collinear sector reappears; we return to it in \secref{sec:irdressing}.

\subsection{The zero mode}\label{sec:irzeromode}

We now extend the integral in \eqref{eqn:softProbability} to $\omega \rightarrow 0$, which is the horizon itself. By \secref{sec:memory}, the extension diverges for a reason that is unrelated to the dynamics.

The operator $N^{\text{rad}}_{\ell m}$ of \eqref{eqn:NandAplusModes} is built from $\lim_{\omega \rightarrow 0}\sqrt{\omega}\,\big(b + b^\dagger\big)$, with the endpoint prescription stated there. On any state of the Fock space of \eqref{eqn:PsiAlgebra} with finitely many photons in smooth wave packets, the factor $\sqrt{\omega}$ annihilates both terms, and $N^{\text{rad}}_{\ell m}$ has vanishing matrix elements between any two such states. The Ward identity \eqref{eqn:WardIdentity}, read through \eqref{eqn:memorySplit}, nevertheless gives
\begin{equation}\label{eqn:memoryMatrixElement}
\langle \text{out} | N^{\text{rad}}_{\ell m} S - S N^{\text{rad}}_{\ell m} | \text{in}\rangle ~ \neq ~ 0 \, , 
\end{equation}
and this is the case whenever the hard dynamics turns a charged leg around; for a process in which every leg keeps its radial branch, $\mathcal N_{\ell m}$ vanishes along with the memory the process adds, as it must, the Coulomb memory of its crossing legs being unchanged. If $S$ mapped the Fock space to itself, the left-hand side would vanish identically. It does not, and therefore $S|\text{in}\rangle$ does not lie in the Fock space built on the vacuum of $|\text{in}\rangle$. This is the horizon version of the argument of \cite{Kapec:2017tkm} at null infinity, and it uses only the existence of a gapless zero mode conjugate to the vacuum label, which is what the corner term of \eqref{eqn:CornerSplit} provides.

The same conclusion follows from the classical field, by reading \eqref{eqn:radiatedStep} backwards. A non-vanishing $N^{\text{rad}}_{\ell m}$ means that $\mathcal{E}_{\ell m}/\ell\left(\ell+1\right)$ along the generator changes by that amount between $w \to - \infty$ and $w \to +\infty$; a function with a finite step has Fourier transform $\simeq i N^{\text{rad}}_{\ell m}/\omega$ as $\omega \rightarrow 0$. By \eqref{eqn:PsiQuantisation} this transform is proportional to $\langle b_{\ell m}\left(\omega\right)\rangle/\sqrt{\omega}$, so that the out-state carries
\begin{equation}\label{eqn:coherentTail}
\langle b^{(\sigma)}_{\ell m}\left(\omega\right) \rangle ~ \propto ~ \dfrac{N^{\text{rad}}_{\ell m}}{\sqrt{\omega}} \qquad \text{as} \qquad \omega \rightarrow 0 \, ,
\end{equation}
which is the $\omega^{-1/2}$ of \eqref{eqn:ampFactorised} seen from the other side: the normalisation factor that produced the apparent pole in the amplitude is what a step in the field requires. At leading soft order the insertions of several soft photons factorise, since by \eqref{eqn:hardChargeOnModes} each acts on the hard state by a charge that commutes with the others, and the out-state is a coherent state of the mode $\left(\ell, m\right)$ with displacement \eqref{eqn:coherentTail}. Its mean photon number is $\int \mathrm{d}\omega \, \left|\langle b \rangle\right|^2 \propto \int \mathrm{d}\omega/\omega$, and its overlap with any state of finite photon number is
\begin{equation}\label{eqn:fockOverlap}
\exp \left[ - \dfrac{1}{2} \sum_{\sigma} \int \dfrac{\mathrm{d}\omega}{2\pi} \, \left|\langle b^{(\sigma)}_{\ell m}\left(\omega\right) \rangle\right|^2 \right] ~ = ~ 0 \, .
\end{equation}
As mentioned below \eqref{eqn:MemoryGoldstoneBracket}, a change in the zero mode leaves the horizon in a \emph{different} member of the degenerate family labelled by $\varphi^{(0)}_{\ell m}$. The members are unitarily inequivalent representations of the algebra \eqref{eqn:PsiAlgebra}. Therefore, to say that ``the horizon ends in a different vacuum" is also to say that ``the $S$-matrix element between undressed Fock states vanishes". The logarithmic divergence of \eqref{eqn:softProbability} at $\omega \rightarrow 0$ is the radiative memory, counted in photons.

On a stretched horizon the same exponent is finite. With the integral cut off at the window edges,
\begin{align}\label{eqn:fockOverlapStretched}
\left| \langle 0 | \, \text{coherent state} \, \rangle \right| ~ &= ~ \left( \dfrac{c^2 \, \mu_\ell}{E_p} \right)^{\gamma_{\ell m}} ~ = ~ \left( \dfrac{c^2 \sqrt{\ell\left(\ell+1\right)}}{E^{\text{loc}}_p \, R} \right)^{\gamma_{\ell m}} \, , \qquad \text{where} \\
 \gamma_{\ell m} ~ &\coloneqq ~ \dfrac{\left|\mathcal{N}_{\ell m}\right|^2}{4\pi \, \ell\left(\ell+1\right)} \, , \nonumber
\end{align}
with $\mathcal{N}_{\ell m}$ understood per unit hard amplitude, so that $\gamma_{\ell m}$ is one half of the coefficient in \eqref{eqn:softProbabilityResult}. The undressed matrix element at a stretched horizon is suppressed by a finite, redshift-invariant power set by the flux-weighted charge, and at fixed Killing energy, since $\mu_\ell \propto \sqrt{f\left(r\right)}$, it decays as $f\left(r\right)^{\gamma_{\ell m}/2}$ as the stretched horizon is taken to the true one, while at fixed local energy it does not depend on the stretching at all. That is the sense in which \eqref{eqn:softProbabilityResult} is finite: finite on every stretched horizon, with a width that does not depend on the stretching at fixed local energy, and divergent on the horizon along the family at fixed Killing energy, where the memory lives.

\subsection{The order of limits}\label{sec:irlimits}

The preceding two subsections may be stated as a single constraint on how the limits are taken. The width of the radiative window is controlled by one invariant parameter, the ratio $E_p/\mu_\ell = E^{\text{loc}}_p R/\sqrt{\ell\left(\ell+1\right)}$. The error of the soft theorem \eqref{eqn:finalAmp} depends on where $\omega$ sits relative to the second scale $1/R$ of \secref{sec:symmetries}. For $\omega R \gg 1$, where the photon is WKB and the momentum-space expressions apply, the error has two sources that pull in opposite directions. At the ultraviolet edge the soft factor is off by $\mathcal{O}\left(\omega /E_p\right)$, through the dependence of the ratio \eqref{eqn:vertexRatioCorrections} and of the hard amplitude on the photon momentum. At the infrared edge the normalisation $\left(2\tilde{\omega}\right)^{-1/2}$ of the external photon in \eqref{eqn:vertex}, replaced by $\left(2\omega\right)^{-1/2}$ in \eqref{eqn:finalAmp}, is off by $\mu^2_\ell/4\omega^2$. They balance at $\omega \sim \left(\mu^2_\ell E_p\right)^{1/3}$, where their common value is\footnote{The two errors are monotonic in opposite directions, $\mu^2_\ell/4\omega^2$ falling and $\omega/E_p$ rising with $\omega$, so the error is minimised where they are equal, $\mu^2_\ell/\omega^2 = \omega/E_p$, that is at $\omega^3 = \mu^2_\ell E_p$; substituting back into either gives \eqref{eqn:softTheoremError}. The exponents are those of the two expansions. The infrared one is quadratic because $\tilde{\omega} = \omega\sqrt{1 - \mu^2_\ell/\omega^2}$ is even in $\mu_\ell$; the ultraviolet one is linear because the leading terms of the collinear ratio cancel and what is left are the subleading terms of \eqref{eqn:kdotp}, which are first order in $\omega/E_n$. On a logarithmic scale the balance point sits one third of the way across the window from its infrared edge. The correction in the non-collinear branch of \eqref{eqn:vertexRatioCorrections}, of order $\mu^2_\ell/\omega E_p$, is the product of the two and so smaller than either. The $\mathcal{O}\left(1\right)$ coefficients are not controlled by this estimate.}
\begin{equation}\label{eqn:softTheoremError}
\left( \dfrac{\mu_\ell}{E_p} \right)^{2/3} ~ = ~ \left( \dfrac{\sqrt{\ell\left(\ell+1\right)}}{E^{\text{loc}}_p \, R} \right)^{2/3} \, .
\end{equation}
This regime is not the near-horizon limit: the balance point has $\omega R \gg 1$ only for $f\left(r\right) \gg \big(\ell\left(\ell+1\right) E^{\text{loc}}_p R\big)^{-2/3}$, the frozen-mass computation on which the collinear bracket rests requires in addition \eqref{eqn:frozenMassRegime}, and the stretched horizon lies well inside both. 

Where \eqref{eqn:frozenMassRegime} fails, at $\omega R \gg 1$ or for $\omega R \ll 1$, the soft factor on the collinear half is fixed by the field-equation computation of \appref{app:interactionTerms} and not by the frozen-mass computation, and the two error sources are unchanged, the ultraviolet one now coming from the hard amplitude alone. For $\omega R \ll 1$ the infrared-edge error is absent, because the photon's radial function is a plane wave of wavenumber $\omega$ with exact normalisation, \eqref{eqn:photonRadialExact}; what remains is $\mathcal{O}\left(\omega/E_p\right)$ from the dependence of the hard amplitude on the photon momentum, which falls with $\omega$, and the correction $R^2 \mu^2_\ell$ to the photon mode, which does not depend on $\omega$ at all. The error therefore decreases monotonically as $\omega \rightarrow 0$, down to a floor that vanishes with $f\left(r\right)$. The exact soft theorem is the limit $\omega \rightarrow 0$ followed by $f\left(r\right) \rightarrow 0$, on the out-states of \secref{sec:externalstates}, and it does not require $E_p/\mu_\ell \rightarrow \infty$: the lower edge of the window \eqref{eqn:softlimit} is where the WKB wavenumber ceases to be real, and for $\omega R \ll 1$ the amplitude does not see it. This is the order that has recently been found to matter at null infinity as well: in odd dimensions the memory integral and the large-$r$ expansion do not commute, and the Coulombic-order memory exists only if the integral is taken on a surface at finite radius before the radius is sent to infinity \cite{Raclariu:2026mfs}. Here, likewise, the zero mode is taken on the stretched horizon before the surface is taken to $\mathcal{H}$.

The radiative logarithm \eqref{eqn:softProbabilityResult} and the soft theorem are therefore not in competition. The three ways of taking the limits differ in what they do to the mean number of radiative photons:
\begin{itemize}
	\item Holding the local energy fixed keeps the logarithm finite, with the width $\ln\big(E^{\text{loc}}_p R\big)$ of \secref{sec:irwindow}. Since $E_p = \sqrt{f\left(r\right)}\,E^{\text{loc}}_p$, the window \eqref{eqn:softlimit} sends $\omega \rightarrow 0$ with $f\left(r\right)$, the regime is $\omega R \ll 1$, and the soft theorem holds to $\mathcal{O}\left(\omega/E_p\right)$. Along this route $E_p R \rightarrow 0$ and $\omega R \rightarrow 0$, and the collinear half is fixed by the field-equation computation of \appref{app:interactionTerms}; the route yields the soft theorem, the complete phase space and the finite count \eqref{eqn:softProbabilityResult} together.
	\item Holding the Killing energy fixed while $r^* \rightarrow -\infty$, which is what a horizon $S$-matrix labelled by conserved Killing energies naturally does, leaves the soft theorem exact and makes the logarithm divergent, since the window widens without bound.
	\item The correlated limit \eqref{eqn:boundaryModeCondition} that defines the boundary mode drives $E_p/\mu_\ell \rightarrow \infty$ along the way, so that the memory becomes sharp as the number of radiative photons diverges.
\end{itemize}
Setting the masses to zero at the outset leaves the amplitude unchanged, since the value fixed in \eqref{appeqn:vertexRatio} does not depend on the masses and the field-equation computation of \appref{app:interactionTerms} does not involve them, and leaves $\int \mathrm{d}\omega/\omega$ with no infrared edge at all; it is the limit of the third route. On every route the full photon number, extended to $\omega \rightarrow 0$, diverges. That divergence is the theorem of \secref{sec:irzeromode}.

\subsection{Dressing by the corner mode}\label{sec:irdressing}

The divergence has a standard remedy which can be made explicit on the horizon phase space. Write the hard charge as a sum over partial waves with an operator coefficient $\mathcal{Q}_{\ell m}$ of \eqref{eqn:DeltaG},
\begin{equation}\label{eqn:hardChargeDensity}
	Q_{\mathrm{hard}}\left[\lambda\right] ~ = ~ \sum_{\ell m} \lambda_{\ell m} \, \mathcal{Q}_{\ell m} \, , 
\end{equation}
so that, by \eqref{eqn:memorySplit},
\begin{equation}
	Q_{\mathrm{soft}}\left[\lambda\right] ~ = ~ Q_{\text{cut}}\left[\lambda\right] - Q_{\mathrm{hard}}\left[\lambda\right] ~ = ~ - \sum_{\ell m} \lambda_{\ell m} \Big( \ell\left(\ell+1\right) N^{\text{rad}}_{\ell m} + \mathcal{Q}_{\ell m} \Big) \, .
\end{equation}
On states localised on the sphere, which are the states that diagonalise the hard charges and the states in which \eqref{eqn:SmatrixPhase} was written, $\mathcal{Q}_{\ell m}$ has the eigenvalue $\sum_n \sigma_n q_n Y_{\ell m}\left(\Omega_n\right)$. The operators $\mathcal{Q}_{\ell m}$ commute among themselves, the residual symmetry algebra being abelian. They also commute with the boundary coordinate $\varphi_{\ell m}$: by \eqref{eqn:AplusBranches} it is a coordinate on the Maxwell phase space, the $w$-independent part of $A^{+}_{\ell m}$, whereas by \eqref{eqn:hardChargeResult} $\mathcal{Q}_{\ell m}$ is bilinear in the scalar modes, and the two sectors are quantised independently in \appref{app:MaxwellinPWs}. Define
\begin{equation}\label{eqn:dressingOperator}
W ~ \coloneqq ~ \exp\left[ i \sum_{\ell m} \varphi_{\ell m} \, \mathcal{Q}_{\ell m} \right] \, ,
\end{equation}
where the sum runs over $\ell \geq 1$. The bracket \eqref{eqn:MemoryGoldstoneBracket} holds for $N^{\text{rad}}_{\ell m}$ as it does for $N_{\ell m}$, since $\mathcal{Q}_{\ell m}$ commutes with $\varphi_{\ell m}$, and gives
\begin{align}\label{eqn:dressedMemory}
W^\dagger N^{\text{rad}}_{\ell m} W ~ &= ~ N^{\text{rad}}_{\ell m} - \dfrac{\mathcal{Q}_{\ell m}}{\ell\left(\ell+1\right)} \, , \qquad \text{and hence} \qquad W^\dagger \, Q_{\mathrm{soft}} \, W ~ = ~ Q_{\text{cut}} \, . 
\end{align}
Let $|\psi\rangle$ be any state of the Fock space of \eqref{eqn:PsiAlgebra} tensored with a hard state, so that $N^{\text{rad}}_{\ell m}|\psi\rangle = 0$, and let $|\psi\rangle\!\rangle \coloneqq W |\psi\rangle$. Then \eqref{eqn:dressedMemory} gives $Q_{\mathrm{soft}}|\psi\rangle\!\rangle = 0$ for every $\left(\ell, m\right)$: the dressed states are the states of vanishing soft charge. Since the Ward identity is the statement that $S$ commutes with the soft charge, the obstruction of \secref{sec:irzeromode} is absent between dressed states, and the dressed $S$-matrix
\begin{equation}\label{eqn:dressedSmatrix}
\langle\!\langle \text{out} | S | \text{in} \rangle\!\rangle ~ \coloneqq ~ \langle \text{out} | \, W^{+\dagger} \, S \, W^{-} \, | \text{in} \rangle \, ,
\end{equation}
with $W^\pm$ built from $\varphi^\pm_{\ell m}$ and $\mathcal{Q}^\pm_{\ell m}$ on the two horizons, is the object to which the soft theorem applies without a vanishing prefactor. On a hard eigenstate the dressed state is an eigenstate of the radiative zero mode, $N^{\text{rad}}_{\ell m}|\psi\rangle\!\rangle = - \mathcal{Q}_{\ell m}\,|\psi\rangle\!\rangle/\ell\left(\ell+1\right)$, equal and opposite to the Coulomb memory \eqref{eqn:coulombMemory} of its own charges, so that its memory $N_{\ell m}$ vanishes. An undressed state already carries the shock of every hard charge, in $\Psi_{\ell m}$; what the dressing adds is the radiative zero mode that the scattering would otherwise have to supply. That is the horizon form of the Faddeev--Kulish prescription \cite{Chung:1965zza, Kibble:1968sfb, Kulish:1970ut}, and it makes the relation to \eqref{eqn:SmatrixPhase} transparent: $W$ is the phase of \eqref{eqn:SmatrixPhase} with the fixed parameter $\lambda_{\ell m}$ replaced by the dynamical boundary coordinate $\varphi_{\ell m}$ and the charges $q_n$ replaced by the fluxes $\sigma_n q_n$. The dressing is the exponential of the pure-gauge boundary mode, and can therefore be thought of as a Wilson line to the corner.

The same operator, written in the radiative modes, creates a coherent state of the zero mode, and this is where the counting of \secref{sec:ircounting} is recovered. Since $W$ shifts $N^{\text{rad}}_{\ell m}$ by a c-number on each hard eigenstate and $N^{\text{rad}}_{\ell m}$ is linear in $b$ and $b^\dagger$, we take $W$ to act on the modes as the displacement $W^\dagger b^{(\sigma)}_{\ell m}\left(\omega\right) W = b^{(\sigma)}_{\ell m}\left(\omega\right) + \alpha^{(\sigma)}_{\ell m}\left(\omega\right)$ that gives the dressed state the coherent tail \eqref{eqn:coherentTail}; the corner bracket fixes only the zero mode of this displacement, and inserting it into \eqref{eqn:NandAplusModes} the shift in \eqref{eqn:dressedMemory} is reproduced by
\begin{equation}\label{eqn:dressingDisplacement}
\alpha^{(\sigma)}_{\ell m}\left(\omega\right) ~ = ~ - \, \dfrac{\mathcal{Q}_{\ell m}}{\sqrt{2 \, \ell\left(\ell+1\right) \, \omega}} \, , \qquad \text{independently of } \sigma \, ,
\end{equation}
with the endpoint prescription of \eqref{eqn:NandAplusModes}. There are three observations to be made about this displacement:
\begin{itemize}
	\item The displacement has the form of the amplitude \eqref{eqn:ampFactorised}, with $\mathcal{N}_{\ell m} \rightarrow \mathcal{Q}_{\ell m}$; this is the Ward identity again, and the relative sign between the $b$ and $b^\dagger$ insertions in \eqref{eqn:branchSum} is the sign that makes the dressing undo the emission rather than double it. 
	\item The photon number of the dressing, $\sum_\sigma \int \left(\mathrm{d}\omega/2\pi\right) \left|\alpha^{(\sigma)}_{\ell m}\right|^2 = \left|\mathcal{Q}_{\ell m}\right|^2/\big(2\pi\ell\left(\ell+1\right)\big) \int \mathrm{d}\omega/\omega$, has the same coefficient as \eqref{eqn:softProbability}, so that the logarithm of \eqref{eqn:softProbabilityResult} is the norm of the relative dressing between the in- and out-states and the exponent $\gamma_{\ell m}$ of \eqref{eqn:fockOverlapStretched} is one half of it, exactly as the exponentiation of the flat-space soft factor would give.
	\item The Ward identity fixes only the branch-symmetric part of the displacement, since $N^{\text{rad}}_{\ell m}$ involves the branch sum alone, and \eqref{eqn:dressingDisplacement} is the symmetric choice. 
	The Coulomb memory of a hard leg sits on that leg's own branch, \eqref{eqn:coulombMemory}, and a dressing that cancels it branch by branch differs from \eqref{eqn:dressingDisplacement} by a branch-antisymmetric displacement with coefficient $\varrho_{\ell m} = \sum_n q_n Y_{\ell m}\left(\Omega_n\right)$, the charge-weighted moment of \eqref{eqn:chargeVsFlux}, which shifts $\Delta A^{+ \, \text{rad}}_{\ell m}$ by $- \varrho_{\ell m}/\ell\left(\ell+1\right)$ and $N^{\text{rad}}_{\ell m}$ by nothing. At leading order in the near-horizon and soft expansions this factor drops out of \eqref{eqn:dressedSmatrix}: $\varrho_{\ell m}$ is conserved by the near-horizon dynamics (see \secref{sec:gaugeinvariance}), so it is the same displacement on the in- and out-sides, and $S$ commutes with the branch-antisymmetric zero mode by \eqref{eqn:noKick}. The two prescriptions therefore give the same dressed $S$-matrix, with the same infrared exponent, and differ only in the dressed states themselves, the per-leg cloud cancelling in addition the electric kick \eqref{eqn:memoryKick} of the state's own charges.
\end{itemize}

Finally, two general remarks are also in order. First, the vanishing \eqref{eqn:fockOverlap} of undressed elements is exact and required no loop computation, being a property of coherent-state overlaps; the finiteness of the dressed element \eqref{eqn:dressedSmatrix} to all orders requires in addition that the virtual soft photons of the near-horizon theory exponentiate into \eqref{eqn:fockOverlap}, as they do in flat space. We have not verified this at loop level; we merely treat it as an expectation. Second, the inclusive alternative to dressing would proceed as in flat space \cite{Bloch:1937pw}. The real emission of soft photons below a resolution scale, whose number \eqref{eqn:softProbability} diverges logarithmically, is to be combined with the virtual soft corrections to the hard process, and it is the combination that is finite. We have not computed the virtual corrections in the near-horizon theory. The count \eqref{eqn:softProbabilityResult} is a different quantity, the mean number of real photons above the scale $\mu_\ell\left(r\right)$, and it is finite without them.

\paragraph{Summing the partial waves.} The exponent $\gamma_{\ell m}$ is stated per partial wave. The full suppression of an undressed element is the product over $\left(\ell, m\right)$, and the sum of the exponents does not converge on its own. On hard states localised on the sphere, with $\varepsilon_n \coloneqq \pm\sigma_n q_n$ for outgoing and incoming legs respectively,
\begin{equation}\label{eqn:partialWaveSum}
\sum_{m} \left| \Delta\mathcal{Q}_{\ell m} \right|^2 ~ = ~ \sum_{n, n'} \varepsilon_n \varepsilon_{n'} \, \dfrac{2\ell+1}{4\pi} \, P_\ell\left(\cos\theta_{nn'}\right) \, ,
\end{equation}
by the addition theorem, where $\Delta\mathcal{Q}_{\ell m}$ is the difference of the out and in eigenvalues. At leading order in the near-horizon limit the dynamics is local on the sphere (\secref{sec:gaugeinvariance}), so the outgoing and incoming legs of one charge sit at the same angle and their cross term carries $P_\ell\left(1\right) = 1$ at every $\ell$. For a charge that keeps its radial branch, $\varepsilon^{\mathrm{out}}_n = - \varepsilon^{\mathrm{in}}_n$ and its four terms cancel; for a charge that is turned around, $\varepsilon^{\mathrm{out}}_n = \varepsilon^{\mathrm{in}}_n$ and they add to $4 q^2_n \left(2\ell+1\right)/4\pi$. Only the charges that change branch contribute, with effective charge $2 q_n$, and $\sum_\ell \left(2\ell+1\right)/\ell\left(\ell+1\right)$ is logarithmically divergent. The sum is cut off by the window itself, since $\mu_\ell \propto \sqrt{\ell\left(\ell+1\right)}$ and the window \eqref{eqn:softlimit} for the $\ell$-th partial wave is empty once $c^2 \mu_\ell > E_p$, i.e. for $\sqrt{\ell\left(\ell+1\right)} \gtrsim E^{\text{loc}}_p R/c^2$. At leading logarithmic accuracy the summed exponent is then
\begin{equation}\label{eqn:doubleLog}
\sum_{\ell m} \gamma_{\ell m} \ln \left( \dfrac{E^{\text{loc}}_p R}{c^2\sqrt{\ell\left(\ell+1\right)}} \right) ~ \simeq ~ \dfrac{1}{4\pi^2} \sum_{n \, \text{turned}} q_n^2 \; \ln^2 \left( \dfrac{E^{\text{loc}}_p R}{c^2} \right) ~ + ~ \mathcal{O}\left( \ln \dfrac{E^{\text{loc}}_p R}{c^2} \right) \, ,
\end{equation}
where the sum runs over the charges that change radial branch and the single logarithm collects the cross terms between distinct charges, whose Legendre sums converge for charges at distinct angles. The near-horizon $S$-matrix between undressed states therefore carries a Sudakov double logarithm rather than the single logarithm of massive flat-space QED. This is the collinear sector of \secref{sec:ircollinear}: each partial wave contributes a soft logarithm, the sum over partial waves is the sum over angular resolutions $\theta \sim 1/\ell$, and the two are cut off by the same invariant, the size of the hole in units of the hard wavelength. The collinear regulator that the mass provides in flat space is here provided by the transverse mass $\sqrt{\ell\left(\ell+1\right)}/R$ of the horizon. The dressing \eqref{eqn:dressingOperator} removes the double logarithm as it removes the single one, since it is exact in $\left(\ell, m\right)$; the estimate \eqref{eqn:doubleLog} is at leading logarithmic accuracy only, the window edges at each $\ell$ not being controlled beyond $\mathcal{O}\left(1\right)$.

\subsection{The state}\label{sec:irstate}

As mentioned in \secref{sec:externalstates}, everything in this paper so far has been said in the vacuum annihilated by the $b^{(\sigma)}_{\ell m}\left(\omega\right)$ of \eqref{eqn:PsiQuantisation}, which are positive frequency in $w$ at fixed $r^{*}$, that is with respect to Killing time. Near the horizon that is the Boulware state. It is not the state of a physical black hole: its renormalised stress tensor diverges on the horizon in a freely falling frame, so it is compatible with a fixed background only at finite $f\left(r\right)$, and it is in that sense the state of a cold stretched horizon. The state that describes a black hole in equilibrium with its radiation is the Hartle--Hawking state, in which the near-horizon modes are thermally populated at the Hawking temperature
\begin{equation}\label{eqn:hawkingTemperature}
T_H ~ = ~ \dfrac{\kappa}{2\pi} ~ = ~ \dfrac{1}{4\pi R} \, , \qquad n\left(\omega\right) ~ = ~ \dfrac{1}{e^{\omega/T_H} - 1} \, ,
\end{equation}
in Killing frequency. For a black hole formed by collapse the appropriate state is the Unruh state, with the complication noted in footnote \ref{fn:unruh}; we confine ourselves to the Hartle--Hawking state. Since the infrared question is a question about soft modes, and $T_H$ is a scale in precisely the frequency variable in which the soft limit is taken, the choice of state is not a refinement of the results above but a change of regime.

\paragraph{Both branches are thermal.} In the Hartle--Hawking state both branches of the radiative mode are populated,
\begin{equation}\label{eqn:HHoccupation}
\big\langle b^{(\sigma)\dagger}_{\ell m}\left(\omega\right) b^{(\sigma')}_{\ell' m'}\left(\omega'\right) \big\rangle_{\text{HH}} ~ = ~ n\left(\omega\right) \, \delta_{\sigma\sigma'}\delta_{\ell\ell'}\delta_{mm'} \, 2\pi\delta\left(\omega - \omega'\right) \, ,
\end{equation}
and in the Boulware state neither is. The two branches, which \secref{sec:memory} needed to be independent data, are independently populated on the states of \secref{sec:externalstates}, before the outgoing branch has been reflected by the potential. In the global Hartle--Hawking state the reflected part of an outgoing mode is correlated with it, and the results below hold for Killing times short compared with the return time $2\left|r^*_0\right|$.

\paragraph{What does not change.} The Ward identity \eqref{eqn:WardIdentity}, the soft theorem \eqref{eqn:finalAmp}, the ratio \eqref{eqn:vertexRatio}, the memory relations \eqref{eqn:QsoftIsMemory} and \eqref{eqn:GaussLawMemory}, and the dressing \eqref{eqn:dressingOperator}--\eqref{eqn:dressedMemory} are operator identities and hold in any state. The dressing acts on a thermal density matrix as a displacement, $\rho_{T_{H}} \rightarrow W \rho_{T_{H}} W^\dagger$, and the mean number of photons it adds, $\sum_\sigma\int \left(\mathrm{d}\omega/2\pi\right)\left|\alpha^{(\sigma)}_{\ell m}\right|^2$, is independent of the occupation; so is the net number emitted by the process. The radiative count \eqref{eqn:softProbabilityResult}, read as the net number of soft photons added to the mode $\left(\ell, m\right)$ at a stretched horizon, is therefore unaffected. What the state changes is everything that is quadratic in the field: overlaps, fluctuations and the distinction between emission and absorption.

\paragraph{The overlap.} The thermal counterpart of the vacuum overlap \eqref{eqn:fockOverlap} is the expectation value of the displacement in the thermal state, $\operatorname{Tr}\left(\rho_{T_{H}} \, W\right)$, and for each mode it carries the factor $2n\left(\omega\right) + 1 = \coth\left(\omega/2T_H\right)$ in the exponent, by \eqref{appeqn:thermalTrace}:
\begin{align}\label{eqn:thermalOverlap}
\left| \operatorname{Tr}\left(\rho_{T_H} \, W\right) \right| ~ &= ~ \exp\left[ - \dfrac{1}{2}\sum_\sigma \int \dfrac{\mathrm{d}\omega}{2\pi} \left|\alpha^{(\sigma)}_{\ell m}\right|^2 \Big( 2 n_\sigma\left(\omega\right) + 1 \Big) \right] \nonumber \\
&= ~ \exp\left[ - \gamma_{\ell m} \int \dfrac{\mathrm{d}\omega}{\omega} \, \coth \dfrac{\omega}{2 T_H} \right]
\end{align}
in the Hartle--Hawking state, with $n_\sigma = n$ on both branches by \eqref{eqn:HHoccupation}. For $\omega \ll T_H$ the integrand is $2T_H/\omega^2$, and the logarithmic divergence of \secref{sec:irzeromode} becomes a linear one, as it does for soft photons in any thermal bath \cite{Weldon:1991eg, Weldon:1993qj, Donoghue:1984zz, Johansson:1985if, Indumathi:1996ec}. On a stretched horizon, with the lower limit $c\mu_\ell \ll T_H$,
\begin{align}\label{eqn:thermalOverlapStretched}
\int_{c\mu_\ell}^{E_p/c} \dfrac{\mathrm{d}\omega}{\omega} \, \coth \dfrac{\omega}{2 T_H} ~ &\simeq ~ \left[ \dfrac{2 T_H}{c \, \mu_\ell} + \ln \dfrac{E_p}{c \, T_H} \right] \nonumber \\
&= ~ \left[ \dfrac{r}{2\pi c R \sqrt{f\left(r\right) \ell\left(\ell+1\right)}} + \ln \Big( 4\pi \sqrt{f\left(r\right)} \, E^{\text{loc}}_p R \Big) - \ln c \right]  ,
\end{align}
up to terms of order one. The suppression of an undressed element is no longer a power of $f\left(r\right)$ but an exponential in $f\left(r\right)^{-1/2}$. This is a statement about the amplitude and not about the states: $\operatorname{Tr}\left(\rho_{T_H} W\right)$ is the thermal expectation of the displacement, the object whose vanishing is the infrared divergence of the undressed $S$-matrix in a thermal state. The fidelity between $\rho_{T_H}$ and $W \rho_{T_H} W^\dagger$ carries $\tanh\left(\omega/2T_H\right) = 1/\left(2n\left(\omega\right)+1\right)$ in place of $\coth$, by \eqref{appeqn:thermalFidelityMode},
\begin{equation}\label{eqn:thermalFidelity}
F\left(\rho_{T_H} \, , \, W \rho_{T_H} W^\dagger\right) ~ = ~ \exp\left[ - \gamma_{\ell m} \int \dfrac{\mathrm{d}\omega}{\omega} \, \tanh \dfrac{\omega}{2 T_H} \right] ~ \simeq ~ \left( \dfrac{c \, T_H}{E_p} \right)^{\gamma_{\ell m}} \, ,
\end{equation}
which reduces to \eqref{eqn:fockOverlap} at $T_H = 0$ and whose integral converges at $\omega \rightarrow 0$. In the Hartle--Hawking state, therefore, the thermal state and its displacement by the memory have finite overlap, independent of where the horizon is stretched, with the temperature as the infrared cutoff in place of $\mu_\ell$. The bath makes the undressed amplitude smaller and the displaced states less distinguishable at the same time; the second is the statement that reappears below as the growth of the variance.

\paragraph{The regimes.} Equation \eqref{eqn:thermalOverlapStretched} also shows that the window \eqref{eqn:softlimit} has acquired a third scale. In local units the three scales at a stretched horizon are the transverse mass $\sqrt{\ell\left(\ell+1\right)}/r$, the Tolman temperature $T^{\text{loc}}_H = T_H/\sqrt{f\left(r\right)} = 1/4\pi R\sqrt{f\left(r\right)}$, and the hard energy $E^{\text{loc}}_p$. The first is fixed by the geometry, the third by the process, and the second grows without bound as the stretched horizon is taken to the true one. At fixed local energy there are accordingly three regimes in $f\left(r\right)$. For $f\left(r\right) \gg 1/16\pi^2\ell\left(\ell+1\right)$ the temperature lies below the threshold, the whole window is vacuum-like and \secref{sec:irwindow}--\secref{sec:irdressing} apply as written; for the low partial waves this regime barely overlaps the near-horizon region $f\left(r\right) \ll 1$ at all. For $\big(4\pi E^{\text{loc}}_p R\big)^{-2} \ll f\left(r\right) \ll 1/16\pi^2\ell\left(\ell+1\right)$ the temperature lies inside the window: soft photons with $\mu_\ell \ll \omega \ll T_H$ are emitted into a bath with occupation $n \simeq T_H/\omega \gg 1$, stimulated emission and absorption each exceed the spontaneous rate by that factor, and only their difference is given by \eqref{eqn:softProbabilityResult}\footnote{The linear divergence has a stronger consequence than sharpening the overlap once it is resummed. In the approximation in which the hard charges do not recoil, which is the approximation of \secref{sec:irzeromode}, Weldon \cite{Weldon:1993qj} summed the thermal soft-photon sector to all orders. Upon resummation, the first-order spectrum $\mathrm{d}P/\mathrm{d}\omega \simeq A\,T/\omega^2$, with $A$ the coefficient of the vacuum $1/\omega$, is replaced by a finite spectrum cut off below $\omega \sim A\,T$. Per partial wave the coefficient here is $A_{\ell m} = 2 \gamma_{\ell m}$ by \eqref{eqn:softProbability}, so the cutoff sits at a Killing frequency of order $\gamma_{\ell m} T_H$, that is at $\gamma_{\ell m} T^{\text{loc}}_H$ in local units, and it rises above the transverse mass $\sqrt{\ell\left(\ell+1\right)}/r$ once $f\left(r\right) \lesssim \gamma^2_{\ell m}/16\pi^2 \ell\left(\ell+1\right)$, which is inside the second regime. Below that radius the sub-threshold part of the window is reorganised by the resummation rather than merely populated, and the three-scale picture acquires a fourth scale. The count \eqref{eqn:softProbabilityResult} and the overlaps \eqref{eqn:thermalOverlap}--\eqref{eqn:thermalOverlapStretched} are fixed-order statements and are unaffected; we do not pursue the resummed spectrum here.}; photons with $T_H \ll \omega \ll E_p$ see the vacuum. The two sub-windows are the regimes $\omega R \ll 1$ and $\omega R \gg 1$ below \eqref{eqn:photonRadialExact}. In this range of $f\left(r\right)$ one has $\mu_\ell R \ll 1$, so \eqref{eqn:frozenMassRegime} is met in neither, the collinear half is fixed by the field-equation computation of \appref{app:interactionTerms} in both, and the soft theorem holds in both. This second range of $f\left(r\right)$ is the one in which the near-horizon soft theorem operates for any process of astrophysical interest, since its lower boundary is exponentially close to the horizon. For $f\left(r\right) \ll \big(4\pi E^{\text{loc}}_p R\big)^{-2}$ the hard energy itself lies below the temperature, the external states are indistinguishable from the bath, and a hierarchy of the form \eqref{eqn:softlimit} no longer exists. The strict limit $f\left(r\right) \rightarrow 0$ at fixed local energy therefore ends in the third regime, and this is the reason why the near-horizon $S$-matrix is a stretched-horizon object.

\paragraph{The memory in a thermal state.} The last effect is on the fluctuations of the zero mode. Let $N^{(T)}_{\ell m}$ be the radiative memory accumulated over a Killing-time interval of length $T$ along the generator, which tends to $N^{\text{rad}}_{\ell m}$ as $T \rightarrow \infty$. Since $\sqrt{\ell\left(\ell+1\right)}\,\Psi_{\ell m}$ is a canonically normalised two-dimensional massless field on the horizon by \appref{app:radiativeModeQuantisation}, its two-point function along the generator is that of a chiral boson on a line, and the thermal excess over the vacuum variance is
\begin{align}\label{eqn:memoryVariance}
\big\langle \big(N^{(T)}_{\ell m}\big)^2 \big\rangle_{\text{HH}} - \big\langle \big(N^{(T)}_{\ell m}\big)^2 \big\rangle_{\text{B}} ~ &= ~ \dfrac{1}{\pi \, \ell\left(\ell+1\right)} \ln \left[ \dfrac{\sinh \left(\pi T_H T\right)}{\pi T_H T} \right] \nonumber \\
&\simeq ~ \begin{cases} \dfrac{\pi T^2_H T^2}{6 \, \ell\left(\ell+1\right)} \, , & T \ll T^{-1}_H \, , \\[1em] \dfrac{T_H \, T}{\ell\left(\ell+1\right)} \, , & T \gg T^{-1}_H \, , \end{cases}
\end{align}
with the vacuum variance $\langle (N^{(T)}_{\ell m})^2 \rangle_{\text{B}} \simeq \ln\left(\Lambda T\right)/\pi\ell\left(\ell+1\right)$, $\Lambda$ the highest frequency the measurement resolves. The overall normalisation is subject to the endpoint convention of \eqref{eqn:NandAplusModes}. In the Boulware state the memory has the slowly growing variance of the zero mode of any two-dimensional massless field; in the Hartle--Hawking state it performs, after a thermal time, a random walk driven by the bath, with diffusion constant $T_H/\ell\left(\ell+1\right)$.

What a measurement of $N^{(T)}_{\ell m}$ resolves is fixed by comparing the memory a process deposits, $\Delta N^{\text{rad}}_{\ell m} = - \Delta\mathcal{Q}_{\ell m}/\ell\left(\ell+1\right)$ by \eqref{eqn:radiatedMemory} or $N^{\rm C}_{\ell m} = \mathcal{Q}_{\ell m}/\ell\left(\ell+1\right)$ by \eqref{eqn:coulombMemory}, with these fluctuations. For a single elementary charge, $\sum_m \left|\mathcal{Q}_{\ell m}\right|^2 = q^2\left(2\ell+1\right)/4\pi$, and the signal is smaller than the vacuum variance for any $\Lambda T \gtrsim 1$, by a factor of order $\ell\left(\ell+1\right)\ln\left(\Lambda T\right)/q^2$; summing the partial waves gains a logarithm of their number. A single measurement of $N^{(T)}_{\ell m}$ therefore resolves the memory of one charge only at order $q^2/\ell\left(\ell+1\right)$ in the Boulware state, and the bath lowers this further. This is a statement about the signal-to-noise ratio of the single observable $N^{(T)}_{\ell m}$ in the states considered; a state squeezed in the zero mode would do better, and optimal discrimination on the whole radiation state is a different question. In the language of \secref{sec:irzeromode}, the coherent state \eqref{eqn:coherentTail} of one charge carries $\bar{n}_{\ell m} \sim q^2\ln\left(\Lambda T\right)/\ell\left(\ell+1\right)$ photons per partial wave, fewer than one for $\ln\left(\Lambda T\right) \ll \ell\left(\ell+1\right)/q^2$; in that range its fidelity with the state without the memory is $1 - \mathcal{O}\big(q^2 \ln\left(\Lambda T\right)/\ell\left(\ell+1\right)\big)$, and the orthogonality \eqref{eqn:fockOverlap} is reached as $T \rightarrow \infty$, over all the soft modes at once, not through the single observable $N^{(T)}_{\ell m}$. The memory of a macroscopic coherent flux, $\left|\Delta\mathcal{Q}_{\ell m}\right|^2 \gg \ell\left(\ell+1\right) \ln\left(\Lambda T\right)/\pi$, exceeds the vacuum variance, and in the Hartle--Hawking state it remains resolvable against the random walk for a Killing time up to
\begin{equation}\label{eqn:memoryCoherenceTime}
T ~ \lesssim ~ \dfrac{\left|\Delta\mathcal{Q}_{\ell m}\right|^2}{\ell\left(\ell+1\right) \, T_H} ~ = ~ \dfrac{4\pi R \, \left|\Delta\mathcal{Q}_{\ell m}\right|^2}{\ell\left(\ell+1\right)} \, ,
\end{equation}
with $\mathcal{Q}_{\ell m}$ in place of $\Delta\mathcal{Q}_{\ell m}$ for a crossing flux. The condition that makes the memory visible is the condition that \eqref{eqn:memoryCoherenceTime} lie in the regime $T \gg T^{-1}_H$ of linear growth, so the estimate holds exactly where it applies. The soft hair of \secref{sec:memory} is thus exact as an operator statement, an $\mathcal{O}\left(q^2\right)$ effect against the zero mode’s own fluctuations for a single charge, and for a macroscopic flux a memory resolvable against the random walk for the Killing time \eqref{eqn:memoryCoherenceTime}. This is a crossover of the signal-to-noise ratio of $N^{(T)}_{\ell m}$; the bath does not remove the mean displacement, and whether it is recoverable from the full radiation field is not addressed here.

\subsection{Comparison with null infinity}\label{sec:irsummary}

There are two distinct comparisons involved. The first is between bases, and it holds at either boundary. In the plane-wave basis, with covariantly normalised oscillators, the soft factor has a pole and the memory is its residue. In the partial-wave basis the soft factor is a number per leg, the divergence is entirely the norm of the external zero mode, and the memory is the zero mode itself. In both, the photon number is logarithmically divergent, by the counting of \secref{sec:ircounting}, and the Faddeev--Kulish dressing is forced. The second comparison is between the boundaries. On the horizon the number per leg is a sign set by the radial branch of the leg. The dressing takes the explicit form \eqref{eqn:dressingOperator}, a Wilson line in the boundary value of the pure-gauge mode, and on a dressed state the zero mode of the radiative field is equal and opposite to the Coulomb memory \eqref{eqn:coulombMemory} of every crossing charge. In its infrared behaviour the horizon differs from null infinity in three further respects. It has an invariant infrared scale, $\sqrt{\ell\left(\ell+1\right)}/R$, which sets the resolution of the radiative count at a stretched horizon and makes the radiative count \eqref{eqn:softProbabilityResult} finite and independent of the stretching, whereas at $\mathcal{I}$ the corresponding scale is zero. Its hard charges are effectively massless, so that the collinear sector is present and, summed over partial waves, converts the soft logarithm into the double logarithm \eqref{eqn:doubleLog}. And it comes with a temperature, so that the zero mode in which the memory lives is thermally driven, with the consequences \eqref{eqn:thermalFidelity} and \eqref{eqn:memoryCoherenceTime}. What it does not do is regulate the infrared divergence of the $S$-matrix. The degeneracy of the horizon vacua and the divergence of the undressed $S$-matrix are the same fact here as at $\mathcal{I}$, and the results of \secref{sec:memory} are results about the former.

\section{Conclusions and outlook}\label{sec:discussion}

We have computed a soft photon theorem near the horizon of a Schwarzschild black hole from the near-horizon dynamics of scalar QED, constructed the covariant phase space of the Maxwell and matter sectors on a stretched horizon, shown that the Noether flux of the residual gauge transformation through it is not conserved by the scattering whereas the soft charge, the flux with its matter part removed, is, and shown that the soft theorem is the Ward identity of the soft charge, which is the horizon memory. In the partial-wave basis the soft factor carries no power of the frequency, and the infrared divergence remains. Both the divergence and the memory are properties of the corner of the phase space, where the memory is the zero mode conjugate to the vacuum label. This is a statement about the basis, and it holds at null infinity too. The main difference from null infinity is that on a stretched horizon both radial branches are independent boundary data. The hard charge is a flux, and with both branches present it weights each leg by the direction of its radial motion. The soft factor is then a sign per leg. The scattering does not conserve this flux, and the field it radiates carries the change: its zero mode enters only the magnetic observable, where it cancels the change in the Coulomb memory of the charges. A charge that merely crosses the horizon leaves the Coulomb memory of its own shock wave on both boundary observables. The horizon also carries a temperature: in the Hartle--Hawking state the memory performs a random walk driven by the bath, the memory of a single charge is small compared with the fluctuations of the zero mode, and that of a macroscopic flux remains resolvable against the random walk of the zero mode for a Killing time of order $R\left|\Delta\mathcal{Q}\right|^2/\ell\left(\ell+1\right)$. 

Our analysis here is on a fixed background, in the linearised Maxwell sector and at leading soft order. The in- and out-states are read before the outgoing branch has been reflected by the potential. On the half of the phase space where photon and hard leg move on the same branch, the momentum-space computation is controlled only in the regime \eqref{eqn:frozenMassRegime}; outside it the soft factor is fixed, at leading order, by the field equation for the radial electric flux, and a complete derivation from the classical field equations in the full potential is left to future work. As at null infinity, the invariance of the $S$-matrix under the large gauge transformation is established through the soft theorem; deriving it from the phase space of both sectors with the corner pair included is left open. The radiative data is taken to settle at the two ends of each generator so that the memory is finite; the late-time tails of Price's law \cite{Price:1971fb, Price:1972pw} are consistent with this\footnote{The late-time tails of Price's law \cite{Price:1972pw}, under which a multipole $\ell \geq 1$ approaches its final value along the horizon at least as fast as $v^{-(2\ell+2)}$, make $\partial_v\Psi_{\ell m}$ integrable along the generator, so the late endpoint value exists. Price's law does not say what that value is; \eqref{eqn:GaussLawMemory} does.}; whether the horizon has an analogue of the Newman--Penrose constants \cite{Newman:1968uj} is not known to us.

Several things follow naturally from here. The gravitational case replaces $A^-_{\ell m}$ and $\Psi_{\ell m}$ by the Regge--Wheeler and Zerilli--Moncrief variables \cite{Regge:1957td, Zerilli:1970se, Moncrief:1974am, Martel:2005ir}, with horizon supertranslations whose charges are known and a vertex that is already the Dray--'t Hooft shock wave \cite{Dray:1984ha, tHooft:1996rdg, Gaddam:2020mwe}. Two things do not carry over automatically: the gravitational vertex has two more powers of momentum than the electromagnetic one, so whether the frequency independence survives needs a computation, and the gravitational Ward identity should relate the soft graviton memory to the shift of 't Hooft's $S$-matrix, which would close the triangle in the sector where the information question is posed. The $\ell = 0$ sector, where the horizon charge is the charge enclosed by the horizon and there is no soft charge, is the electromagnetic counterpart of the $\ell = 0, 1$ sector of the horizon supertranslations, where the charges are the mass and momentum of the hole, there are no radiative gravitons, and the balance is ordinary conservation rather than a soft theorem. The corner pair found here, one canonical pair per radiative partial wave commuting with the radiative modes, is of the same kind as the pair adjoined in the crossed-product construction of the exterior algebra \cite{Witten:2021unn, Chandrasekaran:2022eqq, Chandrasekaran:2022cip, Kudler-Flam:2023qfl}, and whether it sits inside that construction, with the random walk of the memory as the electromagnetic shadow of the fluctuating boost energy, is a question that we have not answered in this paper. Finally, the static patch of de Sitter has a horizon of the same local form, with the observer inside it and the crossing branch carrying the modes that leave the patch, and the same construction there would be the infrared triangle of the cosmological horizon.

\section*{Acknowledgements}

We would like to thank Alok Laddha, Ankit Aggarwal, and Prahar Mitra for various helpful conversations related to this work. Nava would also like to thank Ankit Aggarwal for collaboration on related work that significantly informed these results. We are also grateful to Ana-Maria Raclariu for many helpful comments on a draft of this paper. 

Fabiano gratefully acknowledges the hospitality of the Center for Fundamental Physics at ShanghaiTech University, where part of this work was carried out. He also acknowledges support from the Istituto Nazionale di Fisica Nucleare (INFN) through the Commissione Scientifica Nazionale 4 (CSN4) Iniziativa Specifica ``Quantum Universe'' (QGSKY).

Nava was supported by project RTI4001 of the Department of Atomic Energy, Govt. of India during the initial stages of this work. 

Nava and Nico are currently supported by the Delta-Institute for Theoretical Physics (D-ITP) that is funded by the Dutch Ministry of Education, Culture and Science (OCW).

\appendix

\section{Covariant phase space of the Maxwell sector}\label{app:MaxwellPhaseSpace}

In this appendix, we derive the phase space structure of the Maxwell sector on the horizon using the covariant phase space formalism \cite{Lee:1990nz, Iyer:1994ys, Wald:1999wa, Barnich:2001jy}. Since the horizon is a boundary of the region under consideration, the boundary and corner contributions to the symplectic potential require care; we follow the treatment of \cite{Harlow:2019yfa}.

Since $F_{ab} = - \sum_{\ell m} \epsilon_{ab} \Box_\eta \Psi_{\ell m} Y_{\ell m}$, the reduced action written purely in terms of the Moncrief variable $\Psi_{\ell m}$ is quadratic in $\Box_\eta \Psi_{\ell m}$ and hence higher-derivative; it is therefore not the most convenient starting point for identifying the canonical variables. Instead, we begin directly from the covariant presymplectic potential of Maxwell theory and only afterwards express the result in terms of partial waves and radiative data.

From the gauge transformation in \eqref{eqn:gaugetransfo}, we note that the odd mode $A^-_{\ell m}$ is gauge invariant, while the even mode $A^+_{\ell m}$ carries the gauge degree of freedom. In what follows, we shall also use the orthogonality relations of the eigenvectors of the two-sphere Laplacian
\begin{align}
    \int \mathrm{d}\Omega \, \gamma^{AB}\eta_A^{\ell m}\eta_B^{\ell' m'} ~ &= ~ \ell(\ell+1)\delta_{\ell\ell'}\delta_{mm'} \, , \\
    \int \mathrm{d}\Omega \, \gamma^{AB}\bar{\eta}_A^{\ell m}\bar{\eta}_B^{\ell' m'} ~ &= ~ \ell(\ell+1)\delta_{\ell\ell'}\delta_{mm'} \, , \\
    \int \mathrm{d}\Omega \, \gamma^{AB}\eta_A^{\ell m}\bar{\eta}_B^{\ell' m'} ~ &= ~ 0 \, .
\end{align}
In particular, we see that the $\ell=0$ mode drops out automatically from the vector sector. We will also make use of the solutions to the equations of motion written in terms of the gauge-invariant modes as in \eqref{eqn:AaSoln} and \eqref{eqn:chiaSoln}, the latter expressed through the Moncrief variable $\Psi_{\ell m}$ and the source-determined scalar $\rho_{\ell m}$. That decomposition is one of a two-dimensional one-form into its curl-free and divergence-free parts, and the appearance of $\partial_a \rho_{\ell m}$ there should not be confused with the original Maxwell gauge symmetry: $\chi^{\ell m}_a$, and hence $\rho_{\ell m}$, is gauge invariant. 

\subsection{Covariant presymplectic potential}\label{app:MaxwellPresymplectic}

Varying the Maxwell action in \eqref{eqn:MaxwellAction} yields the pre-symplectic potential current
\begin{equation}
    \Theta^\mu_{\text{Max.}} ~ = ~ - F^{\mu\nu}\delta A_\nu ~ = ~ F^{\nu\mu} \delta A_\nu \, .
\end{equation}
The presymplectic potential on a hypersurface $\Sigma$ is therefore obtained by integrating this current on $\Sigma$. For the null hypersurface of interest, it is convenient to write the result in coordinates adapted to a constant-$r^*$ surface just outside the horizon which corresponds to a surface $r > R$. This would then be a time-like hypersurface which approaches the null horizon of interest in the limit $r^* \rightarrow - \infty$. On this time-like surface, the induced metric is
\begin{align}
    \mathrm{d}s^2_{\Sigma} ~ = ~ - f\left(r\right) \mathrm{d}w^2 + r^2 \mathrm{d}\Omega^2 \, .
\end{align}
The vector normal to this constant $r^*$ surface is outward pointing if $n^\mu \partial_\mu r^* > 0$. Demanding this along with unit-normalisation uniquely determines the vector to be
\begin{align}
    n_\mu ~ = ~ \sqrt{f\left(r\right)} \, \delta^{r^*}_\mu \, .
\end{align}
The induced measure is then given by
\begin{align}
    \mathrm{d}\Sigma_\mu ~ = ~ n_\mu \sqrt{f\left(r\right)} \, r^2 \, \mathrm{d}w \, \mathrm{d}\Omega ~ = ~ f\left(r\right) \delta^{r^*}_\mu \, r^2 \, \mathrm{d}w \, \mathrm{d}\Omega \, .
\end{align}
In the limit where $r^* \rightarrow - \infty$, the induced metric becomes degenerate along the null direction $w$. Therefore, the limiting surface is the horizon. It is meaningful to first compute the symplectic two-form on $\Sigma$ and then take the limit. Hence the presymplectic potential on the time-like surface is
\begin{equation}
    \Theta_{\Sigma}^{\text{Max.}} ~ = ~ - \int_\Sigma \mathrm{d}\Sigma_\mu \, F^{\mu\nu}\delta A_\nu ~ = ~ - \int_\Sigma \mathrm{d}w \, \mathrm{d}\Omega \, r^2 f(r)\, F^{r^* \nu}\delta A_\nu \, .
\end{equation}
Using the inverse metric
\begin{equation}
    g^{ww} ~ = ~ 0 \, , \qquad g^{wr^*} ~ = ~ \frac{\eta^{wr^*}}{f(r)} \, , \qquad g^{r^*r^*} ~ = ~ \frac{1}{f(r)} \, , \qquad g^{AB} ~ = ~ \frac{\gamma^{AB}}{r^2} \, ,
\end{equation}
we find
\begin{align}\label{eqn:MaxwellThetaExact}
    \Theta_{\Sigma}^{\text{Max.}} ~ &= ~ - \eta^{w r^*} \int_\Sigma \mathrm{d}w \, \mathrm{d}\Omega \, \left(\dfrac{r^2}{f\left(r\right)} \eta^{w r^*} F_{w r^*} \delta A_w + \gamma^{AB} \left(F_{w B} + \eta^{w r^*} F_{r^* B}\right) \delta A_A\right) \nonumber \\
    &= ~ \sum_{\ell m} \ell \left(\ell+1\right) \int \mathrm{d}w \, \Bigg[\left(\partial_w \Psi_{\ell m} + \partial^{r^*}\!\rho_{\ell m}\right) \, \delta A^+_{\ell m} - \partial^{r^*}\! A^-_{\ell m} \, \delta A^-_{\ell m}\Bigg]
\end{align}
where in the second equality, we plugged in the field strength tensor in terms of the radiative modes \eqref{eqn:FinTermsofChi_a} and integrated over the sphere. Moreover, we imposed the gauge condition $A_w = 0$. In what follows, we will restrict our attention to the source-free contributions that correspond to the pure Maxwell phase space. Therefore, we will set $\rho_{\ell m} = 0$.

It is instructive to integrate the first term by parts in the null coordinate. In the gauge $A_w = 0$, equation \eqref{eqn:AaSoln} gives $\partial_w A^+_{\ell m} = - \chi^{\ell m}_w = - \partial^{r^*}\!\Psi_{\ell m}$. Discarding a total variation, which does not affect the symplectic two-form, we obtain
\begin{align}\label{eqn:MaxwellThetaPolarised}
    \Theta_{\Sigma}^{\text{Max.}} &= \sum_{\ell m} \ell\left(\ell+1\right) \left\{\Big[\Psi_{\ell m} \, \delta A^+_{\ell m}\Big]^{w = +\infty}_{w = -\infty} - \int \mathrm{d}w \, \left(\partial^{r^*}\!\Psi_{\ell m} \, \delta \Psi_{\ell m} + \partial^{r^*}\! A^-_{\ell m} \, \delta A^-_{\ell m}\right)\right\} \, ,
\end{align}
with the orientation \eqref{eqn:orientation}; the symplectic potential of a surface carries the orientation of the surface, and for a timelike surface parametrised by $w\in\mathbb R$ the induced measure $f\,r^2\,\mathrm dw\,\mathrm d\Omega$ is positive with $w$ increasing on both horizons. 

The bulk term is manifestly symmetric under the exchange $\Psi_{\ell m} \leftrightarrow A^-_{\ell m}$ of the two master variables; indeed, once their relative normalisation is fixed it is invariant under the full $SO(2)$ duality rotation acting on the pair $\left(\Psi_{\ell m}, A^-_{\ell m}\right)$. The entire asymmetry between the electric and magnetic sectors on the horizon therefore resides in the corner term at the two ends of the horizon, which pairs the Moncrief variable against the boundary value of the pure-gauge mode $A^+_{\ell m}$ (isolated later in \eqref{eqn:AplusBranches}). The odd sector admits no such term, since it carries no gauge degree of freedom.

The corner term is precisely the soft charge. Evaluating it on a residual gauge transformation \eqref{lambda}, for which $\delta_\lambda A^+_{\ell m} = \lambda_{\ell m}$ with $\lambda_{\ell m}$ independent of the null coordinate, gives
\begin{align}
    \ell\left(\ell+1\right)\Big[\Psi_{\ell m} \, \delta_\lambda A^+_{\ell m}\Big]^{w = +\infty}_{w = -\infty} ~ = ~ \ell\left(\ell+1\right) \lambda_{\ell m} \, \Delta \Psi_{\ell m} ~ = ~ - \, Q_{\text{soft}}\left[\lambda_{\ell m}\right] \, ,
\end{align}
by \eqref{eqn:softCharge}. The soft charge is thus not an additional structure imposed on the horizon phase space, but the unique boundary pairing that the duality-symmetric bulk symplectic form fails to account for.

\subsection{Symplectic form and canonical brackets}\label{app:symplecticform}
 
We take the exterior variation of the polarised form \eqref{eqn:MaxwellThetaPolarised} where the bulk and corner contributions have already been separated: the corner carries the boundary pairing identified in \appref{app:MaxwellPresymplectic}, while the bulk carries the radiative data. Setting $\rho_{\ell m}=0$ as before,

\begin{align}\label{eqn:MaxwellSymplecticModes}
    \Omega^{\text{Max.}}_\Sigma ~ &= ~ \delta \Theta^{\text{Max.}}_\Sigma \nonumber \\
    &= ~ \sum_{\ell m} \ell \left(\ell+1\right) \Bigg\{
    \Big[\delta \Psi_{\ell m} \wedge \delta A^+_{\ell m}\Big]^{w = +\infty}_{w = -\infty}
    \nonumber \\
    &\hspace{2.4cm} - \int \mathrm{d}w \, \Big(
    \delta \, \partial^{r^*}\!\Psi_{\ell m} \wedge \delta \Psi_{\ell m}
    \; + \; \delta \, \partial^{r^*}\! A^-_{\ell m} \wedge \delta A^-_{\ell m}
    \Big) \Bigg\} \, ,
\end{align}%
where no horizon limit has yet been taken and the orientation is that of \eqref{eqn:orientation}. As in \appref{app:MaxwellPresymplectic} we work on $\Sigma$ and approach the horizon afterwards; the limit is taken in the controlled sense of \eqref{eqn:boundaryModeCondition} and never by imposing a condition at $f\left(r\right)=0$.
 
The normal derivative in the bulk term is where the horizon limit enters. Using $w = t + \eta^{wr^*}r^*$, the operator $\partial^{r^*} = \eta^{wr^*}\partial_w + \partial_{r^*}$ is the derivative normal to the constant-$r^*$ surface, and on the mode functions of \eqref{eqn:PsiQuantisation} it acts as
\begin{align}\label{eqn:drstarOnBranches}
    \partial^{r^*} e^{-i\omega w + i k_{(\sigma)}r^*} ~ = ~ i \, \sigma \tilde{\omega} \,
    e^{-i\omega w + i k_{(\sigma)}r^*} ~ = ~ - \, \sigma \, \dfrac{\tilde{\omega}}{\omega} \,
    \partial_w \, e^{-i\omega w + i k_{(\sigma)}r^*} \, ,
\end{align}
so that in the near-horizon limit $\tilde\omega\to\omega$ the correct replacement is branch resolved,
\begin{align}\label{eqn:drstarLimit}
    \partial^{r^*}\Psi_{\ell m} ~ \longrightarrow ~ - \sum_{\sigma = \pm} \sigma \,
    \partial_w \Psi^{(\sigma)}_{\ell m} \, ,
\end{align}
where $\Psi^{(\sigma)}_{\ell m}$ is the branch-$\sigma$ part of \eqref{eqn:PsiQuantisation}, and identically for $A^-_{\ell m}$. Only on the crossing branch, $\eta^{wr^*}\sigma=-1$, does this reduce to $\eta^{wr^*}\partial_w$; on the branch along the horizon, $\eta^{wr^*}\sigma=+1$, it gives $-\eta^{wr^*}\partial_w$. The replacement $\partial^{r^*}\to\eta^{wr^*}\partial_w$ is therefore the projector onto the crossing branch, and a regularity condition at the horizon is what performs the projection: since $r^*\simeq R\ln f\left(r\right)$ near the horizon, the phase of the branch along the horizon, by \eqref{eqn:ka}, is
\begin{align}\label{eqn:alongPhase}
    e^{i k_{(\sigma)}r^*}\Big|_{\eta^{wr^*}\sigma = +1} ~ \simeq ~
    e^{2 i \eta^{wr^*}\omega r^*} ~ \sim ~ f\left(r\right)^{2 i \eta^{wr^*}\omega R} \, ,
\end{align}
which is bounded and normalisable but not smooth in $r$ at $f\left(r\right)=0$: it oscillates rather than diverges. Discarding it would be inconsistent with the treatment of the matter sector, where both branches are retained throughout: the branch label $\sigma_n$ weighting each hard leg in \eqref{eqn:hardChargeResult} is the same distinction, and the same condition imposed on $\phi_{\ell m}$ would set $\sigma_n = -\eta^{wr^*}$ on every leg and collapse \eqref{eqn:finalAmp} onto \eqref{eqn:A+amplitude}. Applying \eqref{eqn:drstarLimit}, the bulk term of
\eqref{eqn:MaxwellSymplecticModes} becomes
\begin{align}\label{eqn:BulkSymplecticBranches}
    \Omega^{\text{bulk}}_\Sigma ~ = ~ \sum_{\ell m}\ell\left(\ell+1\right)
    \int \mathrm{d}w \sum_{\sigma , \sigma' = \pm} \sigma \Big(
    \delta \partial_w \Psi^{(\sigma)}_{\ell m} \wedge \delta \Psi^{(\sigma')}_{\ell m}
    + \delta \partial_w A^{-(\sigma)}_{\ell m} \wedge \delta A^{-(\sigma')}_{\ell m}
    \Big) \, .
\end{align}
Here, $A^{-}_{\ell m}$ obeys \eqref{eqn:oddEoM}, the same equation as $\Psi_{\ell m}$, hence the use of the same branch decomposition. The duality statement below \eqref{eqn:MaxwellThetaPolarised} is made before any limit and is untouched. 
 
\paragraph{The boundary coordinate.} In the gauge $A_w=0$ used in \eqref{eqn:MaxwellThetaExact}, equation \eqref{eqn:AaSoln} gives $\partial_w A^+_{\ell m} = -\,\partial^{r^*}\!\Psi_{\ell m}$, so that by \eqref{eqn:drstarLimit}
\begin{align}\label{eqn:AplusBranches}
    A^{+}_{\ell m}\left(w , r^*\right) ~ = ~ \sum_{\sigma=\pm} \sigma \,
    \Psi^{(\sigma)}_{\ell m}\left(w , r^*\right) \; + \; \varphi_{\ell m}\left(r^*\right) \, .
\end{align}
The radiative datum $\Psi_{\ell m}=\sum_\sigma\Psi^{(\sigma)}_{\ell m}$ is the branch-symmetric combination of the two branches, and the part of $A^+_{\ell m}$ that varies along the generators is the branch-antisymmetric one. The remaining piece $\varphi_{\ell m}$ is the $w$-independent integration constant. It is not determined by $\Psi_{\ell m}$, and it is the boundary coordinate of the horizon phase space --- the analogue of the boundary Goldstone mode at null infinity. Since $A^{+}_{\ell m}$ exists only for $\ell \geq 1$, so does $\varphi_{\ell m}$. There is no boundary coordinate and no vacuum label at $\ell = 0$. Its leading near-horizon term $\varphi^{(0)}_{\ell m}$ is the one that survives the limit, in the same expansion as \eqref{eqn:AplusNearHor}.
 
The gauge parameter $\lambda_{\ell m}$ of \eqref{lambda} is a fixed, field-independent label: it selects which residual transformation is being performed, it is the argument of $Q_{\text{soft}}\left[\lambda_{\ell m}\right]$, and it satisfies $\delta\lambda_{\ell m}=0$ on phase space. The boundary mode $\varphi_{\ell m}$ of \eqref{eqn:AplusBranches} is a dynamical coordinate on that phase space, with $\delta\varphi_{\ell m}\neq 0$. They are related by the statement that the parameter shifts the coordinate: since $\delta_\lambda\Psi_{\ell m}=0$ and $\delta_\lambda A^+_{\ell m}=\lambda_{\ell m}$ by \eqref{eqn:gaugetransfo}, \eqref{eqn:AplusBranches} gives
\begin{align}\label{eqn:largegaugetrafo}
    \delta_\lambda \varphi_{\ell m} ~ = ~ \lambda_{\ell m} \qquad \text{and} \qquad
    \delta_\lambda \Psi_{\ell m} ~ = ~ 0 \, .
\end{align}
Distinct values of $\varphi^{(0)}_{\ell m}$ therefore label the degenerate horizon vacua of \secref{sec:memory}, and $\lambda^{(0)}_{\ell m}$ is the finite transformation carrying one into another. A small transformation, $\lambda^{(0)}_{\ell m}=0$, leaves $\varphi^{(0)}_{\ell m}$ untouched and is gauge.
 
\paragraph{The corner term is the zero-mode pairing.} Substituting \eqref{eqn:AplusBranches} into the corner term of \eqref{eqn:MaxwellSymplecticModes} and using \eqref{eqn:orientation}, the $w$-independence of $\varphi_{\ell m}$ gives
\begin{align}\label{eqn:CornerSplit}
    \Big[\delta\Psi_{\ell m}\wedge\delta A^+_{\ell m}\Big]^{w=+\infty}_{w=-\infty} ~ = ~
    \Big[\delta\Psi_{\ell m}\wedge \delta\!\!\sum_{\sigma = \pm}\!\sigma
    \Psi^{(\sigma)}_{\ell m}\Big]^{w=+\infty}_{w=-\infty}
    \; + \; \delta\left(\Delta\Psi_{\ell m}\right)\wedge\delta\varphi_{\ell m} \, .
\end{align}
The first term involves only radiative data at the two ends of the horizon. The second pairs the memory $\Delta\Psi_{\ell m}$ of \eqref{eqn:memoryOperator} against the boundary coordinate $\varphi_{\ell m}$: one conjugate pair for each radiative partial wave, rather than a pair at every $w$,
\begin{align}\label{appeqn:MemoryGoldstoneBracket}
    \left[\Delta\Psi_{\ell' m'} \, , \, \varphi_{\ell m}\right] ~ = ~
    \dfrac{i}{\ell\left(\ell+1\right)} \, \delta_{\ell\ell'}\delta_{mm'} \, .
\end{align}
Note that $\Delta\Psi_{\ell m}$ is branch symmetric, receiving a contribution from each $\sigma$, whereas the radiative corner term in \eqref{eqn:CornerSplit} is branch antisymmetric. This is the phase-space origin of the branch sum in \eqref{eqn:softChargeOnHorizons}, and of the absence of a $\sigma$ weight, in contrast to the hard charge \eqref{eqn:hardChargeResult}, which is the flux across $\Sigma$ and is weighted by $\sigma$.
 
\paragraph{What the bulk term does and does not determine.}
$\Omega^{\text{bulk}}_\Sigma$ in \eqref{eqn:BulkSymplecticBranches} is evaluated on a surface of constant $r^*$ with the derivative normal to that surface. It is therefore the symplectic \emph{flux} across $\Sigma$ and not the symplectic form on a Cauchy surface; the two differ by the flux through the boundary of any region they bound. As we establish in \appref{app:MaxwellinPWs}, this is the same distinction drawn for the hard modes in the paragraph following \eqref{eqn:scalarAlgebra}, and it has the same consequence: the weight $\sigma$ in \eqref{eqn:BulkSymplecticBranches} makes the form indefinite on the diagonal and non-diagonal off it, and the mode algebra may not be read off from it. That algebra is fixed in \appref{app:radiativeModeQuantisation} on a Cauchy surface, with the result \eqref{eqn:PsiAlgebra}. The corner term carries no normal derivative and no branch weight. The bracket \eqref{appeqn:MemoryGoldstoneBracket} is read off from it on the phase space in which $\varphi_{\ell m}$ is an independent boundary coordinate, the extension of the radiative phase space by the corner pair in the sense of \cite{Harlow:2019yfa}. That the pair is independent of the bulk data is the content of \eqref{eqn:AplusBranches}, and it is an input: the bulk algebra on a Cauchy surface does not contain $\varphi_{\ell m}$, and we do not derive the extension from it.
 
\paragraph{The phase space.} The phase space on which the charges act is that of the exterior region. Its in-data are the characteristic data on $\mathcal{H}^- \cup \mathcal{I}^-$ and its out-data those on $\mathcal{H}^+ \cup \mathcal{I}^+$. The symplectic current is conserved, so the symplectic form on a surface of constant Schwarzschild time, on which the brackets \eqref{eqn:scalarAlgebra} and \eqref{eqn:PsiAlgebra} are read off in \appref{app:MaxwellinPWs}, is the symplectic form on either characteristic surface up to the contributions of their ends, and the branch along the stretched horizon is the near-horizon part of the data on $\mathcal{I}^\mp$, transmitted through the barrier, with the bracket that current assigns it. The form $\Omega_\Sigma$ of \eqref{eqn:MaxwellSymplecticModes} at finite $r^*$ is the flux of the same current through the timelike surface $\Sigma$. It is used for its corner term, which in the horizon limit sits at the two ends of $\mathcal{H}$ and supplies the pair \eqref{appeqn:MemoryGoldstoneBracket} that the bulk algebra on either surface does not contain. The generator statement $\iota_{X_\lambda}\Omega_\Sigma = \delta Q$ is verified for both sectors in \appref{app:softGenerates} and \appref{app:hardGenerates}. The same residual transformation, acting on the matter as $\delta_\lambda\tilde{\phi} = iq\lambda\tilde{\phi}$ with one phase on both branches, gives $Q_{\text{cut}} = Q_{\mathrm{soft}} + Q_{\mathrm{hard}}$ when contracted into the flux form through $\Sigma$, and the charge on a surface of constant time weighted by $\lambda$, the charge of the pure-gauge statement \eqref{eqn:SmatrixPhase}, when contracted into \eqref{eqn:scalarSymplectic} on that surface. The hard charge \eqref{eqn:hardCharge} is the matter part of the former: the current flux through $\Sigma$ integrated over all $w$, by conservation of the current the net charge carried across $\Sigma$ counted with the sign of the direction of crossing, and on the states of \secref{sec:externalstates} the flux-weighted sum \eqref{eqn:hardChargeResult}. It is this operator, and not the charge on a slice, that the Maxwell equation along $\Sigma$ pairs with $\Delta\Psi_{\ell m}$ in \eqref{eqn:softCharge}--\eqref{eqn:hardCharge}, and the identity of \secref{sec:wardidentity} is a statement about it.
 
\paragraph{The radiative bracket.} For completeness we record what \eqref{eqn:PsiAlgebra} implies along the horizon. Inserting the branch-$\sigma$ part of \eqref{eqn:PsiQuantisation}, in the limit \eqref{eqn:boundaryModeCondition}, into \eqref{eqn:PsiAlgebra} gives
\begin{align}\label{eqn:PsiChiralBracket}
    \left[\partial_w \Psi^{(\sigma)}_{\ell m}\left(w\right) \, , \,
    \Psi^{(\sigma')}_{\ell'm'}\left(w'\right)\right] ~ = ~
    - \, \dfrac{i}{2 \, \ell\left(\ell+1\right)} \,
    \delta_{\sigma\sigma'}\delta_{\ell\ell'}\delta_{mm'} \, \delta\left(w-w'\right) \, ,
\end{align}
diagonal in the branch label, with the factor of one half characteristic of a chiral boson on a null line. Restricted to the crossing branch, \eqref{eqn:BulkSymplecticBranches} reduces to the standard chiral form and its inverse is \eqref{eqn:PsiChiralBracket}; it is only with both branches retained that the flux form ceases to be invertible as a Fock structure. And by \eqref{eqn:CornerSplit} both branches contribute to $\Delta\Psi_{\ell m}$.
 
We conclude that in the even-parity sector the gauge-invariant radiative datum along the horizon is $\partial_w\Psi_{\ell m}$, quantised by \eqref{eqn:PsiAlgebra}, and that the boundary coordinate $\varphi_{\ell m}$ --- by \eqref{eqn:AplusBranches} the only part of $A^+_{\ell m}$ not already carried by $\Psi_{\ell m}$ --- is conjugate to the memory $\Delta\Psi_{\ell m}$. By contrast the odd-parity mode $A^-_{\ell m}$ is already (large) gauge invariant, admits no corner term, and furnishes a radiative variable directly; its algebra follows on a Cauchy surface in the same way and, by the duality of the bulk term of \eqref{eqn:MaxwellThetaPolarised}, coincides with that of $\Psi_{\ell m}$. Nothing below depends on it, since the odd sector decouples from the matter near the horizon by \appref{app:magneticsofttheorem}.

\subsection{Soft charge generates large gauge transformations}\label{app:softGenerates}
 
We now verify that $Q_{\text{soft}}$ is the Hamiltonian generator of \eqref{eqn:largegaugetrafo}. Let $X_\lambda$ be the phase-space vector field implementing the residual transformation, so that by \eqref{eqn:largegaugetrafo}
\begin{equation}
    X_\lambda\left(\varphi_{\ell m}\right) ~ = ~ \lambda_{\ell m} \, ,
    \qquad
    X_\lambda\left(\Psi_{\ell m}\right) ~ = ~ 0 \, ,
    \qquad
    \delta\lambda_{\ell m} ~ = ~ 0 \, .
\end{equation}
The generator is defined by contracting $X_\lambda$ into the symplectic form,
\begin{equation}\label{eqn:pullbacktwoform}
    \iota_{X_\lambda}\Omega^{\text{Max.}}_\Sigma ~ = ~ \delta Q_{\text{soft}}\left[\lambda\right] \, ,
    \qquad \text{with} \qquad
    \iota_{X_\lambda}\Omega^{\text{Max.}}_\Sigma ~ \coloneqq ~
    \Omega^{\text{Max.}}_\Sigma\left(X_\lambda \, , \, \delta\right) \, ,
\end{equation}
the first slot being filled by the gauge transformation and the second by an arbitrary variation on phase space. Every term of \eqref{eqn:MaxwellSymplecticModes} except one is annihilated: the bulk term \eqref{eqn:BulkSymplecticBranches} and the radiative corner term in \eqref{eqn:CornerSplit} carry $\delta\Psi_{\ell m}$ or $\delta A^-_{\ell m}$ in both slots, and $X_\lambda$ annihilates both. Only the zero-mode pairing survives,
\begin{align}
    \iota_{X_\lambda}\Omega^{\text{Max.}}_\Sigma ~ &= ~
    \sum_{\ell m}\ell\left(\ell+1\right) \Big[
    \big(\iota_{X_\lambda}\delta\,\Delta\Psi_{\ell m}\big)\,\delta\varphi_{\ell m}
    - \delta\!\left(\Delta\Psi_{\ell m}\right)
    \big(\iota_{X_\lambda}\delta\varphi_{\ell m}\big) \Big] \nonumber \\
    &= ~ - \sum_{\ell m}\ell\left(\ell+1\right)\lambda_{\ell m} \,
    \delta\!\left(\Delta\Psi_{\ell m}\right) ~ = ~
    \delta\left[- \sum_{\ell m}\ell\left(\ell+1\right)\lambda_{\ell m}\Delta\Psi_{\ell m}\right] ,
\end{align}
where the gauge parameter was pulled through the variation using $\delta\lambda_{\ell m}=0$. Hence
\begin{equation}
    Q_{\text{soft}}\left[\lambda\right] ~ = ~ - \sum_{\ell m}\ell\left(\ell+1\right)
    \lambda_{\ell m}\,\Delta\Psi_{\ell m} \, ,
\end{equation}
up to an additive constant on phase space, in agreement with \eqref{eqn:softCharge}. Conversely, using \eqref{appeqn:MemoryGoldstoneBracket},
\begin{align}
    \left\{\varphi_{\ell'm'} \, , \, Q_{\text{soft}}\left[\lambda\right]\right\}
    ~ = ~ \sum_{\ell m}\ell\left(\ell+1\right)\lambda_{\ell m}
    \left\{\Delta\Psi_{\ell m} \, , \, \varphi_{\ell'm'}\right\}
    ~ = ~ \lambda_{\ell'm'} \, ,
\end{align}
while $\left\{\Delta\Psi_{\ell m},Q_{\text{soft}}\left[\lambda\right]\right\}=0$ since $Q_{\text{soft}}$ depends on the radiative data only through $\Delta\Psi_{\ell m}$ itself. This reproduces \eqref{eqn:largegaugetrafo}. Although $A^+_{\ell m}$ is pure gauge in the bulk, its $w$-independent boundary value $\varphi_{\ell m}$ is a genuine coordinate on the horizon phase space, canonically conjugate to the gauge-invariant memory $\Delta\Psi_{\ell m}$, and the soft charge is the generator that translates it.

\subsection{Hard charge as a Hamiltonian generator}\label{app:hardGenerates}

The matter sector contributes the hard charge to the same contraction. Varying \eqref{eqn:MatterAction} gives the presymplectic potential current 
\begin{equation}
    \Theta^\mu_{\text{matter}} ~ = ~ - \Big[ \left(D^\mu\tilde{\phi}\right)\delta\tilde{\phi}^* + \left(D^\mu\tilde{\phi}\right)^{*}\delta\tilde{\phi} \Big] \, ,
\end{equation}
on which the residual transformation acts as $X_\lambda\big(\tilde{\phi}\big) = i q \lambda \tilde{\phi}$ and $X_\lambda\left(A_\mu\right) = \partial_\mu\lambda$, hence $X_\lambda\big(D_\mu\tilde{\phi}\big) = i q \lambda \, D_\mu\tilde{\phi}$. Contracting into $\omega^\mu_{\text{matter}} = \delta\Theta^\mu_{\text{matter}}$, we have that
\begin{equation}\label{eqn:matterContraction}
    \iota_{X_\lambda}\omega^\mu_{\text{matter}} ~ = ~ - \, i q \lambda \, \delta \Big[ \tilde{\phi}^* D^\mu\tilde{\phi} - \tilde{\phi} \left(D^\mu\tilde{\phi}\right)^* \Big] ~ = ~ - \, \delta \left( \lambda \, J^\mu \right) \, ,
\end{equation}
with $J^\mu$ the current below \eqref{eqn:MatterAction}. No derivative of $\lambda$ appears, because $\tilde{\phi}$ and $D_\mu\tilde{\phi}$ transform by the same phase, and the field equations have not been used. On $\Sigma$, with $\mathrm{d}\Sigma_\mu = f\left(r\right) r^2 \, \delta^{r^*}_\mu \, \mathrm{d}w \, \mathrm{d}\Omega$ as in \appref{app:MaxwellPresymplectic} and $f\left(r\right) J^{r^*} = \eta^{r^* b} J_b = \eta^{w r^*} J_w + J_{r^*} = {\epsilon_w}^{b} J_b$ by \eqref{eqn:currentEpsilonCombination},
\begin{equation}
    \iota_{X_\lambda}\Omega^{\text{matter}}_\Sigma ~ = ~ - \, \delta \sum_{\ell m} \lambda_{\ell m} \, r^2 \int \mathrm{d}w \, {\epsilon_w}^{b} J^{\ell m}_b \, ,
\end{equation}
which is the first term of \eqref{eqn:hardCharge}. The second term comes from the Maxwell sector once the sources are restored. With $\rho_{\ell m} \neq 0$ the potential \eqref{eqn:MaxwellThetaExact} carries the additional term $\ell\left(\ell+1\right) \int \mathrm{d}w \, \partial^{r^*}\!\rho_{\ell m} \, \delta A^+_{\ell m}$, and since $\rho_{\ell m}$ is gauge invariant while $X_\lambda\left(A^+_{\ell m}\right) = \lambda_{\ell m}$, its contraction is $- \, \ell\left(\ell+1\right) \lambda_{\ell m} \int \mathrm{d}w \, \delta \, \partial^{r^*}\!\rho_{\ell m}$. Altogether,
\begin{equation}\label{eqn:fullGenerator}
    \iota_{X_\lambda} \Big( \Omega^{\text{Max.}}_\Sigma + \Omega^{\text{matter}}_\Sigma \Big) ~ = ~ \delta \Big( Q_{\text{soft}}\left[\lambda\right] + Q_{\text{hard}}\left[\lambda\right] \Big) ~ = ~ \delta Q_{\text{cut}}\left[\lambda\right] \, ,
\end{equation}
term by term as in \eqref{eqn:softCharge} and \eqref{eqn:hardCharge}. The seagull part of $J^\mu$ has been dropped at the order of the single-photon identity as below \eqref{eqn:current}. The vector field is the one residual gauge transformation, rotating both branches of $\tilde{\phi}$ by the same phase; the branch weight of \eqref{eqn:hardChargeResult} comes from $\mathrm{d}\Sigma_\mu$, since $\eta^{r^* b}\partial_b$ has the eigenvalue $i \eta^{w r^*}\sigma\Omega_p$ of \eqref{eqn:modeEigen} on the mode functions. The same contraction on a Cauchy surface gives $\lambda$ times the charge on that surface, which by the constraint is the cut charge at its boundary, the charge of \eqref{eqn:SmatrixPhase}. On shell, $F^{\mu\nu}\partial_\nu\lambda - \lambda J^\mu = \nabla_\nu\left(\lambda F^{\mu\nu}\right)$ is minus the Noether current of \secref{sec:covariantCharge}. The charge $Q_{\text{cut}}$ is then the flux of the Noether current through $\Sigma$.

\section{Scalar QED near the horizon in partial waves}\label{app:MaxwellinPWs}

In this appendix, we will derive the Feynman rules and quantisations for the hard scalar modes and the soft radiative photon near the horizon in a partial wave basis.

\subsection{Scalar propagator near the horizon}\label{app:scalarPropagator}
We begin with the quadratic action for a complex scalar field $\tilde{\phi}$ minimally coupled to the photon:
\begin{align}
    S^{(2)}_{\text{Scalar}} ~ &= ~ - \int \mathrm{d}^4 x \sqrt{- g} \left(g^{\mu \nu} \nabla_\mu \tilde{\phi}^* \nabla_\nu \tilde{\phi} + m^2 \tilde{\phi}^* \tilde{\phi} \right) \nonumber \\
    &= ~ \phantom{-} \int \mathrm{d}^4 x \, \sqrt{-g} \, \left(\tilde{\phi}^* \dfrac{1}{\sqrt{-g}} \partial_\mu \left(\sqrt{- g} g^{\mu \nu} \partial_\nu \tilde{\phi}\right) - m^2 \tilde{\phi}^* \tilde{\phi} \right) \nonumber \\
    &= ~ \phantom{-} \int \mathrm{d}^2 x \, \mathrm{d} \Omega \, \tilde{\phi}^* \left[\partial_a \left(r^2 \, \eta^{a b} \partial_b \tilde{\phi} \right) + \dfrac{f\left(r\right) }{\sqrt{\gamma}} \partial_A \left(\sqrt{\gamma} \, \gamma^{A B} \partial_B \tilde{\phi} \right) - r^2 f\left(r\right) m^2 \tilde{\phi} \right] \nonumber \\
    &= ~ \sum_{\ell m} \int \mathrm{d}^2 x \, \tilde{\phi}^*_{\ell m} \left[ \partial_a \left(r^2 \eta^{a b} \partial_b \, \cdot \right) - f\left(r\right) \left( \ell \left(\ell + 1\right) + r^2 m^2 \right) \right] \tilde{\phi}_{\ell m}  \, ,
\end{align}
where in the last step, we expanded the field in partial waves and integrated the sphere out. Since $r$ depends on $r^*$ through $\partial_{r^*} r = f\left(r\right)$,
\begin{equation}\label{eqn:radialOperatorSplit}
    \partial_a \left(r^2 \eta^{ab}\partial_b \tilde{\phi}\right) ~ = ~ r^2 \, \Box_\eta \tilde{\phi} + 2 r f\left(r\right) \partial^{r^*}\!\tilde{\phi} \, ,
\end{equation}
and the first-derivative term is of the same order in $f\left(r\right)$ as the mass terms retained below. Rather than dropping it, we remove it exactly by the standard rescaling
\begin{equation}\label{eqn:scalarRescaling}
    \phi_{\ell m} ~ \eqqcolon ~ r \, \tilde{\phi}_{\ell m} \, .
\end{equation}
Substituting \eqref{eqn:scalarRescaling}, the two first-derivative terms cancel identically by the symmetry of $\eta^{ab}$:
\begin{align}
    \partial_a \left(r^2 \eta^{ab}\partial_b \dfrac{\phi_{\ell m}}{r}\right) ~ &= ~ \cancel{\left(\partial_a r\right) \eta^{ab}\partial_b \phi_{\ell m}} + r \, \Box_\eta \phi_{\ell m} - \cancel{\left(\partial_a \phi_{\ell m}\right) \eta^{ab}\partial_b r} - \phi_{\ell m} \, \Box_\eta r \nonumber \\
    &= ~ r \, \Box_\eta \phi_{\ell m} - \dfrac{f\left(r\right) R}{r^2} \, \phi_{\ell m} \, ,
\end{align}
where we used that $r$ depends only on $r^*$, so that $\Box_\eta r = \partial_{r^*}^2 r = f\left(r\right) f^{\prime}\left(r\right) = f\left(r\right) R/r^2$. No integration by parts is required and no boundary term is generated. The action therefore becomes, exactly,
\begin{align}\label{eqn:scalarActionExact}
    S^{(2)}_{\text{Scalar}} ~ = ~ \sum_{\ell m} \int \mathrm{d}^2 x \, \phi^*_{\ell m} \left[ \Box_\eta - f\left(r\right)\left(\dfrac{\ell\left(\ell+1\right)}{r^2} + m^2 + \dfrac{R}{r^3}\right) \right] \phi_{\ell m} \, .
\end{align}
This is the Regge--Wheeler form: the overall factor of $r^2$ has been absorbed into the field and every remaining term sits inside a single potential, so a near-horizon expansion may now be taken consistently order by order.

A natural step next may be to take a near-horizon limit which will drop the mass terms. However, the momentum-space rules of \appref{app:interactionTerms} need the mass terms on half of the phase space, so we retain them. Therefore, instead, we will first work to first order in the near-horizon limit $r = R + \varepsilon$ and then finally take the $\varepsilon \rightarrow 0$ limit in due course. Now, up to first order, we have
\begin{align}\label{eqn:scalarAction}
    S^{(2)}_{\text{Scalar}} ~ = ~ \sum_{\ell m} \int \mathrm{d}^2 x \, \phi^*_{\ell m} \Big( \Box_\eta - m^2_{\ell} \Big) \phi_{\ell m}
\end{align}
with
\begin{align}\label{eqn:scalarMass}
    m^2_{\ell} ~ \coloneqq ~ f\left(r\right) \left( \dfrac{ \ell \left(\ell + 1\right) + 1 }{R^2} + m^2 \right) \, ,
\end{align}
where $f\left(r\right) = \varepsilon/R + \mathcal{O}\left(\varepsilon^2\right)$. The curvature term $R/r^3$ in \eqref{eqn:scalarActionExact} contributes the $+1$, since $R/r^3 \rightarrow 1/R^2$ on the horizon; it is of the same order as the centrifugal term, and would have been lost had the first-derivative term in \eqref{eqn:radialOperatorSplit} simply been discarded. With the Fourier transform in the near-horizon region given by
\begin{equation}\label{eqn:scalarFourierModes}
\phi_{\ell m}\left(x\right) ~ = ~ \int \dfrac{\mathrm{d}^{2} p}{\left(2\pi\right)^{2}} \, \phi_{\ell m} \left(p\right) \exp\left(i x^{a} p_{a}\right) \, ,
\end{equation}
we read off the propagator to be
\begin{equation}\label{eqn:bhscalarprop}
    \mathcal{P}_{\phi_{\ell m}} ~ = ~ \dfrac{- i }{p^a p_a + m^2_{\ell} - i \epsilon} \, .
\end{equation}

\subsection{Quantisation of the hard modes}\label{app:hardModeQuantisation}
The on-shell hard scalar mode can be written as
\begin{align}
\phi_{\ell m}\left(x\right) ~ &= ~ \int \dfrac{\mathrm{d}^{2} p}{\left(2\pi\right)^{2}} \, \phi_{\ell m}\left(p\right) \, \exp\left(i x^{a} p_{a}\right) \nonumber \\
&= ~ \int \dfrac{\mathrm{d}^{2} p}{2\pi} \, c_{\ell m}\left(p\right) \, \delta\left(p^2 + m^2_{\ell_p}\right) \, \exp\left(i x^{a} p_{a}\right) \, ,
\end{align}
where the mass-shell condition is
\begin{equation}
    0 ~ = ~ \eta^{a b} p_a p_b + m^2_{\ell_p} ~ = ~ \left(p_{r^*}\right)^2 + 2 \eta^{w r^*} p_w p_{r^*} + m^2_{\ell_p} \, .
\end{equation}
We may impose this by parametrising the scalar momentum as
\begin{equation}\label{eqn:scalarMomentum}
    p_a ~ = ~ \left(- E_p , p_{r^*}\right) ~ = ~ \left(-E_p \, , \, \eta^{w r^*} E_p \pm \sqrt{E_p^2 - m^2_{\ell_p}} \right) \, .
\end{equation}
In what follows, the two signs will be labelled by $\sigma$. Then, using the identity
\begin{equation}
    \delta \left(f \left(\xi\right)\right) ~ = ~ \sum_{f \left(\xi_i\right) = 0} \dfrac{\delta \left(\xi - \xi_i\right)}{\left|f^{\prime} \left(\xi_i\right)\right|} \, ,
\end{equation}
the mode expansion becomes
\begin{align}
\phi_{\ell m}\left(x\right) ~ &= ~ \int \dfrac{\mathrm{d}^{2} p}{\left(2\pi\right) \left| 2 \sqrt{E_p^2 - m^2_{\ell_p}} \right|} \, c_{\ell m}\left(p\right) \, e^{-i w E_p + i r^* p_{r^*}} \nonumber \\
&\qquad \left[\delta\left(p_{r^*} - \eta^{w r^*} E_p - \sqrt{E_p^2 - m^2_{\ell_p}}\right) + \delta\left(p_{r^*} - \eta^{w r^*} E_p + \sqrt{E_p^2 - m^2_{\ell_p}}\right) \right] \nonumber \\
&= ~ \sum_{\sigma = \pm} \int_{\left|E_p\right| \geq m_{\ell_p}} \dfrac{\mathrm{d} E_p}{4\pi \Omega_p} c^{(\sigma)}_{\ell m}\left(E_p\right) e^{ - i w E_p + i r^* p^{(\sigma)}_{r^*}\left(E_p\right)} \, , 
\end{align}
where we defined the frequency $\Omega_p \coloneqq \sqrt{E_p^2 - m^2_{\ell_p}}$ and $p^{(\sigma)}_{r^*}\left(E_p\right) \coloneqq \left(\sigma \Omega_p + \eta^{w r^*} E_p\right)$. The additional label $\sigma$ is meant to split the two branches of the mass-shell condition into separate operator modes. 

We will now split the mode expansion into positive and negative frequencies as follows:
\begin{align}
\phi_{\ell m} ~ &= ~ \sum_{\sigma = \pm} \int_{m_{\ell_p}}^\infty \dfrac{\mathrm{d} E_p}{4\pi \Omega_p} \left(c^{(\sigma)}_{\ell m}\left(E_p\right) e^{ - i w E_p + i r^* p^{(\sigma)}_{r^*}\left(E_p\right)} + d^{(-\sigma)\dagger}_{\ell m}\left(E_p\right) e^{ i w E_p + i r^* p^{(\sigma)}_{r^*}\left(-E_p\right)}\right)  \nonumber \\
&\eqqcolon ~ \sum_{\sigma = \pm} \int_{m_{\ell_p}}^\infty \dfrac{\mathrm{d} E_p}{4\pi \Omega_p} \left(c^{(\sigma)}_{\ell m}\left(E_p\right) V^{(\sigma)}_{E_p} + d^{(-\sigma)\dagger}_{\ell m}\left(E_p\right) V^{(-\sigma)*}_{E_p} \right) \, ,
\end{align}
where we made the change of variables $E_p \to - E_p$ in the negative $E_p$ modes. This allowed us to identify $c^{(\sigma)}_{\ell m}\left(-E_p\right) = d^{(-\sigma)\dagger}_{\ell m}\left(E_p\right)$ as the creation operator and $c^{(\sigma)}_{\ell m}\left(E_p\right)$ as the annihilation operator associated with the complex hard scalar. This split is fixed by the identity 
\begin{equation}
    V^{(\sigma)}_{-E_p} ~ = ~ V^{(-\sigma)*}_{E_p} \, .
\end{equation}
Consequently, the conjugate mode is
\begin{align}
\phi^*_{\ell m} ~ &= ~ \sum_{\sigma = \pm} \int_{m_{\ell_q}}^\infty \dfrac{\mathrm{d} E_q}{4\pi \Omega_q} \left(c^{(\sigma)\dagger}_{\ell m}\left(E_q\right) V^{(\sigma)*}_{E_q} + d^{(-\sigma)}_{\ell m}\left(E_q\right) V^{(-\sigma)}_{E_q} \right) \, .
\end{align}
The normalisation of these operators is not an assumption; it is fixed by the symplectic structure. Varying \eqref{eqn:scalarAction} produces the presymplectic potential current
\begin{align}
    \Theta^a_{\text{scalar}} ~ = ~ - \left(\partial^a \phi_{\ell m} \, \delta \phi^*_{\ell m} + \partial^a \phi^*_{\ell m} \, \delta \phi_{\ell m}\right) \, ,
\end{align}
whose symplectic current $\delta \Theta^a_{\text{scalar}}$ is conserved on shell. The symplectic form
\begin{align}\label{eqn:scalarSymplectic}
    \Omega_{\text{scalar}} ~ = ~ - \sum_{\ell m} \int_\Sigma \mathrm{d}\Sigma_a \left( \delta \, \partial^a \phi_{\ell m} \wedge \delta \phi^*_{\ell m} + \delta \, \partial^a \phi^*_{\ell m} \wedge \delta \phi_{\ell m} \right)
\end{align}
is therefore independent of the choice of Cauchy surface $\Sigma$. Evaluating it on a surface of constant Schwarzschild time, the pairing it induces on solutions is the Klein--Gordon inner product
\begin{align}\label{eqn:KGproduct}
    \left( f \, , \, g \right) ~ \coloneqq ~ i \int \mathrm{d} r^* \left( f^* \, \partial_t \, g - g \, \partial_t f^* \right) \, ,
\end{align}
where $\partial_t = \partial_w$ at fixed $r^*$. Using $w = t + \eta^{w r^*} r^*$, the mode functions are
\begin{align}
    V^{(\sigma)}_{E_p} ~ = ~ e^{ - i w E_p + i r^* p^{(\sigma)}_{r^*}\left(E_p\right)} ~ = ~ e^{ - i t E_p + i \sigma \Omega_p r^*} \, ,
\end{align}
so that $\partial_t V^{(\sigma)}_{E_p} = - i E_p V^{(\sigma)}_{E_p}$, with an eigenvalue that does not depend on the branch. Carrying out the radial integral,
\begin{align}\label{eqn:modeNorm}
    \left( V^{(\sigma)}_{E_p} \, , \, V^{(\sigma')}_{E_{p'}} \right) ~ &= ~ \left(E_p + E_{p'}\right) \, 2\pi \, \delta\left(\sigma \Omega_p - \sigma' \Omega_{p'}\right) \nonumber \\
    &= ~ 2 \, \Omega_p \, \delta_{\sigma \sigma'} \, 2\pi \, \delta\left(E_p - E_{p'}\right) \, ,
\end{align}
where the second line uses $\Omega_p \geq 0$. For $\sigma' = - \sigma$ the argument of the delta function is $\Omega_p + \Omega_{p'}$, which is strictly positive unless both frequencies vanish, so the two branches are orthogonal; for $\sigma' = \sigma$ it is $\Omega_p - \Omega_{p'}$, and $\mathrm{d}\Omega_p/\mathrm{d}E_p = E_p/\Omega_p$ converts $\delta\left(\Omega_p - \Omega_{p'}\right)$ into $\left(\Omega_p/E_p\right)\delta\left(E_p - E_{p'}\right)$. The same computation applied to a positive- and a negative-frequency mode returns a prefactor $\left(E_p - E_{p'}\right)$, which vanishes on the support of the delta function, so the two sets are orthogonal as well.

The normalisation \eqref{eqn:modeNorm} is exactly the one implicit in the measure of the mode expansion, which now inverts to
\begin{align}
    c^{(\sigma)}_{\ell m}\left(E_p\right) ~ = ~ \left( V^{(\sigma)}_{E_p} \, , \, \phi_{\ell m} \right) \, , \qquad d^{(-\sigma)\dagger}_{\ell m}\left(E_p\right) ~ = ~ - \left( V^{(-\sigma)*}_{E_p} \, , \, \phi_{\ell m} \right) \, ,
\end{align}
since $\int \mathrm{d}E_q \left(4\pi \Omega_q\right)^{-1} 2\Omega_q \, 2\pi \, \delta\left(E_p - E_q\right) = 1$. Promoting the modes to operators, the bracket induced by \eqref{eqn:scalarSymplectic} is the inner product of the corresponding mode functions, so that the operator modes satisfy the symplectic algebra
\begin{subequations}\label{eqn:scalarAlgebra}
    \begin{align}
    \left[c^{(\sigma)}_{\ell m} \left(E_{p}\right) \, , \, c^{(\sigma')\dagger}_{\ell' m'} \left(E_{p'}\right) \right] ~ &= ~ 2 \Omega_p \delta_{\sigma \sigma'} \delta_{\ell \ell'} \delta_{m m'} 2\pi \delta\left(E_p - E_{p'}\right) \, , \\
    \left[c^{(\sigma)}_{\ell m} \left(E_{p}\right) \, , \, c^{(\sigma')}_{\ell' m'} \left(E_{p'}\right) \right] ~ &= ~ 0 ~ = ~ \left[c^{(\sigma)\dagger}_{\ell m} \left(E_{p}\right) \, , \, c^{(\sigma')\dagger}_{\ell' m'} \left(E_{p'}\right) \right] \, .
\end{align}
\end{subequations}
The phase space is of course also endowed with analogous commutators for the anti-particles.

Two derivatives appear in this discussion. The pairing \eqref{eqn:KGproduct} is evaluated on a Cauchy surface and involves $\partial_t = \partial_w$, whose eigenvalue $- i E_p$ is the same on both branches. Had the same expression been evaluated on a surface of constant $r^*$, the timelike surface used in \appref{app:MaxwellPhaseSpace}, the derivative appearing would instead be $\partial^{r^*}$, with the branch-dependent eigenvalue $i \sigma \Omega_p$, and the pairing would come out as $2 \sigma \Omega_p \, \delta_{\sigma \sigma'} \, 2\pi \delta\left(E_p - E_{p'}\right)$, indefinite in $\sigma$. That surface is not a Cauchy surface and is not where the theory is quantised; it is the surface across which the radial flux is measured. The branch weighting carried by the hard charge \eqref{eqn:hardChargeResult} is precisely that flux structure, and it is physical: a charge crossing outward contributes with the opposite sign to one crossing inward.

\subsection{Calculation of the hard charge}\label{app:hardChargeCalculation}
In this subsection, we will calculate the hard charge \eqref{eqn:hardCharge} using the quantised hard modes derived in the previous subsection. Throughout, $\widetilde{\partial} \coloneqq \partial_w + \eta^{w r^*}\partial_{r^*}$ as defined in the main text.

Since $p^{(\sigma)}_{r^*}\left(E_p\right) = \sigma \Omega_p + \eta^{w r^*} E_p$ and $\left(\eta^{w r^*}\right)^2 = 1$, the mode functions are eigenfunctions of $\widetilde{\partial}$:
\begin{align}\label{eqn:modeEigen}
    \widetilde{\partial}V^{(\sigma)}_{E_p} ~ = ~ i \, \eta^{w r^*} \sigma \Omega_p \, V^{(\sigma)}_{E_p}
    \qquad \text{and} \qquad
    \widetilde{\partial}V^{(\sigma)*}_{E_p} ~ = ~ - i \, \eta^{w r^*} \sigma \Omega_p \, V^{(\sigma)*}_{E_p} \, .
\end{align}
Because $\widetilde{\partial}V^{(-\sigma)*}_{E_p} = + i \eta^{w r^*}\sigma\Omega_p V^{(-\sigma)*}_{E_p}$, both terms in the mode expansion of $\phi_{\ell m}$ carry the same factor. Writing
\begin{align}
    \Phi^{(\sigma)}_{\ell m}\left(E\right) ~ \coloneqq ~ c^{(\sigma)}_{\ell m}\left(E\right) V^{(\sigma)}_{E} + d^{(-\sigma)\dagger}_{\ell m}\left(E\right) V^{(-\sigma)*}_{E} \, ,
\end{align}
so that $\phi_{\ell m} = \sum_{\sigma}\int_{m_\ell}^\infty \frac{\mathrm{d}E}{4\pi\Omega}\Phi^{(\sigma)}_{\ell m}\left(E\right)$, we have
\begin{align}
    \widetilde{\partial}\phi_{\ell m} ~ &= ~ i \, \eta^{w r^*}\sum_{\sigma}\int_{m_\ell}^\infty \frac{\mathrm{d}E}{4\pi\Omega} \, \sigma \Omega \, \Phi^{(\sigma)}_{\ell m}\left(E\right) \nonumber \\
    \widetilde{\partial}\phi^*_{\ell m} ~ &= ~ - i \, \eta^{w r^*}\sum_{\sigma}\int_{m_\ell}^\infty \frac{\mathrm{d}E}{4\pi\Omega} \, \sigma \Omega \, \Phi^{(\sigma)*}_{\ell m}\left(E\right) \, .
\end{align}
The Wronskian combination appearing in the current is therefore
\begin{align}\label{eqn:wronskian}
    \phi_{\ell_1 m_1}\widetilde{\partial}\phi^*_{\ell_2 m_2} - \phi^*_{\ell_2 m_2}\widetilde{\partial}\phi_{\ell_1 m_1} ~ &= ~ - i \, \eta^{w r^*} \sum_{\sigma_1 , \sigma_2 = \pm} \int_{m_{\ell_1}}^\infty \frac{\mathrm{d}E_p}{4\pi\Omega_p} \int_{m_{\ell_2}}^\infty \frac{\mathrm{d}E_q}{4\pi\Omega_q} \nonumber \\
    &\hspace{2cm} \times \left(\sigma_1 \Omega_p + \sigma_2 \Omega_q\right) \Phi^{(\sigma_1)}_{\ell_1 m_1}\left(E_p\right) \Phi^{(\sigma_2)*}_{\ell_2 m_2}\left(E_q\right) \, ,
\end{align}
with $\Omega_p = \sqrt{E_p^2 - m^2_{\ell_1}}$ and $\Omega_q = \sqrt{E_q^2 - m^2_{\ell_2}}$.

The number-violating products $c\,d$ and $d^\dagger c^\dagger$ in \eqref{eqn:wronskian} carry $w$-dependence $e^{\mp i w \left(E_p + E_q\right)}$ and hence produce $\delta\left(E_p + E_q\right)$, which has no support for $E_p , E_q > 0$; we drop them. Writing both products in normal order and discarding the c-number commutators, the surviving contributions are $c^{(\sigma_2)\dagger}_{\ell_2m_2}(E) c^{(\sigma_1)}_{\ell_1m_1}(E)$ and $d^{(-\sigma_1)\dagger}_{\ell_1m_1}(E) d^{(-\sigma_2)}_{\ell_2m_2}(E)$. These products carry $e^{\mp i w \left(E_p - E_q\right)}$, and
\begin{align}
    \int \mathrm{d}w \, e^{\pm i w \left(E_p - E_q\right)} ~ = ~ 2\pi \, \delta\left(E_p - E_q\right) \, .
\end{align}
Performing the $E_q$ integral,
\begin{align}
    &\int\mathrm{d}w \, \left( \phi_{\ell_1m_1} \widetilde{\partial}\phi^*_{\ell_2m_2} - \phi^*_{\ell_2m_2} \widetilde{\partial}\phi_{\ell_1m_1} \right) \nonumber\\
    &\qquad = ~ - i \eta^{wr^{*}}\sum_{\sigma_1,\sigma_2=\pm} \int_{\max(m_{\ell_1},m_{\ell_2})}^\infty \frac{\mathrm{d}E}{8\pi \, \Omega_1 \Omega_2} \left( \sigma_1\Omega_1+\sigma_2\Omega_2 \right) e^{i r^*\left( \sigma_1\Omega_1-\sigma_2\Omega_2 \right)} \nonumber\\
    &\hspace{5cm} \times \Bigg[ c^{(\sigma_2)\dagger}_{\ell_2m_2}(E) c^{(\sigma_1)}_{\ell_1m_1}(E) + d^{(-\sigma_1)\dagger}_{\ell_1m_1}(E)d^{(-\sigma_2)}_{\ell_2m_2}(E) \Bigg] \, ,
\end{align}
where $\Omega_i \coloneqq \sqrt{E^2 - m^2_{\ell_i}}$ at the common energy, and we used that the two radial phases coincide,
\begin{align}
    p^{(\sigma_1)}_{r^*}\left(E\right) - q^{(\sigma_2)}_{r^*}\left(E\right) ~ = ~ - p^{(-\sigma_1)}_{r^*}\left(E\right) + q^{(-\sigma_2)}_{r^*}\left(E\right) ~ = ~ \sigma_1 \Omega_1 - \sigma_2 \Omega_2 \, .
\end{align}
The difference between the two radial frequencies satisfies
\begin{align}
    \Omega_1 - \Omega_2 ~ &= ~ \frac{m_{\ell_2}^2-m_{\ell_1}^2}{\sqrt{E^2-m_{\ell_1}^2} + \sqrt{E^2 - m_{\ell_2}^2}} \, .
\end{align}
Since the effective squared masses are proportional to $f(r)$, this difference vanishes exponentially as $r^*\to-\infty$. In particular,
\begin{align}
    r^* \left(\Omega_1 - \Omega_2\right) ~ \longrightarrow ~ 0 \, .
\end{align}
For $\sigma_2=\sigma_1$, one has
\begin{align}
    \sigma_1\Omega_1 + \sigma_2 \Omega_2 ~ \longrightarrow ~ 2 \sigma_1 E \quad \text{and} \quad e^{ir^*(\sigma_1\Omega_1 - \sigma_2 \Omega_2)} ~ \longrightarrow ~ 1 \, .
\end{align}
For $\sigma_2=-\sigma_1$, one has
\begin{align}
    \sigma_1 \Omega_1 + \sigma_2 \Omega_2 ~ &= ~ \sigma_1 \left(\Omega_1 - \Omega_2\right) ~  \longrightarrow ~ 0 \, .
\end{align}
Thus only the diagonal branch contributions survive, and
\begin{align}
    &\lim_{r^*\to-\infty} \int\mathrm{d}w \, \left( \phi_{\ell_1m_1} \widetilde{\partial}\phi^*_{\ell_2m_2} - \phi^*_{\ell_2m_2} \widetilde{\partial}\phi_{\ell_1m_1} \right) ~ = \nonumber\\
        &\qquad = ~ - i \, \eta^{w r^*} \sum_{\sigma_1=\pm} \int_0^\infty \frac{\mathrm{d}E}{4\pi E} \, \sigma_1 \Big[ c^{(\sigma_1)\dagger}_{\ell_2m_2}(E) c^{(\sigma_1)}_{\ell_1m_1}(E) + d^{(-\sigma_1)\dagger}_{\ell_1m_1}(E) d^{(-\sigma_1)}_{\ell_2m_2}(E) \Big] \, .
\end{align}
Here the lower limit $\max\left(m_{\ell_1},m_{\ell_2}\right)$ has been sent to zero along with $m_\ell$; as in the neglect of $\rho_{\ell m}$ and the replacement $\Omega_i \to E$, the limit $f\left(r\right) \to 0$ is taken at fixed $E$, so that the apparent logarithmic divergence of $\int \mathrm{d}E/E$ at the origin lies outside the regime in which the expression was derived. 

Now, relabelling $\sigma_1\to-\sigma_1$ in the antiparticle contribution and substituting this result into \eqref{eqn:hardChargeCurrentExpansion}, where the factor of $\eta^{wr^*}$ from \eqref{eqn:modeEigen} cancels against the one from \eqref{eqn:r2CurrentInPWs}, gives
\begin{align}\label{appeqn:hardChargeResult}
    Q_{\mathrm{hard}}\left[\lambda^{(0)}_{\ell m}\right] ~ &= ~ q \lambda^{(0)}_{\ell m} \sum_{\substack{\ell_1m_1\\\ell_2m_2}} \sum_{\sigma_1=\pm} \int_0^\infty \frac{\mathrm{d}E}{4\pi E} \, \sigma_1 \nonumber \\
    &\hspace{2.5cm} \times \Big[ c^{(\sigma_1)\dagger}_{\ell_2m_2}(E) c^{(\sigma_1)}_{\ell_1m_1}(E) - d^{(\sigma_1)\dagger}_{\ell_1m_1}(E) d^{(\sigma_1)}_{\ell_2m_2}(E) \Big] C^{(3)} \, .
\end{align}

\subsection{Maxwell theory near the horizon in partial waves}

The pure Maxwell action is given by \eqref{eqn:MaxwellAction}
\begin{align}
    S_{\text{Maxwell}} ~ = ~ \int \mathrm{d}^4 x \, \sqrt{-g} \, \mathcal{L}_{\text{Maxwell}} ~ = ~ - \dfrac{1}{4} \int \mathrm{d}^4 x \sqrt{-g} \, g^{\mu \rho} g^{\nu \sigma} F_{\mu \nu} F_{\rho \sigma} \, .
\end{align}
As we saw in \secref{sec:radMode}, the gauge degree of freedom is in the even sector. In what follows, we will reduce this action to this physical radiative mode in this sector. From \eqref{eqn:FinTermsofChi_a}, we see that the transverse components of the field strength ($F_{AB}$) only contain the odd sector. Therefore, we will ignore those components. Splitting indices into longitudinal (lowercase) and transverse (uppercase) components, the even sector of the above action can be written as
\begin{align}\label{eqn:MaxwellChi1}
    S^{\text{even}}_{\text{Maxwell}} ~ &= ~ - \dfrac{1}{4} \int \mathrm{d}^4 x \sqrt{-g} \, \left(g^{a c} g^{b d} F_{a b} F_{c d} + 2 g^{a b} g^{A B} F_{a A} F_{b B}\right) \nonumber \\
    &= ~ - \dfrac{1}{2} \sum_{\substack{\ell_1 m_1 \\ \ell_2 m_2}} \int \mathrm{d}^2 x \mathrm{d}\Omega \, \dfrac{r^2}{f\left(r\right)} \eta^{ac} \eta^{bd} \left(\partial_a \chi^{\ell_1 m_1}_b \partial_c \chi^{\ell_2 m_2}_d - \partial_a \chi^{\ell_1 m_1}_b \partial_d \chi^{\ell_2 m_2}_c\right) Y_{\ell_1 m_1} Y_{\ell_2 m_2} \nonumber \\
    &\qquad - \dfrac{1}{2} \sum_{\substack{\ell_1 m_1 \\ \ell_2 m_2}} \int \mathrm{d}^2 x \mathrm{d}\Omega \, \eta^{ab} \gamma^{A B} \left(\chi^{\ell_1 m_1}_a \partial_A Y_{\ell_1 m_1}\right) \left(\chi^{\ell_2 m_2}_b \partial_B Y_{\ell_2 m_2}\right) \, ,
\end{align}
where we used the orthogonality properties of the spherical harmonics to integrate over the sphere. Written in terms of the reduced field strength, this is 
\begin{align}\label{eqn:MaxwellChi2}
    S^{\text{even}}_{\text{Maxwell}} ~ = ~ - \dfrac{1}{4} \sum_{\ell m} \int \mathrm{d}^2 x \, \dfrac{r^2}{f\left(r\right)} \left[F^{\ell m}_{ab}F^{ab}_{\ell m} + \dfrac{2 \ell \left(\ell + 1\right) f\left(r\right)}{r^2} \chi^a_{\ell m} \chi^{\ell m}_a \right] \, .
\end{align}
The two-vector $\chi^{\ell m}_a$ carries a single physical degree of freedom, as the constraint in \eqref{eqn:AaSoln} makes explicit: in the source-free case $\eta^{ab}\partial_a \chi^{\ell m}_b = 0$. In two dimensions this constraint is solved exactly by the Moncrief variable, $\chi^{\ell m}_a = \epsilon_{ab}\partial^b \Psi_{\ell m}$ as in \eqref{eqn:chiaSoln}, so that the physical content of the even sector is the single scalar $\Psi_{\ell m}$ obeying \eqref{eqn:MoncriefEoM},
\begin{align}\label{eqn:PsiKG}
    \left(\Box_\eta - \mu^2_\ell\right)\Psi_{\ell m} ~ = ~ 0 \qquad \text{with} \qquad \mu^2_\ell ~ \coloneqq ~ \dfrac{f\left(r\right)\ell\left(\ell+1\right)}{r^2} \, .
\end{align}
As in the case of the scalar field, the transverse curvature presents itself as an effective mass for the two-dimensional radiative mode near the horizon.

We do not reduce \eqref{eqn:MaxwellChi2} further by substituting $\chi^{\ell m}_a = \epsilon_{ab}\partial^b \Psi_{\ell m}$. Doing so would give a Lagrangian quadratic in $\Box_\eta \Psi_{\ell m}$, whose Euler--Lagrange equation, 
\begin{equation}
    \Box_\eta\left[\dfrac{r^2}{f\left(r\right)} \Box_\eta \Psi_{\ell m} -\ell \left( \ell + 1 \right) \Psi_{\ell m} \right] ~ = ~ 0 \, ,
\end{equation}
is strictly weaker than \eqref{eqn:MoncriefMaster} and admits spurious solutions. This is the degeneracy noted in \appref{app:MaxwellPhaseSpace}; we therefore work with \eqref{eqn:PsiKG} together with the symplectic form derived there.

Unlike the scalar of \appref{app:scalarPropagator}, \eqref{eqn:PsiKG} needs no rescaling: it is already in Regge--Wheeler form, with no $r^2/f\left(r\right)$ prefactor and no curvature term, and $\mu^2_\ell \propto f\left(r\right)$ vanishes exponentially as $r^* \rightarrow -\infty$.

\subsection{Quantisation of the radiative mode}\label{app:radiativeModeQuantisation}
By \eqref{eqn:PsiKG}, $\Psi_{\ell m}$ is a two-dimensional scalar of mass $\mu_\ell$, of exactly the form treated in \appref{app:hardModeQuantisation} with $m_\ell \rightarrow \mu_\ell$. Its on-shell expansion is therefore
\begin{align}
    \Psi_{\ell m}\left(x\right) ~ &= ~ \int \dfrac{\mathrm{d}^{2} k}{2\pi} \, a_{\ell m}\left(k\right) \, \delta\left(k^2 + \mu^2_\ell\right) \, \exp\left(i x^{a} k_{a}\right) \, ,
\end{align}
with no polarisation vector to introduce. With the momentum two-vector $k_a = \left(- \omega, k_{(\sigma)}\right)$, the mass-shell condition can be written as
\begin{equation}\label{eqn:photonMassShell}
    k_{(\sigma)} ~ = ~ \sigma \sqrt{\omega^2 - \mu^2_\ell} + \eta^{w r^*} \omega ~ \eqqcolon ~ \sigma \tilde{\omega} + \eta^{w r^*} \omega \, ,
\end{equation}
where the additional label $\sigma$ keeps track of the two branches of the solution similar to the scalar field. The polarisation vector is not an independent choice here; it is generated by $\epsilon_{ab}\partial^b$ acting on the scalar mode. For $\Psi_{\ell m} \propto e^{i k \cdot x}$,
\begin{align}\label{eqn:polarisation}
    \chi^{\ell m}_a ~ &= ~ \epsilon_{ab}\partial^b \Psi_{\ell m} ~ = ~ - \, i \, \varepsilon^{(\sigma)}_a\left(\omega\right) \Psi_{\ell m} \, ,
    \nonumber \\
    \varepsilon^{(\sigma)}_a\left(\omega\right) ~ &\coloneqq ~ - \, \epsilon_{ab}k^b_{(\sigma)} ~ = ~ \left(- \sigma \tilde{\omega} \, , \, \eta^{w r^*}\sigma\tilde{\omega} + \omega\right) \, .
\end{align}
Transversality is automatic by antisymmetry, $k^a \varepsilon^{(\sigma)}_a = - \epsilon_{ab}k^a k^b = 0$, and the normalisation follows on shell rather than being imposed,
\begin{align}
    \varepsilon^{(\sigma)}_a \varepsilon^{(\sigma)a} ~ = ~ \left(\epsilon_{ab}k^b\right)\left({\epsilon^{a}}_{c}k^c\right) ~ = ~ - \, k \cdot k ~ = ~ \mu^2_\ell \, .
\end{align}
The vector is regular as $\mu_\ell \rightarrow 0$ with norm $\mu_\ell$ rather than unity. Its contravariant radial component,
\begin{align}
    \varepsilon^{(\sigma) r^*} ~ = ~ \eta^{r^* w}\varepsilon^{(\sigma)}_w + \eta^{r^* r^*}\varepsilon^{(\sigma)}_{r^*} ~ = ~ \omega \, ,
\end{align}
is independent of the branch label.

Following \appref{app:hardModeQuantisation}, the two roots of the mass-shell condition are related by
\begin{align}
    k_{(\sigma)}\left(-\omega\right) ~ = ~ - \, k_{(-\sigma)}\left(\omega\right) \, ,
\end{align}
so the negative-frequency mode inherited from the root labelled $\sigma$ is the complex conjugate of the positive-frequency mode on the opposite branch. Since $\Psi_{\ell m}$ is real, relabelling $\sigma \rightarrow -\sigma$ in the negative-frequency term gives
\begin{align}\label{eqn:PsiRawModeExpansion}
    \Psi_{\ell m} ~ = ~ \sum_{\sigma = \pm} \int_{\mu_\ell}^\infty \dfrac{\mathrm{d}\omega}{4\pi\tilde{\omega}} \left[a^{(\sigma)}_{\ell m}\left(\omega\right) e^{- i \omega w + i r^* k_{(\sigma)}\left(\omega\right)} + a^{(\sigma)*}_{\ell m}\left(\omega\right) e^{i \omega w - i r^* k_{(\sigma)}\left(\omega\right)}\right] \, .
\end{align}
At this stage the coefficients $a^{(\sigma)}_{\ell m}$ are the Fourier coefficients appearing in the mass-shell decomposition; their normalisation has not yet been fixed. The normalisation is fixed by the symplectic structure, and it is important to separate the two ingredients that go into doing so, since they are established on different surfaces.

The first is the overall factor. The even-sector symplectic current follows from $\Theta^\mu_{\text{Max.}} = -F^{\mu\nu}\delta A_\nu$ reduced with \eqref{eqn:FinTermsofChi_a}; in the source-free case it is given in terms of 
\begin{equation}
	\Theta^{a}_{\text{even}} ~ = ~ - \ell \left(\ell + 1\right) \partial^{a} \Psi_{\ell m} \delta \Psi_{\ell m} \, ,
\end{equation}
by an exterior variation
\begin{align}\label{eqn:PsiSymplecticCurrent}
    \omega^a_{\text{even}} ~ = ~ \delta \Theta^{a}_{\text{even}} ~ = ~ - \ell\left(\ell+1\right) \,
    \delta \, \partial^a \Psi_{\ell m} \wedge \delta \Psi_{\ell m} \, .
\end{align}
Its integral over a surface of constant $r^*$ is the even bulk term of \eqref{eqn:MaxwellSymplecticModes}. Being a current, \eqref{eqn:PsiSymplecticCurrent} is independent of the surface on which it is integrated, so the factor $\ell(\ell+1)$ it carries is a property of the reduction to partial waves and not of the horizon. It differs from the current of a canonically normalised two-dimensional scalar only by that factor, so that
\begin{align}
    \widehat{\Psi}_{\ell m} ~ \coloneqq ~ \sqrt{\ell\left(\ell+1\right)} \, \Psi_{\ell m}
\end{align}
is canonically normalised.
 
The second ingredient is the surface. The algebra must be read off on a Cauchy surface, exactly as for the hard modes in \appref{app:hardModeQuantisation}: integrating \eqref{eqn:PsiSymplecticCurrent} over a surface of constant Schwarzschild time gives the Klein--Gordon product \eqref{eqn:KGproduct} with $\partial_t$, whose eigenvalue $-i\omega$ is the same on both branches, so that \eqref{eqn:modeNorm} applies verbatim with $E_p \to \omega$ and $\Omega_p \to \tilde{\omega}$,
\begin{align}\label{eqn:PsiModeNorm}
    \left(e^{-i\omega w + i k_{(\sigma)}r^*} \, , \,
    e^{-i\omega' w + i k_{(\sigma')}r^*}\right) ~ = ~
    2 \, \tilde{\omega} \, \delta_{\sigma\sigma'} \, 2\pi \,
    \delta\left(\omega - \omega'\right) \, ,
\end{align}
positive and independent of the branch. Had the same expression been evaluated on the constant-$r^*$ surface of \appref{app:MaxwellPhaseSpace}, the derivative appearing would instead be $\partial^{r^*}$, with the branch-dependent eigenvalue $i\sigma\tilde\omega$ by \eqref{eqn:drstarOnBranches}, and the pairing would come out as $2\sigma\tilde{\omega} \,\delta_{\sigma\sigma'}2\pi\delta(\omega-\omega')$, indefinite in $\sigma$. That surface is not a Cauchy surface and is not where the theory is quantised; it is the surface across which the radial flux is measured, and the branch weighting it carries is the same flux structure that appears in the hard charge \eqref{eqn:hardChargeResult} and in the bulk term \eqref{eqn:BulkSymplecticBranches} of the horizon symplectic form. This is the same distinction drawn below \eqref{eqn:scalarAlgebra} for $\phi_{\ell m}$.
 
The standard scalar quantisation therefore applies to $\widehat{\Psi}_{\ell m}$ directly. Writing $b^{(\sigma)}_{\ell m}\left(\omega\right)$ for its annihilation operators,
\begin{align}\label{appeqn:PsiQuantisation}
    \Psi_{\ell m} ~ = ~ \dfrac{1}{\sqrt{\ell\left(\ell+1\right)}} \sum_{\sigma = \pm}
    \int_{\mu_\ell}^\infty \dfrac{\mathrm{d}\omega}{2\pi} \, \dfrac{i}{\sqrt{2\tilde{\omega}}}
    \left[b^{(\sigma)}_{\ell m}\left(\omega\right)
    e^{- i \omega w + i k_{(\sigma)} r^*} - \text{h.c.}\right] \, ,
\end{align}
the overall phase being a convention, fixed here for later convenience. Comparing with \eqref{eqn:PsiRawModeExpansion}, whose coefficients were left unnormalised, the two expansions describe the same field with
\begin{align}
    a^{(\sigma)}_{\ell m}\left(\omega\right) ~ = ~ i \,
    \sqrt{\dfrac{2 \tilde{\omega}}{\ell \left(\ell + 1\right)}} \,
    b^{(\sigma)}_{\ell m}\left(\omega\right) \, ,
\end{align}
and inverting \eqref{appeqn:PsiQuantisation} against \eqref{eqn:PsiModeNorm} gives
\begin{align}\label{appeqn:PsiAlgebra}
    \left[b^{(\sigma)}_{\ell m}\left(\omega\right) \, , \,
    b^{(\sigma')\dagger}_{\ell' m'}\left(\omega'\right)\right] ~ = ~
    \delta_{\sigma \sigma'} \, \delta_{\ell \ell'} \, \delta_{m m'} \, 2\pi \,
    \delta\left(\omega - \omega'\right) \, .
\end{align}
The measure $\left(2\tilde\omega\right)^{-1/2}$ in \eqref{appeqn:PsiQuantisation} together with \eqref{eqn:PsiAlgebra} is the unit normalisation, whereas the hard scalars were written with the measure $\left(4\pi\Omega_p\right)^{-1}$ and the covariant algebra \eqref{eqn:scalarAlgebra}. These are two conventions for distributing the same Klein--Gordon norm \eqref{eqn:modeNorm} between the measure and the commutator, not two different inner products. This is the quantisation of the source-free field. In the presence of the current, the field with this expansion at early and late times is $\mathcal{E}_{\ell m}/\ell\left(\ell+1\right)$ of \eqref{eqn:radialFlux}, which obeys \eqref{eqn:PsiKG} wherever the current vanishes, and the $\Psi_{\ell m}$ of \eqref{eqn:chiaSoln} differs from it by the piecewise constant of \eqref{eqn:MoncriefMaster}.
 
The factor $\left(2\tilde{\omega}\right)^{-1/2}$ enters the soft theorem in two places. Acting with $\partial_w$ on \eqref{appeqn:PsiQuantisation} it supplies the factor of $\sqrt{\omega}$ in the charge \eqref{eqn:softChargeOnHorizons}; appearing as the external-leg wavefunction in \eqref{eqn:vertex} it supplies the factor of $\omega^{-1/2}$ in the emission amplitude \eqref{eqn:finalAmp}. Since the Ward identity \eqref{eqn:WardIdentity} is linear in the soft operator on both sides, a rescaling $b^{(\sigma)}_{\ell m}\rightarrow\kappa\left(\omega\right)b^{(\sigma)}_{\ell m}$ moves the factor between the two and leaves the identity unchanged. In the covariant normalisation, for instance, that charge would read
\begin{equation}
    - i \lim_{\omega \rightarrow 0}\left(a^{(\sigma)}_{\ell m}
    - a^{(\sigma)\dagger}_{\ell m}\right)
\end{equation}
with no explicit power of $\omega$, and \eqref{eqn:finalAmp} would carry none either.
 
This argument covers rescalings that are independent of $\sigma$. It does not by itself fix the relative normalisation of the two branches, which is what the branch sum in \eqref{eqn:softChargeOnHorizons} and \eqref{eqn:branchSum} depends on. That is fixed instead by \eqref{eqn:PsiModeNorm}, which is branch independent, and independently by the observation that at $\mu_\ell=0$ the near-horizon theory is invariant under $r^*\to-r^*$, which exchanges the two roots $k_{(\sigma)}\to k_{(-\sigma)}$ of \eqref{eqn:photonMassShell}. Both statements are properties of the free field and are independent of any prescription for approaching the horizon.
 
The $\omega^{-1/2}$ of \eqref{eqn:finalAmp} is therefore a feature of the external state and not of the dynamics. The dynamical soft factor \eqref{eqn:vertexRatio} is independent of $\omega$, whereas its flat-space plane-wave counterpart $\varepsilon\cdot p_n/k\cdot p_n$ scales as $\omega^{-1}$. In the covariant normalisation, in which the plane-wave soft theorem exhibits its pole, the partial-wave emission amplitude is finite as $\omega \rightarrow 0$. The factor of $\omega$ between the two is the Jacobian from plane waves to partial waves, and it is the same in flat space. As noted at the top of \appref{app:MaxwellPhaseSpace}, these modes are physical only for $\ell>0$.
 
Applying $\epsilon_{ab}\partial^b$ to \eqref{appeqn:PsiQuantisation} and using \eqref{eqn:polarisation}, the canonically normalised expansion of the radiative mode is
\begin{align}\label{appeqn:chiaQuantisation}
    \chi_a^{\ell m} = \sum_{\sigma=\pm} \int_{\mu_\ell}^{\infty}\frac{\mathrm{d}\omega}{2\pi}
    \left[ U_{a,\omega}^{(\sigma)} b_{\ell m}^{(\sigma)}
    + U_{a,\omega}^{(\sigma)*} b_{\ell m}^{(\sigma)\dagger} \right]
    ~ \text{with} ~ U_{a,\omega}^{(\sigma)} =
    \frac{\varepsilon_a^{(\sigma)}(\omega)}{\sqrt{2\ell(\ell+1)\widetilde\omega}} \,
    e^{-i\omega w+i k_{(\sigma)}r^*} .
\end{align}
In particular,
\begin{align}
    U_{w,\omega}^{(\sigma)} ~ = ~ - \sigma
    \sqrt{\frac{\widetilde\omega}{2\ell(\ell+1)}} \, e^{-i\omega w+i k_{(\sigma)}r^*} \, .
\end{align}

\paragraph{Displacement of a thermal mode.} Two quantities are needed in \secref{sec:irstate}: the thermal expectation value of a displacement, and the fidelity between a thermal state and its displacement. Both follow from a single expectation value, and it is simplest to compute it for one mode of the algebra \eqref{eqn:PsiAlgebra}, normalised for the moment to $\left[b, b^\dagger\right] = 1$, in the thermal state
\begin{equation}\label{appeqn:thermalMode}
\rho ~ = ~ \left(1 - x\right) \sum_{k = 0}^{\infty} x^k \, |k\rangle\langle k| \, , \qquad x ~ \coloneqq ~ e^{-\omega/T_H} \, , \qquad n ~ = ~ \dfrac{x}{1 - x} \, ,
\end{equation}
with $D\left(\alpha\right) \coloneqq \exp\left(\alpha b^\dagger - \bar{\alpha} b\right)$. Adjoin a second mode $\tilde{b}$ and let
\begin{equation}\label{appeqn:thermalPurification}
|\Omega\rangle ~ \coloneqq ~ \sqrt{1 - x} \, \sum_{k = 0}^{\infty} x^{k/2} \, |k\rangle \otimes |k\rangle ~ = ~ S\left(r\right) \, |0\rangle \otimes |0\rangle \, , 
\end{equation}
where
\begin{equation}
S\left(r\right) ~ \coloneqq ~ \exp\left[ r \left( b^\dagger \tilde{b}^\dagger - b \tilde{b} \right) \right] \quad \text{and} \quad \tanh r ~ = ~ e^{-\omega/2T_H} \, ,
\end{equation}
so that $\sinh^2 r = n$ and $\cosh 2r = 2n + 1 = \coth\left(\omega/2T_H\right)$; the partial trace of $|\Omega\rangle\langle\Omega|$ over $\tilde{b}$ is $\rho$. On the horizon this purification is not a device: $|\Omega\rangle$ is the Hartle--Hawking state restricted to the mode and to its partner of the same Killing frequency in the other exterior. Since $S^\dagger b S = b \cosh r + \tilde{b}^\dagger \sinh r$ and $S^\dagger \tilde{b} S = \tilde{b} \cosh r + b^\dagger \sinh r$, the product $D\left(\alpha\right) \tilde{D}\left(\beta\right)$ of displacements of the two modes becomes, when conjugated by $S$, a displacement of $b$ by $\alpha \cosh r - \bar{\beta} \sinh r$ and of $\tilde{b}$ by $\beta \cosh r - \bar{\alpha} \sinh r$, and its expectation value is
\begin{equation}\label{appeqn:twoModeOverlap}
\Big| \langle \Omega | \, D\left(\alpha\right) \tilde{D}\left(\beta\right) \, | \Omega \rangle \Big| ~ = ~ \exp\left[ - \dfrac{1}{2} \left| \alpha \cosh r - \bar{\beta} \sinh r \right|^2 - \dfrac{1}{2} \left| \beta \cosh r - \bar{\alpha} \sinh r \right|^2 \right] \, .
\end{equation}
At $\beta = 0$ the partner is untouched and the expectation value is the thermal trace,
\begin{align}\label{appeqn:thermalTrace}
\Big| \operatorname{Tr}\left( \rho \, D\left(\alpha\right) \right) \Big| ~ &= ~ \exp\left[ - \dfrac{1}{2} \left|\alpha\right|^2 \left( \cosh^2 r + \sinh^2 r \right) \right] \nonumber \\
&= ~ \exp\left[ - \dfrac{1}{2} \left|\alpha\right|^2 \left( 2n + 1 \right) \right] \nonumber \\
&= ~ \exp\left[ - \dfrac{1}{2} \left|\alpha\right|^2 \coth \dfrac{\omega}{2 T_H} \right] \, ,
\end{align}
which is also the Gaussian identity $\langle e^{X} \rangle = e^{\langle X^2 \rangle / 2}$ for $X = \alpha b^\dagger - \bar{\alpha} b$, with $\langle X^2 \rangle = - \left|\alpha\right|^2 \langle b^\dagger b + b b^\dagger \rangle$. The fidelity $F\left(\rho, \tilde{\sigma}\right) \coloneqq \operatorname{Tr} \sqrt{\sqrt{\rho} \, \tilde{\sigma} \sqrt{\rho}}$ between $\rho$ and $\tilde{\sigma} = D \rho D^\dagger$ is, by Uhlmann's theorem, the maximum of $|\langle \Omega | D\left(\alpha\right) U | \Omega \rangle|$ over unitaries $U$ acting on the partner alone, since $D\left(\alpha\right) |\Omega\rangle$ purifies $\sigma$ and $U |\Omega\rangle$ ranges over the purifications of $\rho$. Among the displacements $U = \tilde{D}\left(\beta\right)$ the exponent of \eqref{appeqn:twoModeOverlap} is extremal at $\bar{\beta} = \alpha \tanh 2r$, where its two terms are $\left|\alpha\right|^2 \cosh^2 r / \cosh^2 2r$ and $\left|\alpha\right|^2 \sinh^2 r / \cosh^2 2r$, and
\begin{align}\label{appeqn:thermalFidelityMode}
F\left( \rho \, , \, D \rho D^\dagger \right) ~ &= ~ \exp\left[ - \dfrac{\left|\alpha\right|^2}{2 \cosh 2r} \right] \nonumber \\
&= ~ \exp\left[ - \dfrac{\left|\alpha\right|^2}{2 \left( 2n + 1 \right)} \right] \nonumber \\
&= ~ \exp\left[ - \dfrac{1}{2} \left|\alpha\right|^2 \tanh \dfrac{\omega}{2 T_H} \right] \, .
\end{align}
That the maximum over all unitaries on the partner is attained on a displacement is the known form of the fidelity between Gaussian states of equal covariance \cite{Marian:2012lqm, Banchi:2015rmr}; equivalently, the Bures metric that \eqref{appeqn:thermalFidelityMode} implies, $F^2 \simeq 1 - \left|\alpha\right|^2 / \left(2n+1\right)$, is the quantum Fisher information $4/\left(2n+1\right)$ of a displaced thermal state, which is exact. The optimal partner displacement is $\beta = \bar{\alpha} \tanh 2r = \bar{\alpha} / \cosh\left(\omega/2T_H\right)$: the purification closest to the displaced state displaces the partner by the reflected amplitude, weighted by $\operatorname{sech}\left(\omega/2T_H\right)$. At $T_H \rightarrow 0$ it vanishes and \eqref{appeqn:thermalTrace} and \eqref{appeqn:thermalFidelityMode} both reduce to the vacuum overlap $e^{-\left|\alpha\right|^2/2}$; at high temperature it tends to $\bar{\alpha}$. For the continuum modes of \eqref{eqn:PsiAlgebra} the exponents of independent modes add, $\left|\alpha\right|^2 \rightarrow \sum_\sigma \int \left(\mathrm{d}\omega/2\pi\right) \left|\alpha^{(\sigma)}_{\ell m}\left(\omega\right)\right|^2$ with the occupation of each branch, which gives \eqref{eqn:thermalOverlap} and \eqref{eqn:thermalFidelity}; on a cold branch $r = 0$ and both factors are the vacuum one.

\subsection{Interaction terms near the horizon}\label{app:interactionTerms}
The interaction term in the action \eqref{eqn:MatterAction} can be written as
\begin{align}\label{eqn:totinteraction}
    S_{\text{int.}} ~ &= ~ - \int \mathrm{d}^4 x \sqrt{-g} \, g^{\mu\nu} A_\mu \Bigr[ i q \left( \tilde{\phi}^* \nabla_\nu \tilde{\phi} - \tilde{\phi} \nabla_\nu \tilde{\phi}^* \right) + q^2 A_\nu \left|\tilde{\phi}\right|^2 \Bigr] \nonumber \\
    &= ~ - \int \mathrm{d} r^* \, \mathrm{d} w \, \mathrm{d}\Omega \, \Biggl(r^2 \, \eta^{ab} A_a \Bigl[ i q \left( \tilde{\phi}^* \nabla_b \tilde{\phi} - \tilde{\phi} \nabla_b \tilde{\phi}^* \right) + q^2 A_b \left|\tilde{\phi}\right|^2 \Bigr] \nonumber \\
    &\hspace{4cm} + f\left(r\right) \, \gamma^{AB} A_A \Bigl[ i q \left( \tilde{\phi}^* \nabla_B \tilde{\phi} - \tilde{\phi} \nabla_B \tilde{\phi}^* \right) + q^2 A_B \left|\tilde{\phi}\right|^2 \Bigr]\Biggl) \, ,
\end{align}
where we split the indices into longitudinal and transverse directions using the metric \eqref{eqn:metric} and used the relations $g^{ab} = \eta^{ab} f\left(r\right)^{-1}$ and $g^{A B} = r^{-2} \gamma^{A B}$. It is evident that the transverse part of the action is proportional to $f\left(r\right)$. The odd mode of the photon couples only through this part, and we show in \appref{app:magneticsofttheorem} that its soft factor vanishes in the near-horizon soft limit. This is consistent with the observation that the magnetic charge \eqref{eqn:magneticChargeNearHor} has no hard component. 

In what follows, we will focus on the three-vertex because our interest in this paper is confined to the single-soft photon theorem. Moreover, we saw in \eqref{eqn:gaugetransfo} that the transverse mode $A^+_{\ell m}$ is pure gauge in the bulk.\footnote{It contains a physical boundary mode as discussed around \eqref{eqn:AplusBranches}.} It is therefore convenient to gauge fix this mode to zero in the calculation of the amplitudes; the amplitude summed over the external legs does not depend on this choice. The choice affects only how the emission of a single leg is distributed between the two branches, which the Ward identity does not fix. In the gauge $A_w = 0$ of the rest of the paper, in which $A^+_{\ell m}$ is fixed by \eqref{eqn:AplusBranches}, the photon couples through $A_{r^*}$ alone, $\eta^{ab} A_a j_b = A_{r^*} \, {\epsilon_w}^{b} j_b$, and $A_{r^*}$ is fixed by the field strength: $F_{w r^*} = \partial_w A_{r^*} = - \Box_\eta \Psi_{\ell m}$ by \eqref{eqn:PsiIdentities}, so that on a mode of Killing frequency $\omega$, by \eqref{eqn:PsiKG},
\begin{align}\label{eqn:ArstarGauge}
    A_{r^*} ~ = ~ \dfrac{\mu^2_\ell\left(r^*\right)}{i \omega} \, \Psi_{\ell m} ~ + ~ \varphi'_{\ell m}\left(r^*\right) \, .
\end{align}
The radiative mode then couples with an explicit factor of $\mu^2_\ell$. The non-collinear emission is then suppressed by $\mu^2_\ell/\omega^2$, and the emission of order one from a leg comes from the collinear configuration alone, which sits on the leg's own branch, as the shock \eqref{eqn:coulombMemory} does. The two gauges distribute the emission of a single leg differently between the branches and agree on the sum over legs on each branch, the difference being the charge-weighted combination of \eqref{eqn:chargeVsFlux}, which vanishes on shell; see also the third remark below \eqref{eqn:dressingDisplacement}.

We will now use the equations of motion \eqref{eqn:AaSoln} to reduce the three-vertex part of the above action to the contribution from the radiative mode in the even sector. Relabelling the dummy indices $\ell_1 m_1 \leftrightarrow \ell_2 m_2$ in the second term, so that the scalar bilinear appears in Wronskian form, we may use \eqref{eqn:currentRescaling} to pass to the rescaled field $\phi_{\ell m} = r \, \tilde{\phi}_{\ell m}$. The explicit factor of $r^2$ then cancels, as it did in the hard charge, and
\begin{align}\label{eqn:chiavertex}
    S_{\text{int.}} ~ &= ~ - i q \sum_{\substack{\ell_i m_i \\ i \in \left(1, 2, 3\right)}} \int \mathrm{d} r^* \, \mathrm{d} w \, \eta^{ab} \chi^{\ell_3 m_3}_a \left( \phi^*_{\ell_1 m_1} \partial_b \phi_{\ell_2 m_2} - \phi_{\ell_2 m_2} \partial_b \phi^*_{\ell_1 m_1} \right) \, C^{(3)} \, ,
\end{align}
where we defined 
\begin{subequations}\label{eqn:Yintegrals}
\begin{align}
    C^{(3)} ~ &\coloneqq ~ \int \mathrm{d} \Omega \, Y_{\ell_1 m_1} Y_{\ell_2 m_2} Y_{\ell_3 m_3}  \\
    \tilde{C}^{(3)}_{jn} ~ &\coloneqq ~ \int \mathrm{d} \Omega \, \gamma^{A B} Y_{\ell_k m_k} \partial_A Y_{\ell_j m_j} \partial_B Y_{\ell_n m_n} \\
    \hat{C}^{(3)}_{jn} ~ &\coloneqq ~ \int \mathrm{d}\Omega \gamma^{AB} {\epsilon_A}^C Y_{\ell_k m_k} \nabla_C Y_{\ell_j m_j} \nabla_B Y_{\ell_n m_n}\\
    C^{(4)} ~ &\coloneqq ~ \int \mathrm{d} \Omega \, Y_{\ell_1 m_1} Y_{\ell_2 m_2} Y_{\ell_3 m_3} Y_{\ell_4 m_4} \\
    \tilde{C}^{(4)} ~ &\coloneqq ~ \int \mathrm{d} \Omega \, \gamma^{A B} Y_{\ell_1 m_1} Y_{\ell_2 m_2} \partial_A Y_{\ell_3 m_3} \partial_B Y_{\ell_4 m_4} \, .
\end{align}
\end{subequations}
Substituting the Moncrief variable, $\chi^{\ell_3 m_3}_a = \epsilon_{ab}\partial^b \Psi_{\ell_3 m_3}$, and integrating by parts in the two-dimensional plane, the second-derivative terms drop out by the antisymmetry of $\epsilon^{ab}$ and the vertex takes the compact form
\begin{align}\label{eqn:Psivertex}
    S_{\text{int.}} ~ &= ~ - 2 i q \sum_{\substack{\ell_i m_i \\ i \in \left(1, 2, 3\right)}} \int \mathrm{d} r^* \, \mathrm{d} w \; \Psi_{\ell_3 m_3} \; \epsilon^{ab} \, \partial_a \phi^*_{\ell_1 m_1} \, \partial_b \phi_{\ell_2 m_2} \; C^{(3)} \, .
\end{align}
The discarded term is a total derivative in the $\left(w, r^*\right)$ plane. Its $w$-endpoints vanish with the adiabatic switching of the interaction, and with $\epsilon_{w r^*} = 1$ its $r^*$-endpoints are
\begin{align}\label{eqn:PsivertexBoundary}
    S_\partial ~ = ~ i q \sum_{\substack{\ell_i m_i \\ i \in \left(1, 2, 3\right)}} C^{(3)} \int \mathrm{d} w \, \Big[ \Psi_{\ell_3 m_3} \left( \phi^*_{\ell_1 m_1} \partial_w \phi_{\ell_2 m_2} - \phi_{\ell_2 m_2} \partial_w \phi^*_{\ell_1 m_1} \right) \Big]^{r^* \rightarrow + \infty}_{r^* \rightarrow - \infty} \, ,
\end{align}
evaluated on the external wavefunctions and the internal propagator of the diagram in which the vertex is inserted. With plane-wave external states and the momentum-space propagator, \eqref{eqn:PsivertexBoundary} is the oscillatory surface term at $r^* \rightarrow \pm\infty$ that the $i\epsilon$ of \eqref{eqn:bhscalarprop} sets to zero, so that \eqref{eqn:chiavertex} and \eqref{eqn:Psivertex} give the same Feynman rule; in momentum space the two differ by the term proportional to $\epsilon^{ab} k_a k_b$, which vanishes identically.

Inserting the mode expansion \eqref{appeqn:PsiQuantisation}, we may read off the Feynman rule for this vertex as
\begin{align}\label{eqn:vertex}
    - i \mathcal{V}^{(\sigma, \sigma_{p_n})} ~ = ~ \dfrac{ - 2 i q}{\sqrt{2 \tilde{\omega} \ell \left(\ell + 1\right)}} \, \epsilon^{ab} \, p_{a, n} \, p_{b, j} \, C^{(3)} ~ = ~ \dfrac{2 i q}{\sqrt{2 \tilde{\omega} \ell \left(\ell + 1\right)}} \, \varepsilon_{(\sigma)}\left(\omega\right) \cdot p_{n} \, C^{(3)} \, , 
\end{align}
where we inserted the momentum conserving delta function to write $p_{a, j} = p_{a, n} + k_a$, with $p_j$ and $p_n$ the scalars participating in the interaction, and labelled the photon momentum as $k_a = \left(-\omega, k_{(\sigma)}\right)$, determined by the mass-shell condition \eqref{eqn:photonMassShell}. The outgoing scalar is created by $\phi^*_{\ell_1 m_1}$, whose wavefunction carries $p_n$ in the first slot of $\epsilon^{ab}$, and the internal line is contracted with $\phi_{\ell_2 m_2}$. The second equality uses $\epsilon^{ab}p_{a, n} p_{b, j} = \epsilon^{ab}p_{a, n} k_b = - \varepsilon_{(\sigma)}\left(\omega\right)\cdot p_n$, by \eqref{eqn:polarisation}; transversality needs no separate appeal here, since the term proportional to $\epsilon^{ab}k_a k_b$ vanishes identically. There is of course an analogous vertex for the insertion of the creation operator $b^{(\sigma)\dagger}$ of the photon.

Two conventions are implicit in \eqref{eqn:vertex}. First, $\mathcal{V}^{(\sigma, \sigma_{p_n})}$ is the coefficient obtained by substituting the annihilation part of \eqref{appeqn:PsiQuantisation}, without the factor of $i$ from $e^{i S_{\text{int.}}}$. Second, the emission of an outgoing soft photon contracts instead with $b^{(\sigma)\dagger}$, whose coefficient in \eqref{appeqn:PsiQuantisation} differs by a sign. The Feynman rule for soft emission is therefore $- i \, \mathcal{V}^{(\sigma, \sigma_{p_n})}$, which is pure imaginary and combines with the propagator \eqref{eqn:bhscalarprop} to give a real soft factor, as it must for the amplitude to reproduce the Ward identity \eqref{eqn:WardIdentity}.

Using the mass-shell condition for the scalar 
\begin{align}
    p_{b, n} ~ = ~ \left( - E_{n} \, , \sigma_{p_n} \Omega_{p_n} + \eta^{w r^*} E_{n}\right) \, ,
\end{align}
we can now calculate the numerator in the vertex for a fixed branch of the photon to find
\begin{align}
    \varepsilon_{(\sigma)}\left(\omega\right) \cdot p_{n} ~ &= ~ - \left( \sigma\widetilde\omega E_n - \sigma_{p_n}\omega\Omega_{p_n} \right) \nonumber \\
    &= ~ - \left[ \left(1-\sigma\sigma_{p_n}\right) \sigma \omega E_n - \sigma \frac{E_n \mu_\ell^2}{2\omega} + \sigma_{p_n} \frac{\omega m_{\ell_n}^2}{2 E_n} \right] + \mathcal{O}\left(m^4\right) \, ,
\end{align}
where in the second equality, we expanded the quantities $\tilde\omega$ and $\Omega_{p_n}$ to first sub-leading order in the masses. This is justified so long as $\mu_\ell \ll \omega$ and $m_{\ell_n} \ll E_n$. The former is the soft window \eqref{eqn:softlimit}, inside which the expansion is carried out with the masses retained; they are sent to zero only in the last step, as the collinear configurations below require. The latter is guaranteed in the near-horizon limit because the scalars in question are assumed to be hard particles. 

Similarly, as it will appear in the virtual scalar propagator in the soft theorem, we also compute the quantity
\begin{align}\label{eqn:kdotp}
    k\cdot p_n ~ &= ~ \sigma\sigma_{p_n} \widetilde\omega\Omega_{p_n} - \omega E_n \nonumber \\
    &= ~ \left(\sigma\sigma_{p_n}-1\right)\omega E_n - \sigma\sigma_{p_n} \left( \frac{E_n \mu_\ell^2}{2\omega} + \frac{\omega m_{\ell_n}^2}{2 E_n} \right) + \mathcal{O} \left(m^4\right) \, .
\end{align}
There are two configurations with $\sigma \sigma_{p_n} = 1$ and two terms with $\sigma \sigma_{p_n} = -1$. 

In the soft theorem, we will encounter the ratio of the above quantities. When $\sigma \sigma_{p_n} = -1$, we find
\begin{align}
    \left.\dfrac{\varepsilon_{(\sigma)}\left(\omega\right) \cdot p_{n}}{k \cdot p_n + \frac{1}{2}\left(m^2_{\ell_j} - m^2_{\ell_n} - \mu^2_\ell\right)}\right|_{\sigma \sigma_{p_n} = -1} ~ \approx ~ - \sigma_{p_n} \, ,
\end{align}
where we have dropped sub-leading terms (proportional to the masses). Whereas when $\sigma \sigma_{p_n} = 1$, the hard scalars and the photon are collinear and the apparently leading terms cancel but the sub-leading terms conspire to produce a leading-order result
\begin{align}
    \left.\dfrac{\varepsilon_{(\sigma)}\left(\omega\right) \cdot p_{n}}{k \cdot p_n + \frac{1}{2}\left(m^2_{\ell_j} - m^2_{\ell_n} - \mu^2_\ell\right)}\right|_{\sigma \sigma_{p_n} = +1} ~ \approx ~ - \sigma_{p_n} \, .
\end{align}
To arrive at this equality, we first observe that $\omega m_{\ell_n}^2/2 E_n$ is small compared with $E_n\mu^2_\ell/2\omega$ in the limit $\omega \ll E_n$, the squared masses all being proportional to $f\left(r\right)$. Moreover, in the denominator we see that the masses can all be ignored in comparison to the first term $E_{n} \mu^{2}_{\ell}/2\omega$ which is again dominant in the same limit. Therefore, the ratio is only a ratio of the first terms of the numerator and the denominator which is just the sign $-\sigma_{p_{n}}$.

Therefore, for arbitrary $\sigma \sigma_{p_n}$, we have
\begin{align}\label{appeqn:vertexRatio}
    \dfrac{\varepsilon_{(\sigma)}\left(\omega\right) \cdot p_{n}}{k \cdot p_n + \frac{1}{2}\left(m^2_{\ell_j} - m^2_{\ell_n} - \mu^2_\ell\right)} ~ \approx ~ - \sigma_{p_n} \, .
\end{align}

Two further remarks on the collinear configurations are in order. First, had the masses been dropped in the action, the ratio in \eqref{appeqn:vertexRatio} would be undefined for $\sigma \sigma_{p_n} = +1$: the masses are needed to define the computation on this half of the phase space, and the value they fix it to is independent of them. Second, the computation holds the masses at their values at the radius $r_0$ of the scattering and uses plane waves on an unbounded path, whereas by \eqref{eqn:masses} the masses vary on the scale $R$. For $\sigma \sigma_{p_n} = +1$ the internal line is on shell and moves with the leg, and the $r^*$ integral of the vertex converges along the path of the leg over the length on which the mismatch of wavenumbers, $\Omega_{p_j} - \Omega_{p_n} - \tilde{\omega} \simeq \mu^2_\ell / 2 \omega$, accumulates a phase of order one,
\begin{align}\label{eqn:collinearLength}
    \Delta r^* ~ \sim ~ \dfrac{2 \omega}{\mu^2_\ell\left(r_0\right)} \, .
\end{align}
The frozen masses are therefore justified when $\Delta r^* \ll R$, which together with the window \eqref{eqn:softlimit} is
\begin{align}\label{eqn:frozenMassRegime}
    \mu_\ell\left(r_0\right) ~ \ll ~ \omega ~ \ll ~ R \, \mu^2_\ell\left(r_0\right) \, , \qquad \text{requiring} \qquad R \, \mu_\ell\left(r_0\right) ~ \simeq ~ \sqrt{f\left(r_0\right) \ell\left(\ell+1\right)} ~ \gg ~ 1 \, .
\end{align}
This is a condition on the stretched horizon on which the scattering is placed. It fails as $r_0$ approaches the horizon at fixed $\ell$, and it fails for every $r_0$ in the zero-mode regime $\omega R \ll 1$. On a stretched horizon with $\mu_\ell\left(r_0\right) R \ll 1$, which is where \secref{sec:irwindow} places it, it is empty for every $\omega$. Where it fails, the phase accumulated by a leg that crosses the horizon along its path is $R \mu^2_\ell\left(r_0\right) / 2 \omega \ll 1$, and the momentum-space computation with frozen masses does not control the collinear half of the phase space. There we compute the radiated field from the field equation \eqref{eqn:sourcedEquations} for $\mathcal{E}_{\ell m}$. By \eqref{eqn:GaussFunction} its source is $\Box_\eta G_{\ell m}$. The hard legs are null by \eqref{eqn:scalarMass}, so with the origin of $u$ and $v$ at the process, as in \eqref{eqn:radiatedStep}, they move on the rays $u = 0$ and $v = 0$. The function $G_{\ell m}$ is then constant in each of the four regions into which these two rays divide the plane, and it jumps across each ray by the charges moving on it. With $\Box_\eta = - 4 \, \partial_u \partial_v$, and with $\mu^2_\ell$ neglected near the process,
\begin{equation}\label{eqn:sourceAtProcess}
    \Box_\eta \, \mathcal{E}_{\ell m} ~ = ~ \Box_\eta \, G_{\ell m} ~ = ~ 2 \left(\mathcal{Q}^{\text{out}}_{\ell m} - \mathcal{Q}^{\text{in}}_{\ell m}\right) \delta\left(u\right) \delta\left(v\right) \, ,
\end{equation}
where $\mathcal{Q}^{\text{out}}_{\ell m}$ and $\mathcal{Q}^{\text{in}}_{\ell m}$ are the eigenvalues of \eqref{eqn:DeltaG} on the out- and in-states. A leg that keeps its radial branch does not contribute. The out-field minus the in-field is the retarded minus the advanced solution of this equation,
\begin{equation}\label{eqn:classicalStep}
    \mathcal{E}^{\text{out}}_{\ell m} - \mathcal{E}^{\text{in}}_{\ell m} ~ = ~ - \, \dfrac{\mathcal{Q}^{\text{out}}_{\ell m} - \mathcal{Q}^{\text{in}}_{\ell m}}{4} \, \Big[ \mathrm{sgn}\left(u\right) + \mathrm{sgn}\left(v\right) \Big] \, ,
\end{equation}
which is \eqref{eqn:radiatedStep} per unit hard amplitude. It is a step of the same height on each branch, and the masses do not enter it. The same rays fix the change of $G_{\ell m}$. By \eqref{eqn:GaussFunction}, and with the sign fixed by the eigenvalue quoted below \eqref{eqn:DeltaG}, a charge $q_n$ on the ray $v = 0$ contributes $- q_n Y_{\ell m}\left(\Omega_n\right) \theta\left(-v\right)$ to it, and a charge on the ray $u = 0$ contributes $- q_n Y_{\ell m}\left(\Omega_n\right) \theta\left(u\right)$. With the conservation of the charge at each angle, \eqref{eqn:angularLocality}, the difference between the out- and in-configurations is
\begin{equation}\label{eqn:GaussStep}
    G^{\text{out}}_{\ell m} - G^{\text{in}}_{\ell m} ~ = ~ - \, \dfrac{\mathcal{Q}^{\text{out}}_{\ell m} - \mathcal{Q}^{\text{in}}_{\ell m}}{4} \, \Big[ \mathrm{sgn}\left(u\right) + \mathrm{sgn}\left(v\right) \Big] \, ,
\end{equation}
which is the right-hand side of \eqref{eqn:classicalStep}. Since $\ell\left(\ell+1\right)\Psi_{\ell m} = \mathcal{E}_{\ell m} - G_{\ell m}$ by \eqref{eqn:fluxVariable}, the out- and in-expansions of $\Psi_{\ell m}$ coincide, which gives \eqref{eqn:DeltaPsiConserved}. This is the leading term of a derivation from the classical field equations in the full potential, which we leave to future work.

\subsection{Pure gauge vertex}
In this subsection, we work out the pure-gauge part of the vertex that arises from the $A^+_{\ell m}$ term in \eqref{eqn:totinteraction}. This vertex has two sources: one from the longitudinal term in $A_a$ owing to the equation of motion \eqref{eqn:AaSoln} and the other from the transverse component. The longitudinal term is easy to write down as it has the same structure as \eqref{eqn:chiavertex}:
\begin{align}
    S^{A^+}_{\text{long.}} ~ &= ~ - i q \sum_{\substack{\ell_i, m_i \\ i \in \left(1, 2, 3\right)}} \int \mathrm{d} r^* \, \mathrm{d} w \, \eta^{ab} \, \partial_a A^+_{\ell_3 m_3} \left( \phi^*_{\ell_1 m_1} \partial_b \phi_{\ell_2 m_2} - \phi_{\ell_2 m_2} \partial_b \phi^*_{\ell_1 m_1} \right) \, C^{(3)} \, .
\end{align}
The transverse contribution arises from the last line of \eqref{eqn:totinteraction}
\begin{align}
    S^{A^+}_{\text{trans.}} ~ &= ~ - i q \sum_{\substack{\ell_i, m_i \\ i \in \left(1, 2, 3\right)}} \int \mathrm{d} r^* \, \mathrm{d} w \, \dfrac{f\left(r\right)}{r^2} A^+_{\ell_3 m_3} \left(\phi^*_{\ell_1 m_1} \phi_{\ell_2 m_2}\, \tilde{C}^{(3)}_{23} - \phi_{\ell_1 m_1} \phi^*_{\ell_2 m_2} \, \tilde{C}^{(3)}_{23} \right) \nonumber \\ 
    &= ~ - i q \sum_{\substack{\ell_i, m_i \\ i \in \left(1, 2, 3\right)}} \int \mathrm{d} r^* \, \mathrm{d} w \, \dfrac{f\left(r\right)}{r^2} A^+_{\ell_3 m_3} \phi^*_{\ell_1 m_1} \phi_{\ell_2 m_2}\, \left(\tilde{C}^{(3)}_{23} - \tilde{C}^{(3)}_{13} \right) \, ,
\end{align}
where in the second line, we interchanged the dummy summed labels $\ell_1 , m_1$ and $\ell_2 , m_2$. Next, we evaluate the angular integral:
\begin{align}
    2 \tilde{C}^{(3)}_{j n} ~ &= ~ 2 \int \mathrm{d}\Omega \gamma^{AB} Y_{\ell_k m_k} \nabla_A Y_{\ell_j m_j} \nabla_B Y_{\ell_n m_n} \nonumber \\
    &= ~ \int \mathrm{d}\Omega \gamma^{AB} Y_{\ell_k m_k} \nabla_A \nabla_B \left(Y_{\ell_j m_j} Y_{\ell_n m_n}\right) - \int \mathrm{d}\Omega \gamma^{AB} Y_{\ell_k m_k} \left(\nabla_A \nabla_B Y_{\ell_j m_j}\right) Y_{\ell_n m_n} \nonumber \\
    &\qquad- \int \mathrm{d}\Omega \gamma^{AB} Y_{\ell_k m_k} Y_{\ell_j m_j} \left(\nabla_A \nabla_B  Y_{\ell_n m_n}\right) \nonumber \\
    &= ~ \int \mathrm{d}\Omega \gamma^{AB} \nabla_A \nabla_B \left( Y_{\ell_k m_k} \right) Y_{\ell_j m_j} Y_{\ell_n m_n} - \int \mathrm{d}\Omega \gamma^{AB} Y_{\ell_k m_k} \left(\nabla_A \nabla_B Y_{\ell_j m_j}\right) Y_{\ell_n m_n} \nonumber \\
    &\qquad- \int \mathrm{d}\Omega \gamma^{AB} Y_{\ell_k m_k} Y_{\ell_j m_j} \left(\nabla_A \nabla_B  Y_{\ell_n m_n}\right) \nonumber \\
    &= ~ \Big[ - \ell_k \left(\ell_k + 1\right) + \ell_j \left(\ell_j + 1\right) + \ell_n \left(\ell_n + 1\right) \Big] C^{(3)} \, ,
\end{align}
where we used the product rule in the second equality, partial integration in the first term of the third equality, and carried out the integral in the last equality. 

Plugging this result back into the transverse contribution gives 
\begin{align}
    S^{A^+}_{\text{trans.}} &= - i q \sum_{\substack{\ell_i, m_i \\ i \in \left(1, 2, 3\right)}} \int \mathrm{d} r^* \, \mathrm{d} w \, \dfrac{f\left(r\right)}{r^2} A^+_{\ell_3 m_3} \phi^*_{\ell_1 m_1} \phi_{\ell_2 m_2}\, \Big[ \ell_2 \left(\ell_2 + 1\right) - \ell_1 \left(\ell_1 + 1\right) \Big] C^{(3)}\, ,
\end{align}
Together with the longitudinal contribution, we may then read off the vertex as
\begin{align}\label{eqn:A+vertex}
    \mathcal{V}_{A^+ \phi^* \phi} ~ &= ~ - i q \Bigg( k \cdot \left(2 p_{n} + k\right) + \dfrac{f\left(r\right)}{R^2} \Big[ \ell_2 \left(\ell_2 + 1\right) - \ell_1 \left(\ell_1 + 1\right) \Big] \Bigg) C^{(3)} \nonumber \\
    &= ~ - i q \Big( k \cdot \left(2 p_{n} + k\right) + m^2_{\ell_j} - m^2_{\ell_n} \Big) C^{(3)} \, ,
\end{align}
where in the second line, we used the relation between the scalar masses and their partial wave number \eqref{eqn:scalarMass}. Using $p_j = p_n + k$, the expression in the brackets can be rearranged as $\left(p^2_j + m^2_{\ell_j} \right) - \left(p^2_n + m^2_{\ell_n}\right)$ which is precisely what a pure-gauge vertex must be, in order for the pure-gauge mode $A^+_{\ell m}$ to decouple on-shell. For $\ell = 0$ there is no $A^{+}_{00}$. The pure-gauge configuration is $A^{00}_a = \partial_a \lambda_{00}$, the transverse contribution is absent, and \eqref{eqn:A+vertex} holds with $\lambda_{00}$ in place of $A^{+}_{00}$ and $\ell_j = \ell_n$.

\subsection{Magnetic interactions}\label{app:magneticsofttheorem}
As is evident from the action \eqref{eqn:totinteraction}, the odd-sector interactions arise purely from the transverse contribution:
\begin{align}
    S_{A^-} ~ &= ~ - \int \mathrm{d} r^* \, \mathrm{d} w \, \mathrm{d}\Omega \, \Biggl( f\left(r\right) \, \gamma^{AB} A_A \Bigl[ i q \left( \tilde{\phi}^* \nabla_B \tilde{\phi} - \tilde{\phi} \nabla_B \tilde{\phi}^* \right) + q^2 A_B \left|\tilde{\phi}\right|^2 \Bigr]\Biggl) \nonumber \\
    &= ~ i q \sum_{\substack{\ell_i m_i \\ i \in \{1, 2, 3\}}} \int \mathrm{d} r^* \, \mathrm{d} w \, \dfrac{f\left(r\right)}{r^2} \, A^-_{\ell_3 m_3} \phi^*_{\ell_1 m_1} \phi_{\ell_2 m_2} \left( \hat{C}^{(3)}_{23} - \hat{C}^{(3)}_{13} \right) \, .
\end{align}
We see that the vertex is therefore proportional to $f\left(r\right)$ and appears to vanish in the near-horizon limit. However, in the soft theorem, the relevant contribution contains a propagator of the virtual scalar leg in the denominator. Therefore, the relevant combination is
\begin{align}
    \dfrac{f\left(r\right)}{R^2 \, k \cdot p} ~ &= ~ \dfrac{f\left(r\right)}{R^2 \left[\left(\sigma\sigma_{p_{n}}-1\right)\omega E_n - \sigma\sigma_{p_{n}} \left( \frac{E_n \mu_{\ell}^2}{2\omega} + \frac{\omega m_{\ell}^2}{2E_n} \right) + \mathcal{O}\left(m^4\right)\right]}
\end{align}
where we used \eqref{eqn:kdotp}. It is evident that this contribution vanishes when the photon and the hard scalars are not collinear $\sigma \sigma_{p_n} = -1$. When they are collinear, on the other hand, the first term in the denominator vanishes. We then note that the squared mass is proportional to $f\left(r\right)$. Now, we employ the soft limit \eqref{eqn:softlimit}. In this limit, the dominant term is 
\begin{align}
    \left. \dfrac{f\left(r\right)}{R^2 \, k \cdot p} \right|_{\sigma \sigma_{p_{n}} = 1} ~ &\sim ~ \dfrac{\omega}{E_n} \, ,
\end{align}
which again vanishes in the limit. This estimate uses the frozen dispersion. Outside the regime \eqref{eqn:frozenMassRegime} the $r^*$ integral of the vertex along the path of the leg is again at most of this order, since the vertex carries $f\left(r^*\right)$ and no derivative, and since the odd sector has no gauge freedom by \eqref{eqn:gaugetransfo}, there is no boundary mode and no field carried by the external states that could supply what the vertex does not. The magnetic soft factor is therefore of order $\omega/E_n\ell\left(\ell+1\right)$ for a crossing leg, and for a leg along the horizon with $\omega R \ll \ell\left(\ell+1\right)$, and of order $1/E_n R$ for a leg along the horizon with $\omega R \gtrsim \ell\left(\ell+1\right)$, which leaves the near-horizon region before the collinear phase accumulates. It vanishes in the near-horizon soft limit, and this is what we mean by the vanishing of the magnetic soft theorem.

\printbibliography

\end{document}